%% file: main.tex
\documentclass[%
reprint,
longbibliography,
amsmath,amssymb,
aps,
]{revtex4-2}

\usepackage{graphicx}
\usepackage{dcolumn}
\usepackage{bm}

\usepackage[colorlinks=true,allcolors=blue,bookmarks]{hyperref}

\usepackage{dutchcal} 

\usepackage{filecontents}

\usepackage{orcidlink}

\usepackage{ulem} 

\newcommand{\HSout}[1]{\ifmmode\text{\color{red}\sout{\ensuremath{#1}}}\else{\color{red}\sout{#1}}\fi}

\newcommand{\HS}[1]{\ifmmode\text{\color{red}{\ensuremath{#1}}}\else{\color{red}~\textsf{#1}}\fi} 

\newcommand{\pd}{\partial}
\newcommand{\R}{\mathcal{R}} 
\newcommand{\rr}{\mathcal{r}} 
\newcommand{\sigmac}{\sigma_{\! c}} 
\newcommand{\Rm}{R_m} 
\newcommand{\UniverseMachine}{UniverseMachine }

\newcommand\aap{Astron.~Astrophys.}
\newcommand\aj{Astron.~J.}
\newcommand\apjl{Astrophys.~J.~Lett.}
\newcommand\apss{Astrophys.~Space~Sci.}
\newcommand\mnras{Mon.~Not.~R.~Astron.~Sco.}
\newcommand\jcap{J.~Cosmo.~Astropart.~Phys.}
\newcommand\sovast{Sov.~Astron.}

\begin{document}


\title{What triggers type Ia supernovae: Prompt detonations from primordial black holes or companion stars?}

\author{H. Steigerwald \orcidlink{0000-0001-6747-7389}}
\email{heinrich@steigerwald.name}
\affiliation{PPGCosmo \& Cosmo-ufes, Universidade Federal do Esp\'{i}rito Santo, 29075-910, Vit\'{o}ria, ES, Brazil}

\date{May 27, 2025}

\begin{abstract}
We set up and perform collision rate simulations between dark matter in the form of asteroid-mass primordial black holes (PBHs) and white dwarf stars. These encounters trigger prompt detonations and could be the key to solving the ignition mystery of type Ia supernovae. Our framework is flexible enough to cover the full range of progenitor white dwarf masses, host galaxy stellar masses, galactocentric radial offsets, and cosmic time. The rate distribution pattern is consistent with exhaustive literature observational determinations for a slightly extended log-normal PBH mass spectrum. Most strikingly, the so far unexplained brightness distribution comes out without fine-tuning. We find no severe contradictions, except that the inferred PBH mass scale is unpredicted from first principles.
\end{abstract}

\maketitle


\paragraph*{Introduction.}---It is broad consensus that 
type Ia supernovae (SNe Ia) are the explosive burning of electron-degenerate carbon-oxygen white dwarf (WD) stars, where the synthesized amount of radioactive 
$^{56}$Ni
accounts for the peak brightness 
\cite{1960ApJ...132..565H,*1969Ap&SS...5..180A}.
An empirical tight correlation between peak brightness and post-maximum lightcurve decline rate 
\cite{1977SvA....21..675P,*1993ApJ...413L.105P}, also known as the width-luminosity relation, permits to standardize `normal' 
SNe Ia as extragalactic distance indicators 
and formed the cornerstone for 
the discovery of the late-Universe accelerated expansion 
\cite{1998AJ....116.1009Rmax10,*1999ApJ...517..565Pmax10}. 
The ignition is traditionally believed to occur in binary systems, but the exact progenitors as well as the ignition mechanism are open problems, and currently no single published model is able to consistently explain all observational features 
(reviewed in \cite{2023RAA....23h2001L}).

\footnotetext[1]{According to table~1 of \cite{2023RAA....23h2001L}, the DD scenario is the only with straightforward rate distribution predictions that reproduce the observed at order-of-magnitude level. However, 
we caution that the DD scenario is not the current standard model of SN Ia ignition, as there is no consensus in the community. Other scenarios explain certain observational aspects better than the DD scenario.
}

In recent years, with improving SN Ia statistics, rate distributions have come up as benchmark tests for progenitor scenarios.
Notably, observations of the delay time distribution (DTD)---i.e.~rate per solar mass of stars formed as a function of delay time since a single hypothetical star formation burst 
\cite{2008MNRAS.383.1121M,2018MNRAS.479.3563F}---have almost ruled out non-degenerate companion stars to trigger the ignition, even though these could still ignite the underluminous subclass of SNe Iax 
\cite{2023Natur.615..605G}.
So far, only double-degenerate scenarios where  the primary, sub-Chandrasekhar WD explodes in a prompt detonation (hereafter simply DD scenario),
have been shown to reproduce the width-luminosity relation and to cover the observed ranges of $^{56}$Ni masses and delay times 
\cite{2023RAA....23h2001L}. 
Yet, there are not enough DD systems to account for the observed rate \cite{2011MNRAS.417..408R,2012A&A...546A..70T} 
and, in addition, the $^{56}$Ni-mass distribution (NMD)---i.e. volumetric SN Ia rate as a function of $^{56}$Ni mass synthesized in the explosion 
\cite{2022MNRAS.509.5275S,2014MNRAS.438.3456P}
---is skewed towards dim events 
\cite{2022MNRAS.515..286G}, while observations suggest a mean of $0.6M_{\odot}$ 
\cite{2022MNRAS.509.5275S}.

Prompt detonations can also be triggered in solitary sub-Chandrasekhar WDs by the passage of a light compact object such as a primordial black hole (PBH) 
\cite{2015PhRvD..92f3007G,2019JCAP...08..031M,2021PhRvL.127a1101S}. This would explain the general absence of binary companion signatures \cite{2015Natur.521..332O,*2022ApJ...933L..31S}.
Initial studies \cite{2015PhRvD..92f3007G,2019JCAP...08..031M} focused on subsonic runaway ignition (deflagration) which suffers quenching from Kelvin-Helmholtz instabilities \cite{2019JCAP...08..031M} and the outcome does not reproduce SNe Ia. However, our reanalysis \cite{2021PhRvL.127a1101S} showed that nuclear burning starts supersonically right out of hydrostatic equilibrium (direct detonation) which bypasses potential quenching from instabilities and reproduces the observed spectra and lightcurves 
\cite{2010ApJ...714L..52S,*2017MNRAS.470..157B,2018ApJ...854...52S,2021ApJ...909L..18S} and line polarization 
\cite{2022MNRAS.511.2994L} of `normal' SNe Ia (see, however, 
 \cite{2020MNRAS.499.4725K}).
The existence of PBHs is motivated to play the role of dark matter (DM) and 
the currently unconstrained asteroid-mass window, $10^{-16}M_{\odot}\lesssim m_{\bullet} \lesssim  
10^{-10}M_{\odot}$
\cite{2020ARNPS..70..355C,*2021RPPh...84k6902C,*2020PhRvD.101f3005S,*2023JCAP...05..054K}, allows for collisions with WDs to account for the Milky-Way SN Ia rate and average brightness on order-of-magnitude level \cite{2021PhRvL.127a1101S}, but detailed rate distributions are so far unknown.

In this Letter, we perform, for the first time, full-scale collision rate simulations across WD masses, host galaxy stellar masses, galactocentric radial offsets, and cosmic time. 
For this purpose, we developed a framework to model star formation histories (SFHs) in galactocentric radial zones (companion paper \cite{2025arXiv250521260S}), which is essential because the encounter rate scales with DM density and hence with galactocentric radial offset. We compare our rate predictions alongside those of the DD scenario with exhaustive observational data compilations from the literature.
We work in cosmology with $H_0=70$~km~s$^{-1}$~Mpc$^{-3}$, $\Omega_m=0.3$, and $\Omega_{\Lambda}=0.7$.

\paragraph*{Simulation setup.}---Let $[\pd^2\!N(t;m_w,r,M_{\star})/\pd m_w\pd r]$ $\times dm_w dr$ be the number of WDs with mass in the range $m_w\to m_w\!+\!dm_w$ within a galactocentric spherical shell delimited by radii $r\to r\!+\!dr$ in a model galaxy with present-day stellar mass $M_{\star}$ at cosmic time $t$ \cite{2025arXiv250521260S}. In the SN Ia progenitor mass range ($0.9M_{\odot}\lesssim m_w\lesssim 1.1M_{\odot}$), the evolution equation maybe setup as 
\begin{align}\label{eq:white-dwarf-evolution}
    \frac{d}{dt}\Big(\frac{\pd^2\!N}{\pd m_w \pd r}\Big)+\Gamma_{\rm Ia}(t)\frac{\pd^2\!N}{\pd m_w \pd r}(t) = f(t)\,, \end{align}
where the right-hand side term accounts for WD formation (given by eq.~(A.23) in \cite{2025arXiv250521260S}) and the second term on the left-hand side 
for WD destruction,
where $\Gamma_{\!\rm Ia}(t) = \Gamma_{\!\rm Ia}(t;m_w,r,M_{\star})$ is the average per WD collision rate with PBHs that leads to detonation ignition (we omit parameter dependencies on $m_w$, $r$, and $M_{\star}$ whenever obvious for clarity),
\begin{align}\label{eq:encounter-rate}
    \Gamma_{\!\rm Ia}(t) = \frac{\pi\;\!\rho_{\bullet}(t)}{v_{\bullet w}(t)} \!\int\! \frac{\Rm^2\;\!v(\Rm)^2\;\! \varphi(m_{\bullet})}{m_{\bullet}} \, dm_{\bullet}\,,
\end{align}
where $\rho_{\bullet}(t)=\rho_{\bullet}(t;r,M_{\star})$ is the DM density at radial offset $r$ in a galaxy with present-day stellar mass $M_{\star}$ at cosmic time $t$, $v_{\bullet w}(t)= v_{\bullet w}(t;r,M_{\star})$ is the corresponding mean relative encounter velocity between WDs and PBHs far from each other,  $\Rm=\Rm(m_w,m_{\bullet},X_{\rm C})$ is the ignition cross section radius \cite{2021PhRvL.127a1101S},
$v(\Rm)$ is the PBH's velocity inside the WD at $\Rm$ (see, e.g., eq.~(15) of \cite{2021PhRvL.127a1101S}),  and $\varphi(m_{\bullet})$ is the PBH mass function. 

\footnotetext[10]{See Supplemental Material (after the bibliography).}

We assume the universal NFW halo profile and isotropic halo velocity (see, e.g., appendices B and C of  \cite{2022MNRAS.510.4779S}), where halo masses 
are calculated based on stellar mass histories (see \S~III-1 of \cite{2025arXiv250521260S}) and abundance matching stellar-to-halo mass relation 
\cite{2010ApJ...710..903M,*2020A&A...634A.135G}. We note that the effect of the cusp-core uncertainty on the DM-induced SNe Ia rate in a cosmological volume is only $\sim 3\%$ \cite{2022MNRAS.510.4779S} and can be neglected here. In addition, while there is a $\sim 50\%$ scatter on the stellar-to-halo mass relation \cite{2010ApJ...710..903M,*2020A&A...634A.135G}, the net effect on the SN Ia rate in a large cosmological volume encompassing many galaxies is only a few per cent and may be neglected as well.
We adopt the semi-analytical ignition cross section of \cite{2021PhRvL.127a1101S}, 
slightly improved with more realistic input physics (see \S~SI of \cite{Note10} for details), 
notably (i) allowing for detonation velocity triggering threshold at quasi-nuclear statistical equilibrium 
\cite{2014ApJ...785...61S}, 
(ii) adopting ignition efficiency $\alpha=5$  
\cite{2021PhRvL.127a1101S,1992ApJ...396..649T} and (iii) accounting for the WD composition transition from CO to ONe in the range $1.08M_{\odot}\lesssim m_w\lesssim 1.12M_{\odot}$ 
\cite{2013ApJ...779...58R,2018MNRAS.480.1547L,2023ApJ...950..115B}.
We discuss possible variations of the normalization of $\Gamma_{\rm Ia}$ due to systematic uncertainties of $\alpha$ and the overall number density of PBHs later on.
Finally, we allow for an extended PBH mass spectrum, with log-normal density 
\cite{1993PhRvD..47.4244D}
\begin{align}\label{eq:lognormal}
    \varphi(m_{\bullet}) = \frac{1}{\sqrt{2\pi}\;\!\sigmac\;\!m_{\bullet}}\exp\!\Big[\!-\!\frac{\ln(m_{\bullet}/m_c)^2}{2\sigmac^2}\Big]\,,
\end{align}
where $m_c$ is the peak mass of $m_{\bullet}\varphi(m_{\bullet})$ 
and $\sigmac$ its width, which is generic for slow-roll inflationary models 
\cite{2016PhRvD..94f3530G,*2017JCAP...09..020K} and otherwise a sufficient approximation for most smooth single-peaked distributions such as critical collapse where $\sigmac\simeq 0.26$ 
\cite{2017PhRvD..96b3514C}. 

The simplicity of eq.~\eqref{eq:white-dwarf-evolution} requires some justification. Notably, we have neglected the flow of WDs between adjacent radial zones. This assumption is valid whenever the radial migration timescale, $\tau_{\rm mig}$, is much longer than the SN Ia ignition delay time, $\tau_{\rm Ia} \sim \Gamma_{\!\rm Ia}^{-1}$. 
For the Solar System environment, which is typical, $\Gamma_{\!\rm Ia}^{-1} \simeq 0.3~$Gyr and $\tau_{\rm mig} \simeq 8$~Gyr~$(\delta r_{\rm mig}/3.6~$kpc$)^2$ 
\cite{2018ApJ...865...96F}, where $\delta r_{\rm mig}$ is the radial migration distance. Thus, for migrations of the order of the 
local DM density scale height (which is $\simeq 7$~kpc)
the migration timescale is much longer than the Hubble time. Therefore, the approximation holds for most parts,
but breaks down 
in the outskirts of galaxies beyond a few half-mass radii, $R_{1/2}$, because (i) $\tau_{\rm Ia}$ scales inversely with DM density, (ii) net outward migration of stellar populations \citep{2021MNRAS.508.4484J} and contributions from dwarf satellite galaxies become relevant.  Since most of SNe Ia explode within a few $R_{1/2}$ 
\cite{2017ApJ...836...60Lmax10,2018MNRAS.481.2766H,2020ApJ...905...58Dmax10}, we expect globally accurate predictions, but will underestimate rates at large galactocentric offsets.

\begin{figure*}[htb]
    \centering
    \includegraphics[width=\linewidth]{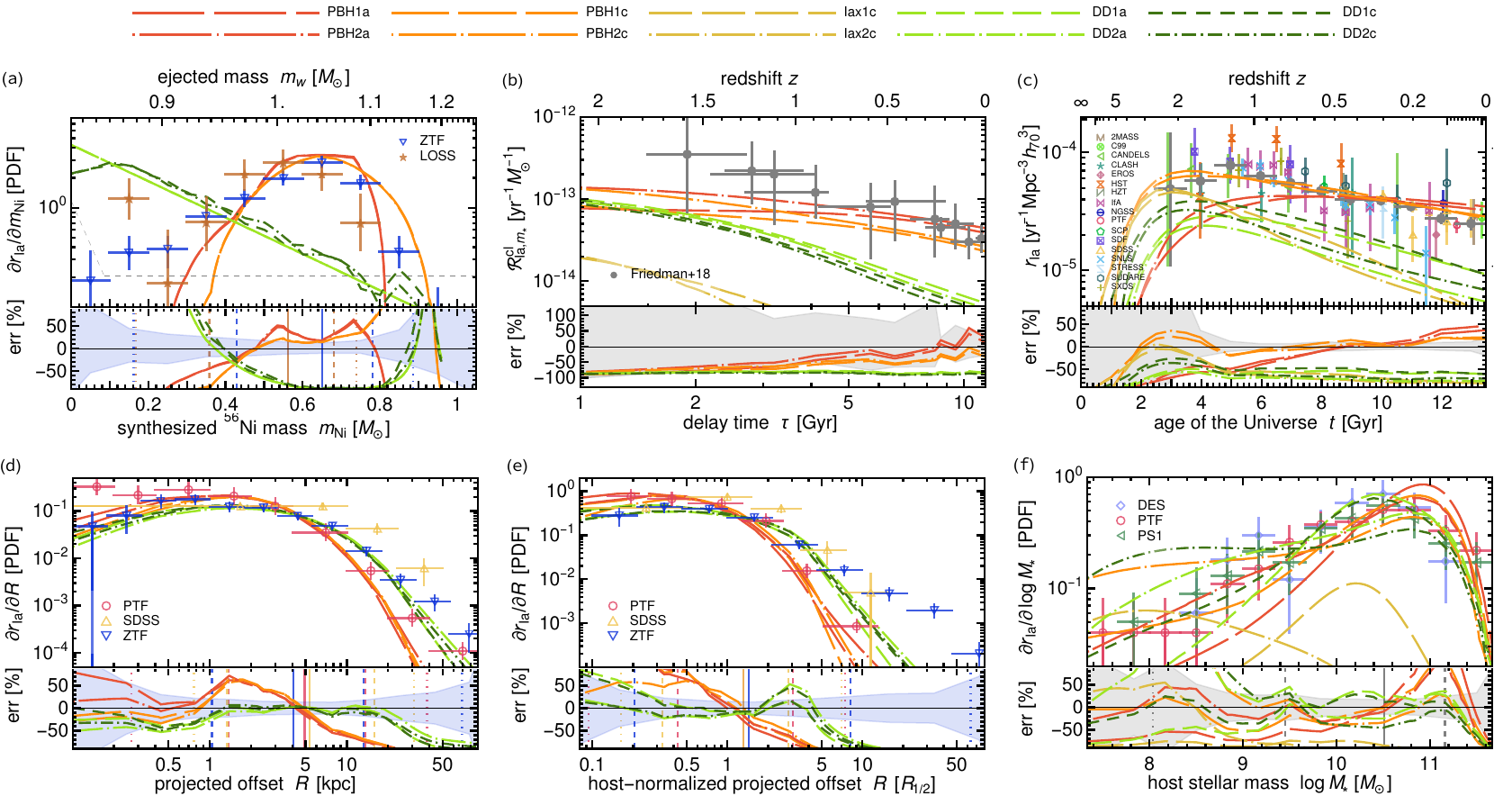}
    \caption{SN Ia rate distribution predictions for PBH ignition (red-orange curves) and for DD ignition (green curves) with possible contributions from SNe Iax (yellow curves) for a selection of modeling assumptions (see table~\ref{tab:models}) compared to observational data points (see \S~SII of \cite{Note10}
    for details).     
    Mini bottom panels show relative errors (same line styles) with respect to the data mean and compared to 1$\sigma$ statistical errors of the data 
    (light blue for ZTF, light gray for surveys combined);
    vertical full, dashed and dotted lines indicate data median, 16$^{\rm th}$ and 84$^{\rm th}$ percentiles, and 2$^{\rm nd}$ and 98$^{\rm th}$ percentiles, respectively (gray if binned, colored if for each survey). (a) NMD, (b) DTD, (c) rate history, (d)--(e) host offset distributions, (f) host-mass distribution.
    }
    \label{fig:1}
\end{figure*}

\paragraph*{Rate distributions.}---Time-integrating eq.~\eqref{eq:white-dwarf-evolution}, the second term on the left-hand side is 
\begin{align}\label{eq:differential-rate}
    \frac{\pd^2\R_{\rm Ia}}{\pd m_w\;\!\pd r}(t) = 
    \Gamma_{\rm Ia}(t)\;\! \exp\!\Big[\!-\!\!\int_0^t\!\!\Gamma_{\!\rm Ia}(t')\;\!dt'\Big]\nonumber \\
    \times \int_0^t\!\! f(t')\;\! \exp\!\Big[\int_0^{t'}\!\!\!\Gamma_{\!\rm Ia}(t'')\;\!dt''\Big]\;\!dt'\,,
\end{align}
where $[\pd^2\R_{\rm Ia}(t;m_w,r,M_{\star})/\pd m_w\pd r]dm_w dr$ represents the SN Ia rate history with explosion mass in the range $m_w\to m_w\!+\!dm_w$ in a radial shell within $r\to r\!+\!dr$ in a galaxy with present-day stellar mass $M_{\star}$.
We calculate the NMD history by integrating eq.~\eqref{eq:differential-rate} further over radial shells and over the galaxy stellar mass function, $\phi(M_{\star})$, 
\begin{align}\label{eq:NMD}
    &\frac{\pd\rr_{\rm Ia}}{\pd m_{\rm Ni}}(t;m_{\rm Ni}) = \frac{dm_w}{dm_{\rm Ni}} \!\iint\!\! \frac{\pd^2\R_{\rm Ia}}{\pd m_w\;\!\pd r}\;\!\phi(M_{\star})\;\!dM_{\star}\,dr\,,
\end{align}
where $\phi(M_{\star})\;\!dM_{\star}$ is the number density of galaxies with stellar mass in the range $M_{\star}\to M_{\star}\!+dM_{\star}$ (see \cite{2025arXiv250521260S} \S~III-3 for details), and $m_{\rm Ni}=-5.0335+8.4376\;\!m_w-2.86945\;\!m_w^2$ fits the $^{56}$Ni yield for given WD mass of hydrodynamical simulation results 
\cite{2018ApJ...854...52S,2015A&A...580A.118M} with core carbon mass fraction $X_{\rm C}=0.3$ (see 
\S~SI of \cite{Note10} for details). The NMD, shown in Fig.~\ref{fig:1}a compared to data at redshift $0.025$ 
\cite{2022MNRAS.509.5275S},  corresponds to the `snapshot' $t=13.1$~Gyr of eq.~\eqref{eq:NMD},
normalized such that $\int(\pd \rr_{\rm Ia}/\pd m_{\rm Ni})\;\!dm_{\rm Ni}=1$. 

In order to calculate the remaining rate distributions shown in Fig.~\ref{fig:1}b--f, we first calculate two intermediate quantities, the three-dimensional offset distribution history per galaxy with present-day stellar mass $M_{\star}$, from integrating eq.~\eqref{eq:differential-rate} over WD masses,
\begin{align}\label{eq:galaxy-shell-rate}
    \frac{\pd\R_{\rm Ia}}{\pd r}(t;r,M_{\star}) = \int \!\!\frac{\pd^2\R_{\rm Ia}}{\pd m_w\;\!\pd r}(t;m_w,r,M_{\star})\,dm_w\,,
\end{align}
and the total rate history per galaxy with present-day stellar mass $M_{\star}$, from integrating eq.~\eqref{eq:galaxy-shell-rate} over radial shells,
\begin{align}\label{eq:galaxy-rate}
    \R_{\rm Ia}(t;M_{\star}) = \int \frac{\pd \R_{\rm Ia}}{\pd r}(t;r,M_{\star})\;\!dr\,.
\end{align}

We calculate the rate history in massive early-type cluster galaxies in units of stellar mass formed,  $\R_{\rm Ia,m_{\star}}^{\rm cl}$, shown in Fig.~\ref{fig:1}b compared to data from \cite{2018MNRAS.479.3563F}, by averaging the galaxy rates of eq.~\eqref{eq:galaxy-rate} over the cluster galaxy mass function, $\varphi_{\rm cl}(M_{\star})$,
\begin{align}\label{eq:DTD}
    \R_{\rm Ia,m_{\star}}^{\rm cl}(t) = \!\int\!\!\frac{\R_{\rm Ia}(t;M_{\star})}{\int_0^t\!\Psi(t';M_{\star})\;\!dt'}\;\!\varphi_{\rm cl}(M_{\star})\;\!dM_{\star}\,,
\end{align}
where $\Psi(t;M_{\star})$ is the average SFH of galaxies with present-day stellar mass $M_{\star}$ (see \cite{2025arXiv250521260S} \S~III-2), and $\varphi_{\rm cl}(M_{\star})\;\!dM_{\star}$ represents the fraction of cluster galaxies with mass in the range $M_{\star}\to M_{\star}\!+\!dM_{\star}$. 
Since these galaxies form their stars in a relatively short burst at  $t_{\rm cl}\simeq 2.11~$Gyr (or $z \simeq 3$  
\cite{2018MNRAS.479.3563F}), $\R_{\rm Ia,m_{\star}}^{\rm cl}\!(t_{\rm cl}\!+\!\tau)$ is usually considered a proxy of the true underlying DTD, $\Phi_{\rm Ia}(\tau)$.
We assume a log-normal distribution  for $\varphi_{\rm cl}(M_{\star})$, defined as in eq.~\eqref{eq:lognormal},
with peak $\log M_{\rm cl}/M_{\odot} = 11.2$ and width $\sigma_{\rm cl}=0.2$,
which fits the population of low-redshift cluster galaxies 
\cite{2008MNRAS.383.1121M}.

The total volumetric rate history, shown in Fig.~\ref{fig:1}c compared to a literature data compilation taken from 
\cite{2020ApJ...890..140S}, is given by
\begin{align}\label{eq:VRH}
    \rr_{\rm Ia}(t) = \int \!\!\R_{\rm Ia}(t;M_{\star})\;\!\phi(M_{\star})\,dM_{\star}\,.
\end{align}

In order to derive the host offset distribution---i.e. volumetric rate as a function of sky-projected offset from its host galaxy 
\cite{2017ApJ...836...60Lmax10,2018MNRAS.481.2766H,2020ApJ...905...58Dmax10}---, we first generate, for given $M_{\star}$ and $t$, three-dimensional random offsets $r_j=\{r_1,r_2,r_3,\ldots\}$ with distribution specified by eq.~\eqref{eq:galaxy-shell-rate}; then, we define associated projected offsets $R_j=(r_j^2-z_j^2)^{1/2}$ where $z_j\in[-r_j,r_j]$ is a random number representing the line-of-sight component of the three-dimensional offset, and 
bin the projected offsets $R_j=\{R_1,R_2,R_3,\ldots\}$ to obtain the projected offset distribution history $\pd\R_{\rm Ia}(t;R,M_{\star})/\pd R$ for a given present-day stellar mass $M_{\star}$. Finally, integrating over the galaxy stellar mass function $\phi(M_{\star})$, we have 
\begin{align}\label{eq:HOD}
    \frac{\pd\rr_{\rm Ia}}{\pd R}(t;R) = \int\! \frac{\pd\R_{\rm Ia}}{\pd R}(t;R,M_{\star}) \;\!\phi(M_{\star})\,dM_{\star}\,,
\end{align}
of which the `snapshot' $t= 13.1$~Gyr [normalized such that $\int (\pd \rr_{\rm Ia}/\pd R)\;\!dR=1$]
is shown in Fig.~\ref{fig:1}d compared to data at effective survey redshift $z\simeq 0.025$ 
\cite{2017ApJ...836...60Lmax10,2018MNRAS.481.2766H,2020ApJ...905...58Dmax10}.

Defining $\tilde{R} \equiv R/R_{\rm 1/2}$, where $R_{1/2}=R_{1/2}(t;M_{\star})$ is the circularized (projected) half-mass radius history of a galaxy with present-day stellar mass $M_{\star}$ (see eq.~(8) of \cite{2025arXiv250521260S}), we have $dR/d\tilde{R} = R_{1/2}$; thus, the projected host offset distribution history in units of galaxy half-mass radii, is given by
\begin{align}\label{eq:HND}
    \frac{\pd\rr_{\rm Ia}}{\pd\tilde{R}}(t;\tilde{R}) = \!\int\! \!\frac{\pd\R_{\rm Ia}}{\pd R}(t;R,M_{\star})\;\!R_{1/2}(t;M_{\star})\;\!\phi(M_{\star})\,dM_{\star}\,,
\end{align}
of which the `snapshot' $t\simeq 13.1$~Gyr (normalized to unity) is shown in Fig.~\ref{fig:1}e compared to data at effective survey redshift $z\simeq 0.025$ 
\cite{2017ApJ...836...60Lmax10,2018MNRAS.481.2766H,2020ApJ...905...58Dmax10}.

Finally, we also have the volumetric host-mass distribution history, 
\begin{align}\label{eq:HMD}
    \frac{\pd\rr_{\rm Ia}}{\pd\log M_{\star}}(t;\log M_{\star}) = \R_{\rm Ia}(t;M_{\star})\;\!\phi(M_{\star})\;\!M_{\star}\,, 
\end{align}
of which the `snapshot' $t=12.2$ [normalized such that $\int (\pd\rr_{\rm Ia}/\pd\log M_{\star})\;\!d\log M_{\star}=1$], 
is shown in Fig.~\ref{fig:1}f, compared to data at effective survey redshift $z=0.1$ 
\cite{2020MNRAS.495.4040Wmax10}.

\footnotetext[3]{We note that the reason we can't apply the simple DTD formalism of eqs.~\eqref{eq:galaxy-rate-binary} and \eqref{eq:galaxy-shell-rate-binary} to the PBH ignition scenario is that the encounter rate, $\Gamma_{\rm Ia}(t)$, depends on cosmic time. If it didn't, the DTD would be that of a simple Poisson point process with constant rate, delayed by the main-sequence lifetime $\tau_{\star}$ (neglecting here, for the purpose of illustration, possible merger delay times),
$\Phi_{\rm Ia}(\tau) \simeq \Gamma_{\rm Ia} \exp[-\Gamma_{\rm Ia}(\tau\!-\!\tau_{\star})]$, 
for $\tau>\tau_{\star}$ and zero otherwise, as can be verified setting $\Gamma_{\rm Ia}=~$const in 
eq.~\eqref{eq:differential-rate},
and replacing $f(t)$ by eq.~(A.23) of \cite{2025arXiv250521260S}.}

For the DD scenario, the total rate history, equivalent to eq.~\eqref{eq:galaxy-rate},
is given by the convolution of the (true) DTD, $\Phi_{\rm Ia}(\tau)$, with the SFH 
\begin{align}\label{eq:galaxy-rate-binary}
    \R_{\rm Ia}(t;M_{\star}) = \!\int_0^t\!\! \Phi_{\rm Ia}(\tau)\;\!\Psi(t\!-\!\tau;M_{\star})\;\!d\tau\,,
\end{align}
and the rate history differentiated with respect to radial shells, equivalent to eq.~\eqref{eq:galaxy-shell-rate}, is
\begin{align}\label{eq:galaxy-shell-rate-binary}
    \frac{\pd \R_{\rm Ia}}{\pd r}(t;r,M_{\star}) = \!\int_0^t\!\! \Phi_{\rm Ia}(\tau)\;\!\frac{\pd \Psi}{\pd r}(t\!-\!\tau;r,M_{\star})\;\!d\tau\,,
\end{align}
where $[\pd \Psi(t;r,M_{\star})/dr] dr$ is the SFH within a radial zone between radii $r\to r\!+\!dr$ 
(see \S~IV of \cite{2025arXiv250521260S} for details).
We adopt $\Phi_{\rm Ia}(\tau)$ of \cite{2011MNRAS.417..408R}, table~1, model A1, DDS, which is consistent with \cite{2012A&A...546A..70T} as well as other authors (see, e.g., fig.~7 of  \cite{2023RAA....23h2001L}). 
The DD rate distributions shown in Fig.~\ref{fig:1}b--f are calculated following 
eqs.~\eqref{eq:DTD} to \eqref{eq:HMD}.  
For the DD NMD shown in Fig.~\ref{fig:1}a, we use the fitting $\log(\pd r_{\rm Ia}/\pd m_{\rm Ni}) = -3.63\!-\!1.83\;\! m_{\rm Ni}/M_{\odot}$ to the results of 
\cite{2022MNRAS.515..286G}.

\paragraph*{Modeling assumptions.}---Most of the assumptions underlying rate predictions affect both scenarios (PBH and DD) in a similar way. In order not to favor one or the other, we set up a list of eight assumption combinations, summarized in table~\ref{tab:models}, to cover the extent of systematic uncertainties as much as possible.  
\begin{table}[htb]
    \caption{Modeling assumptions underlying PBH and DD scenarios. For details on the SFH models, IMFs, and IFMRs, see \cite{2025arXiv250521260S}.}
    \label{tab:models}
    \begin{ruledtabular}
    \begin{tabular}{llllrr}
        Model  &  SFH & IMF & IFMR & Iax & $\sigma_{\rm Ni}$ \\
        \hline\\[-3mm]
        1a  & \UniverseMachine & canonical & canonical & -- & -- \\
        1b  & \UniverseMachine & $Z$-dep. & $Z$-dep. & -- & -- \\
        1c  & \UniverseMachine & $Z$-dep. & $Z$-dep. & 15\% & 0.15$M_{\odot}$ \\
        1d  & \UniverseMachine & $(Z,\Psi)$-dep. & $Z$-dep. & 15\% & 0.15$M_{\odot}$ \\
        2a  & mass assembly & canonical & canonical & -- & -- \\
        2b  & mass assembly & $Z$-dep. & $Z$-dep. & -- & -- \\
        2c  & mass assembly & $Z$-dep. & $Z$-dep. & 15\% & 0.15$M_{\odot}$ \\
        2d  & mass assembly & $(Z,\Psi)$-dep. & $Z$-dep. & 15\% & 0.15$M_{\odot}$ \\[-0.5mm]
    \end{tabular}
    \end{ruledtabular}
\end{table}
Star formation histories are the most important ingredient, and we adopt (i) average SFHs of \UniverseMachine 
\cite{2013ApJ...770...57B,*2013ApJ...762L..31B} and (ii) SFHs calculated with a simple mass assembly model  
\cite{2011ApJ...734...48L} which has been used in the past to calculate SN Ia rates \cite{2014MNRAS.445.1898C}. We have decomposed these SFHs into radial shells as explained in \cite{2025arXiv250521260S}.

\footnotetext[4]{We calculated the effects of non-universal IMFs on the WD mass function, but not yet on SFHs. We are not aware of published SFH models assuming non-universal IMFs.}

There are strong theoretical and empirical reasons to believe that the stellar initial mass function (IMF) is not universal, but depends on gas phase metallicity ($Z$), mainly at lower stellar mass, $m_{\star}\lesssim 1M_{\odot}$ 
\cite{2012MNRAS.422.2246M,*2020A&A...637A..68Y}, and on star formation rate ($\Psi$) at higher stellar mass, $m_{\star}\gtrsim 1M_{\odot}$ 
\cite{2003ApJ...598.1076K,*2013MNRAS.436.3309W,*2017MNRAS.464.3812F}. We test different possibilities (table~\ref{tab:models} column 3 \cite{Note4}).
We also allow for a $Z$-dependent initial-final mass relation (IFMR) based on simulation results of 
\cite{1999ApJ...513..861U}, for the models where metallicity evolution is taken into account (table~\ref{tab:models} column 4), and we use the metallicity-stellar mass-redshift relation of 
\cite{2014ApJ...791..130Z} (see Appendix of \cite{2025arXiv250521260S} for details). 

For both scenarios (PBH and DD), we also include possible contributions from SNe Iax to the rate distributions where these peculiar events haven't been sorted out from observational data (those shown in panels b, c and f in Fig.~\ref{fig:1}). 
Their fractional contribution in low-redshift volumetric surveys is subdominant, but varies strongly depending on weather subluminous events are counted or not \cite{2022MNRAS.511.2708Smax10}.
However, due to a short DTD, their contribution is more important at high redshift and in low-mass galaxies. 
We calculate SN Iax rate distributions as 
for the DD scenario (not worrying about the exact ignition mechanism),
adopting $\Phi_{\rm Iax}(\tau)=(\tau/\tau_0)^{-\beta}$ for $\tau>\tau_0$ and zero otherwise, with empirical parameters $\tau_0=0.04$~Gyr, $\beta=2.52$ from 
\cite{2020MNRAS.493..986T} and normalized such that the local  volumetric contribution is either 0 or 15\% \cite{2022MNRAS.511.2708Smax10} of the total low-redshift volumetric rate $\rr_{\rm Ia}\!+\!\rr_{\rm Iax}$, as indicated in table~\ref{tab:models} column 5. 

Since the $^{56}$Ni mass inferred from observations 
\cite{2022MNRAS.509.5275S} might differ from that predicted for given $m_w$ \cite{2010ApJ...714L..52S,*2017MNRAS.470..157B,2018ApJ...854...52S,2021ApJ...909L..18S} from event to event for many different reasons (e.g. core-compositional variations, line-of-sight effects \cite{2014ApJ...785..105M} due to $^{56}$Ni clumps in the ejecta and/or asymmetric explosions, extinction effects, etc.), we also allow for a $^{56}$Ni-mass `dispersion' ($\sigma_{\rm Ni}$), zero in some models and $0.15M_{\odot}$ in others, as indicated in table~\ref{tab:models} column 6. 
These $0.15M_{\odot}$ account for adding quadratically $0.10M_{\odot}$ for dispersion due to core compositional variations (see Fig.~S5 in \cite{Note10}), estimating $0.10M_{\odot}$ for dispersion due to explosion asymmetries and $^{56}$Ni clumps in the ejecta, and estimating $0.05M_{\odot}$ for dispersion due to extinction effects.

\paragraph*{Parameter fitting and scenario comparison.}---With these preparations, we performed simulations running through log-normal parameter space $\{\log m_c$,~$\sigmac\}$ on a grid with range [19,~23]$\times$[0,~3], where $m_c$ is in gram, and spacing 0.2 in either parameter. 
We set up two data sample selections: the `complete', and the `restricted' where specific data points were removed for different reasons (see 
\S~SII of \cite{Note10} for details). 
The log-normal parameters that best-fit the data under standard $\chi^2$-minimization (see 
\S~SII of \cite{Note10} for details) are as follows ($m_c$ in gram):
\begin{table}[h!]
    \begin{ruledtabular}
    \begin{tabular}{lrrrrrrrr}
    Param. $\diagdown$ Model & 1a & 1b & 1c & 1d & 2a & 2b & 2c & 2d \\
    \hline\\[-3mm]
    $\log m_c$ {\rm(complete)} & 
    $21.2$ & $21.3$ & $21.1$ & $21.0$ & $21.2$ & $21.3$ & $21.1$ & $21.0$ \\
    \\[-4mm]
    $\log m_c$ {\rm(restricted)} & 
    $21.0$ & $21.3$ & $21.0$ & $21.0$ & $21.2$ & $21.3$ & $21.1$ & $21.0$ \\
\hline\\[-3mm]
    $\sigmac$ {\rm(complete)} & 
    $0.5$ & $0.6$ & $0.3$ & $0.2$ & $0.6$ & $0.6$ & $0.4$ & $0.4$ \\
    \\[-4mm]
    $\sigmac$ {\rm(restricted)} & 
    $0.0$ & $0.6$ & $0.0$ & $0.2$ & $0.6$ & $0.6$ & $0.4$ & $0.4$ \\
    \\[-4mm]
    \end{tabular}
    \end{ruledtabular}
\end{table}\\
Assuming that these are Gaussian distributed, the overall best-fit  parameters are
\begin{align}\label{eq:parambest}
    \log m_c =&\; 21.13\pm 0.12{\rm(sys)}\pm 0.05{\rm(stat)}\,, \nonumber \\ \sigmac =&\; 0.40\pm 0.20{\rm (sys)}\pm 0.10{\rm (stat)}\,,
\end{align}
where $m_c$ is in gram, and where systematic errors are at $1\sigma$ and statistical at $2\sigma$ confidence level, respectively. 

We use the Akaike information criterion (AIC) to assess the quantitative comparison between scenarios. The estimator penalizes additional degrees of freedom to avoid overfitting the data and is defined ${\rm AIC} = \chi^2+2\;\!\nu$, 
where $\nu$ is the number of free parameters to be adjusted, $\nu=0$ for the DD scenario and $\nu=2$ for log-normal PBHs.
The better theory has smallest AIC, and the likelihood difference between theories can be estimated calculating the $p$-value, $p\equiv\exp(\Delta $AIC$/2)$, with differences of more than 5 standard deviations ($\sigma$) being considered decisive evidence.
For all modeling assumptions and data compilation selections, PBH ignition is favored over DD ignition at more than $21\sigma$ confidence levels 
(see \S~SII of  \cite{Note10} for details).

Let us briefly discuss systematic uncertainties on the overall rate normalization of eq.~\eqref{eq:encounter-rate}. The main uncertainty is the ignition efficiency, which was limited to $1 \lesssim \alpha \lesssim 10$ in \cite{2021PhRvL.127a1101S}. Here we adopted $\alpha = 5$ for theoretical reasons given in \S~SI of \cite{Note10}, nevertheless a smaller variation around this value is possible.
Then, the best-fit peak mass $m_c$ decreases (increases) by 0.18~dex per unit increment (decrement) of $\alpha$. In all cases of the ignition efficiency $\alpha \in [1,~10]$, PBH ignition is still favored over the DD scenario at more than $5.4\sigma$ 
(see \S~SII of  \cite{Note10} for details).
A minor variation of the normalization of eq.~\eqref{eq:encounter-rate} comes from the uncertain amount of DM in galaxies. An increment (decrement) of 0.05~dex, which corresponds to the systematic uncertainty of the stellar-to-halo mass relation \cite{2010ApJ...710..903M,*2020A&A...634A.135G}, would induce an 0.05~dex decrement (increment) on $m_c$.
The same effect would occur if the fraction of DM in the form of asteroid-mass PBHs wasn't equal to unity.
The strong variation of the best-fit $m_c$ with alterations of the ignition efficiency
stems from fact that the minimum PBH mass that yields an ignition (see Fig.~S1 of \cite{Note10}) lies within the extended mass spectrum for most encounter situations.
In other words, for higher ignition efficiency, the average minimum PBH mass that yields an ignition decreases, and the best-fit $m_c$ decreases even more in such a way that only a smaller upper fraction of the spectrum contributes to ignitions, statistically speaking, because the number density of PBHs is inversely proportional to $m_c$. 
We find no significant variations of 
the best-fit mass spectrum width 
$\sigma_c$
related to systematic uncertainties of the normalization of eq.~\eqref{eq:encounter-rate}.

\paragraph*{Summary and discussion.}---Collisions between WDs and asteroid-mass PBHs offer a plausible explanation for the ignition of normal SNe Ia, for the following five reasons: 
First, the remarkable coincidence that a compact object falling into the WD potential marginally exceeds the plasma detonation velocity offers a straightforward trigger mechanism \cite{2021PhRvL.127a1101S}.
Second, rate distributions are consistent with observations (Fig.~\ref{fig:1})---which is the main result of this work---and allow for simple interpretations: the $^{56}$Ni-mass distribution (Fig.~\ref{fig:1}a) has a high-mass cut-off caused by WD composition transition and a low-mass "tail" because lighter WDs require heavier PBHs to ignite; the DTD (Fig.~\ref{fig:1}b)  and host-offset distributions (Fig.~\ref{fig:1}d--e) owe their shape to the effect of progenitor population depletion \cite{2025arXiv250521260S}. 
Third, the DD scenario---and hence all known binary-star ignition mechanisms---are disfavored in overall statistical comparison with PBHs to reproduce the observed rate distributions at the level of more than $21\sigma$ for any of the evaluated modeling assumptions (Table~\ref{tab:models}), mainly due to failing with the distribution of $^{56}$Ni masses and the total rate histories (Fig.~\ref{fig:1}a--c). 
These shortcomings of the DD scenario are unlikely to be alleviated with systematic effects of SFH modeling, stellar IMF, IFMR, contributions from SNe Iax, or a $^{56}$Ni-mass `dispersion', suggesting that, perhaps, a radically new mechanism is necessary.
Forth, observations of ejecta and remnants also favor prompt detonations in single WDs  
\cite{2010ApJ...714L..52S,*2017MNRAS.470..157B,2018ApJ...854...52S,2021ApJ...909L..18S,2022MNRAS.511.2994L,2015Natur.521..332O,*2022ApJ...933L..31S} (see, however, 
 \cite{2020MNRAS.499.4725K}).
Lastly, the inferred PBH mass spectrum, eq.~\eqref{eq:parambest}, is so far consistent with formation from critical collapse 
\cite{2017PhRvD..96b3514C}, and otherwise consonant with all current observational bounds 
\cite{2020ARNPS..70..355C,*2021RPPh...84k6902C,*2023JCAP...05..054K,*2020PhRvD.101f3005S}.

These empirical arguments in favor of asteroid-mass PBHs (or any equivalent population of compact objects) to constitute the cosmological DM call for a prediction of their very existence from first principles. 
In the meantime, independent, more direct observational proof (or disproof) can be gathered with parallax microlensing probes of gamma-ray bursts. The projected twin satellite \textit{Daksha} 
\cite{2022arXiv221112052Bmax10}, possibly in cross-calibration with the planned 
\textit{MoonBEAM} 
\cite{2021AAS...23731502H,*2022HEAD...1930505H}, would have the necessary detection efficiency for spectrum~\eqref{eq:parambest}, but only if one of the satellites was launched into lunar orbit or set on Sun-Earth L2 
\cite{2020PhRvR...2a3113J,*2024MNRAS.527.3306G}. 
Future replication studies might take into account net radial stellar population migration \cite{2021MNRAS.508.4484J} and contributions from dwarf satellite galaxies, which affect the small fraction of events at very large offsets. 
Finally, a deeper understanding of the ignition mechanism  \cite{2021PhRvL.127a1101S} can be gained with multidimensional hydrodynamical simulation; we commend such simulations to the community.\\

The author thanks the anonymous referees for their very constructive comments that improved the manuscript significantly. The author is also grateful to Emilio Tejeda, Doron Kushnir, Silvia Toonen, Kurtis Williams, María E. Camisassa, Amir Sharon, Louis-Gregory Strolger, Kishalay De, Valerio Marra, Ragnhild Lunnan and Peter Behroozi for useful correspondence and discussions. The work was initiated with financial support of FAPES/CAPES DCR Grant No. 009/2014.



\bibliography{biblio}

\newpage
\include{SM}


\end{document}

%% file: SM.tex
\thispagestyle{empty} 
\onecolumngrid
\begin{center}
\textbf{\large Supplemental Material: What triggers type Ia supernovae: Prompt detonations from primordial black holes or companion stars?}
\vspace{4ex}
\end{center}

\twocolumngrid

\setcounter{equation}{0}
\setcounter{figure}{0}
\setcounter{table}{0}
\setcounter{page}{1}
\setcounter{section}{0}
\makeatletter
\renewcommand{\theequation}{S\arabic{equation}}
\renewcommand{\thefigure}{S\arabic{figure}}
\renewcommand{\thetable}{S\Roman{table}}
\renewcommand{\thesection}{S\Roman{section}}

\section{Explosion modeling}\label{sec:explosion-modeling}

We follow the prescriptions of our previous work \cite{2021PhRvL.127a1101S} with several improvements, as detailed below.

\begin{figure}[htb]
    \centering
    \includegraphics[width=0.95\linewidth]{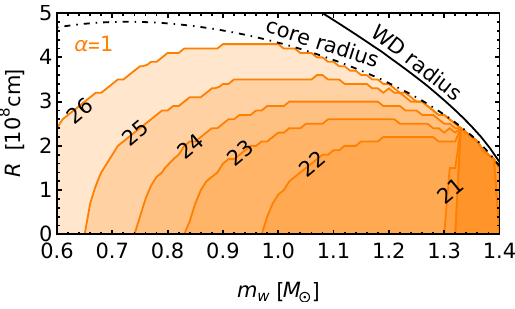}
    \includegraphics[width=0.95\linewidth]{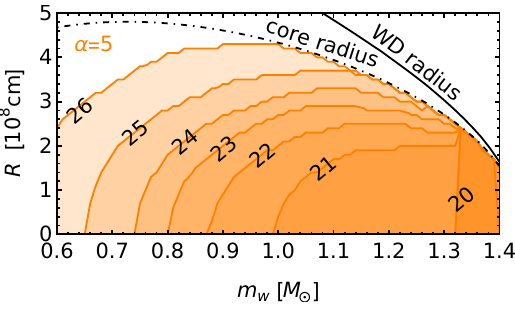}
    \includegraphics[width=0.95\linewidth]{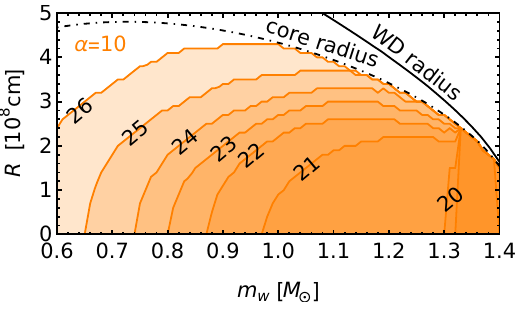}
    \caption{Detonation ignition cross section radius $R_m$ (orange filled contours, not to be confused with sky-projected galactocentric radial offset) 
    as a function of WD mass and for a range of PBH masses, $\log m_{\bullet}$~[g], as indicated by numbers, calculated as detailed in \protect\cite{2021PhRvL.127a1101S}, 
    but assuming the actual detonation velocity for quasi nuclear statistical equilibrium conditions (see text). Top, middle, and bottom panels show different assumptions of the ignition efficiency parameter $\alpha$ (see \protect\cite{2021PhRvL.127a1101S}
    for details).
    }
    \label{fig:detonation-cross-section}
\end{figure}

\paragraph*{Velocity threshold.}---Zeldovich's criterion of prompt detonation ignition requires the propagating shock wave to exceed the Chapman-Jouguet velocity, i.e. the velocity for which the post-shock velocity at the end of the reaction zone (where nuclear statistical equilibrium is attained) is sonic. The Chapman-Jouguet velocity in WDs is $\sim 1.2\times 10^4$~km~s$^{-1}$. 
However, real detonations in WDs first burn the carbon-oxygen material to a state of quasi nuclear statistical equilibrium, consisting of isotopes with binding energies near that of $^{28}$Si 
\cite{2014ApJ...785...61S}.  
Since the heat release up to this state is only part of the full heat release, the real detonation velocity in WDs is slightly lower, $\sim 1.0\times 10^4$~km~s$^{-1}$ \cite{2014ApJ...785...61S}. 
Requiring the velocity threshold for quasi-nuclear statistical equilibrium, the detonation ignition cross section increases slightly for given PBH mass. This can be seen in Fig.~\ref{fig:detonation-cross-section}, which has been calculated as detailed in \cite{2021PhRvL.127a1101S}, 
and in comparison with fig.~2 there. The difference is small but noticeable: very roughly, comparable detonation ignition cross sections, i.e. same $m_w$ and same $\alpha$-value, allow for a few times lighter PBHs.

\begin{figure}
    \centering
    \includegraphics[width=0.95\linewidth]{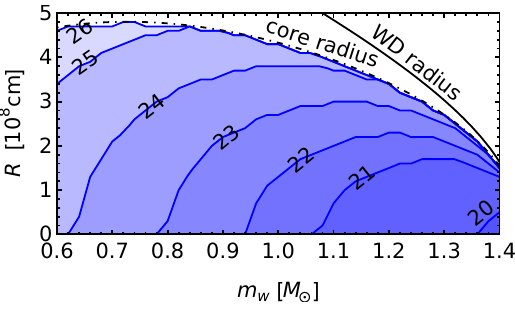}
    \caption{Deflagration ignition cross section radius (blue filled contours) as a function of WD mass and for a range of PBH masses, $\log m_{\bullet}$~[g], as indicated by numbers, calculated based on analytical shock modeling as described in 
    \protect\cite{2021PhRvL.127a1101S} 
    and satisfying the runaway criteria of 
    \protect\cite{1992ApJ...396..649T}.
    }
    \label{fig:deflagration-cross-section}
\end{figure}

\paragraph*{Ignition efficiency.}---Since the shock strength weakens with distance from the PBH, 
we \cite{2021PhRvL.127a1101S} introduced the $\alpha$ parameter to set a cut-off to the region (in units of the accretion radius) within which the detonation ignition conditions must be satisfied, with $\alpha \sim 1$ ($\alpha \sim 10$) representing a conservative (optimistic) value.
Here, we found a theoretical argument to settle the value of $\alpha$, which goes as follows:
in the regime where the velocity criteria are overly satisfied---i.e. in the centers of very massive WDs where infalling objects acquire speeds far beyond the detonation velocity---the remaining questioning reduces to the runaway criteria for deflagration ignition.
Using the analytical shock description detailed in  
\cite{2021PhRvL.127a1101S}
and requisitioning the hot spot runaway criteria of 
\cite{1992ApJ...396..649T},
we established the cross section for deflagration ignition (see Fig.~\ref{fig:deflagration-cross-section}).

A close comparison between Fig.~\ref{fig:detonation-cross-section} and Fig.~\ref{fig:deflagration-cross-section} reveals that the deflagration ignition cross section is most similar to that for detonation ignition with $\alpha = 5$ in the regime specified above, i.e. in the centers of high-mass WDs.  
Based on this argumentation, we adopt the value $\alpha=5$ in this work.

\begin{figure}
    \centering
    \includegraphics[width=0.9\linewidth]{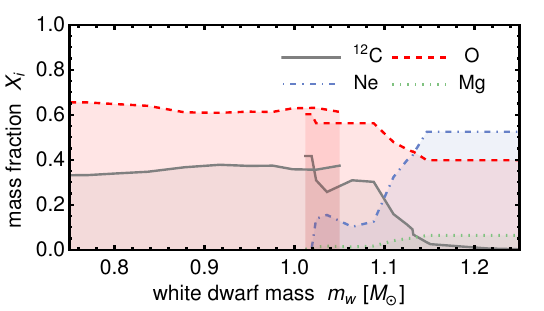}
    \caption{Elemental abundances in WD cores as a function of WD mass according to numerical simulations for $m_w<1.05M_{\odot}$ \protect\cite{2013ApJ...779...58R} 
    and $m_w >1.01M_{\odot}$ \cite{2018MNRAS.480.1547L}.
    Overlapping is visible.}
    \label{fig:pWDelements}
\end{figure}

\paragraph*{Core composition.}---Numerical simulations 
\cite{2018MNRAS.480.1547L,2013ApJ...779...58R} 
suggest that the core carbon mass fraction is $X_{\rm C} \approx 0.3$ for WDs with mass $m_w \lesssim 1.08M_{\odot}$ and then drops to $X_{\rm C} \lesssim 0.05$ for WDs with mass $m_w \approx 1.15M_{\odot}$ (see Fig.~\ref{fig:pWDelements}). 

\begin{figure}[ht]
\centering
\includegraphics[width=0.49\linewidth]{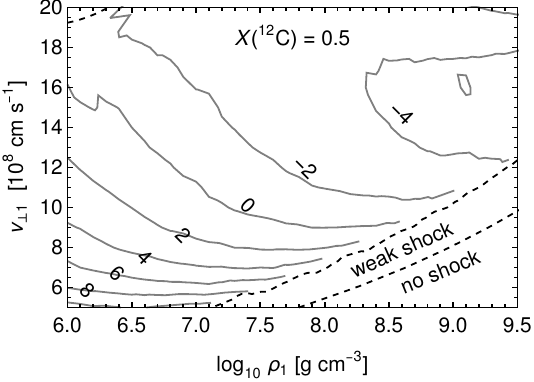}
\includegraphics[width=0.49\linewidth]{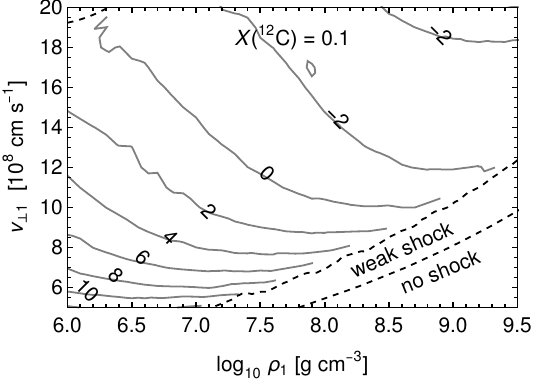}
\includegraphics[width=0.49\linewidth]{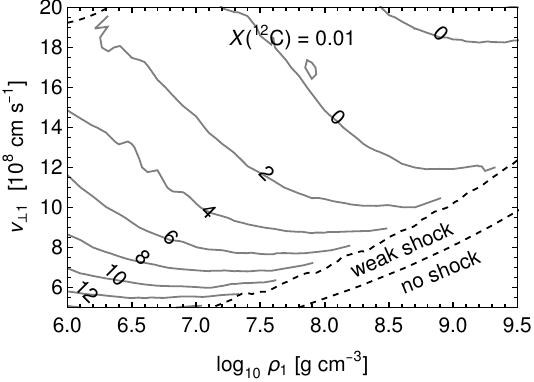}
\includegraphics[width=0.49\linewidth]{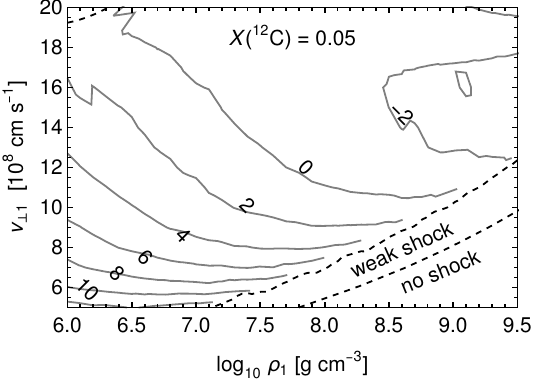}
\caption{Induction length, $\log(\xi_{\perp}/{\rm cm}$), as a function of pre-shock density $\rho_1$ and orthogonal velocity $v_{\perp 1}$ for four different upstream carbon mass fractions $X_{\rm C}=(0.5,0.1,0.05,0.01)$, and assuming uniformly temperature $T_{\!1}=10^7$K. It can be noted that $\xi_{\perp} \propto X_{\rm C}^{\,-2}$. The post-shock state in the upper left corner delimited by the black dashed line is supersonic. For details see 
\cite{2021PhRvL.127a1101S}.}
\label{fig:pInd}
\end{figure}

Since the induction length $\xi_{\perp}$ (see  
\cite{2021PhRvL.127a1101S} 
for details) is inversely proportional to the square of the carbon mass fraction, $\xi_{\perp} \propto {X_{\rm C}}^{-2}$,---we have verified this with 1D numerical simulations (see Fig.~\ref{fig:pInd})---the detonation ignition cross section vanishes quickly beyond $m_w\gtrsim 1.1M_{\odot}$. 
To account for the transition in a simple way, we first performed all rate calculations (as detailed in the main Letter) assuming the ignition cross section shown in Fig.~\ref{fig:detonation-cross-section} (with $\alpha=5$ as justified above), and then suppress the rate with a linear transition from $m_w = 1.08M_{\odot}$ to $m_w = 1.12M_{\odot}$, such that the rate is zero for $m_w > 1.12M_{\odot}$.

\begin{figure}
    \centering
    \includegraphics[width=0.9\linewidth]{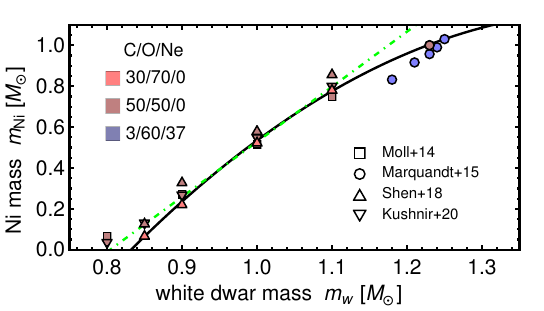}
    \caption{Synthesized $^{56}$Ni mass in pure or direct carbon detonations (i.e. without initial He-shell detonation) as a function of WD mass for solar metallicities and various core compositions as indicated by colors according to simulation results 
    \protect\cite{2014ApJ...785..105M,2015A&A...580A.118M,2018ApJ...854...52S,2020MNRAS.499.4725K}. 
    We fit the 30/70-data of 
    \citet{2018ApJ...854...52S}
    combined with the data point of \citet{2015A&A...580A.118M} 
    at second order: $m_{\rm Ni}=-5.0335 + 8.4376\;\!m_w - 2.86945\;\!m_w^{\,2}$
    (full black line).
    Also shown is the linear fit used by 
    \cite{2022MNRAS.515..286G}  
    (green dot-dashed line).
    }
    \label{fig:MwdMNi56}
\end{figure}

\paragraph*{Nucleosynthesis.}---According hydrodynamical simulation studies 
\cite{2010ApJ...714L..52S,*2017MNRAS.470..157B,2018ApJ...854...52S,2021ApJ...909L..18S},
the amount of $^{56}$Ni produced in a prompt detonation of a sub-Chandrasekhar WD in hydrostatic equilibrium is determined primarily by its total mass. Secondary parameters, such as core carbon mass fraction are also important (see Fig.~\ref{fig:MwdMNi56}). 
We fit the results of 
\cite{2018ApJ...854...52S}
with C/O equal 30/70, which is most representative of the expected core compositions of WDs in the mass range $[0.9,1.1]M_{\odot}$ 
\cite{2018MNRAS.480.1547L,2013ApJ...779...58R} 
(see, however, 
\cite{2023ApJ...950..115B},
where a 1-$M_{\odot}$ WD has 40/60 C/O mass fractions). 
Since 
\cite{2018ApJ...854...52S}
does not provide the $^{56}$Ni masses, we use their provided $B$-band peak brightnesses, $M_B$, that we convert back to $^{56}$Ni masses, using the relation $M_B = -2.5\log m_{\rm Ni}-19.841\pm 0.020$. 
Throughout this Letter, we use this $m_{\rm Ni}(m_w)$ relation (full black line in Fig.~\ref{fig:MwdMNi56}), which is comparable to the fit used by 
 \cite{2022MNRAS.515..286G}
(green dot-dashed line in Fig.~\ref{fig:MwdMNi56}).

\section{Data analysis}\label{sec:data-analysis}

In the following, we describe in detail the data selection and analysis, distribution by distribution.
We let $\vec{\theta}$ be a vector containing the free parameters of each scenario. Log-normal PBHs have two free parameters, $\vec{\theta}= \{\log m_c,\sigmac\}$, 
and the DD scenario has none, $\vec{\theta}=\{\}$. 
As mentioned in the main Letter, we set up two data compilations: one referred to as `complete', and the other as `restricted', where we apply cuts as and when required and as detailed below. We also bin data adequately if necessary.

\begin{figure*}[htb!]
    \centering
    \includegraphics[width=0.952\linewidth]{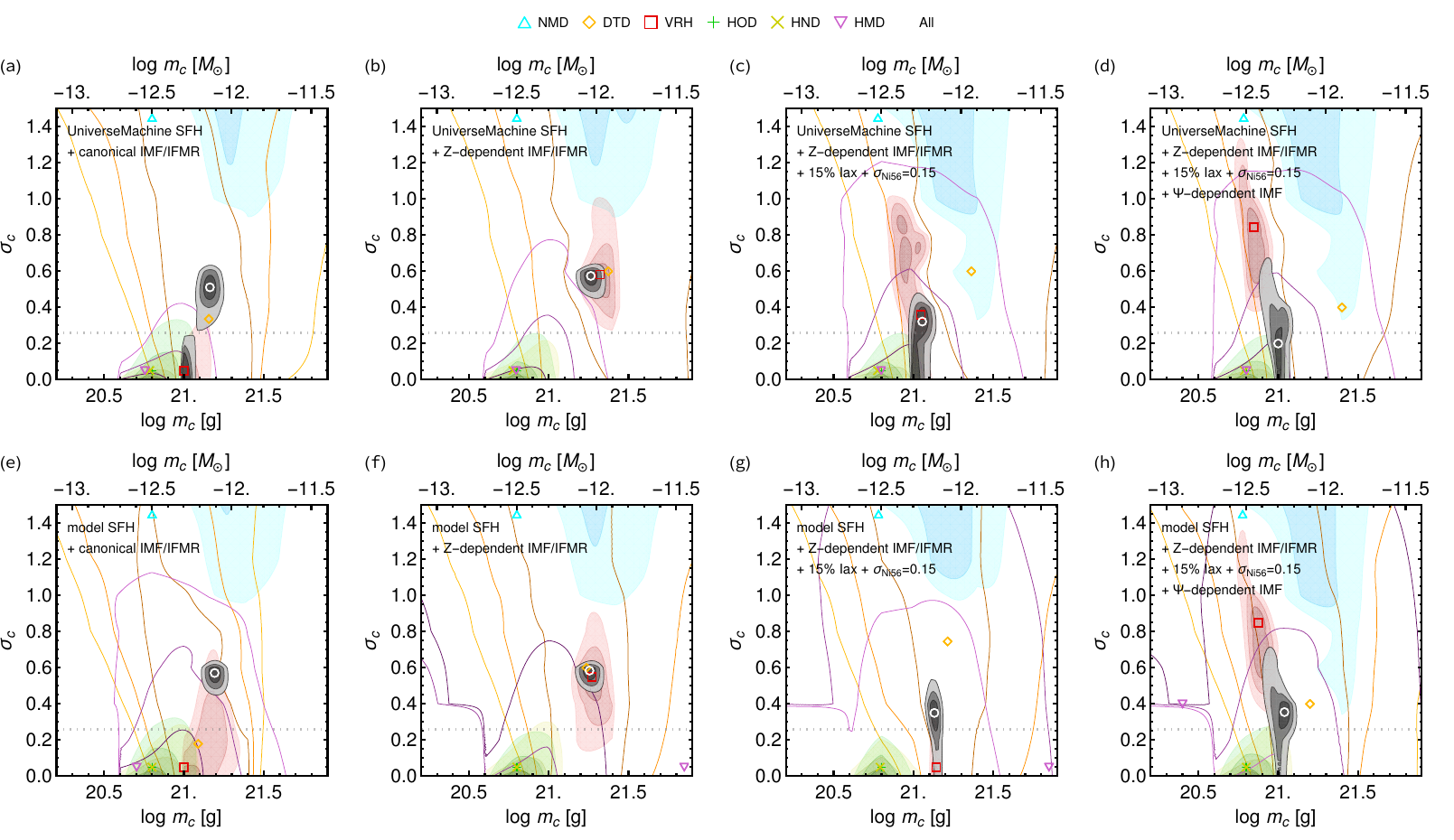}
    \caption{Best-fit values (colored plot markers) and confidence intervals (shaded regions or lines with corresponding colors) on log-normal PBH mass function parameters from the observed rate distributions shown in 
    Fig.~\ref{fig:1} 
    and for the modeling assumption combinations of 
    table~I of the main text, respectively. The confidence intervals from all rate distributions combined are shown in shades of gray and the best-fit values are marked by white circles. The gray dotted line marks the value predicted for formation from critical collapse 
    \cite{2017PhRvD..96b3514C}.
    }
    \label{fig:2}
\end{figure*}

\begin{figure*}[htb!]
    \centering
    \includegraphics[width=0.952\linewidth]{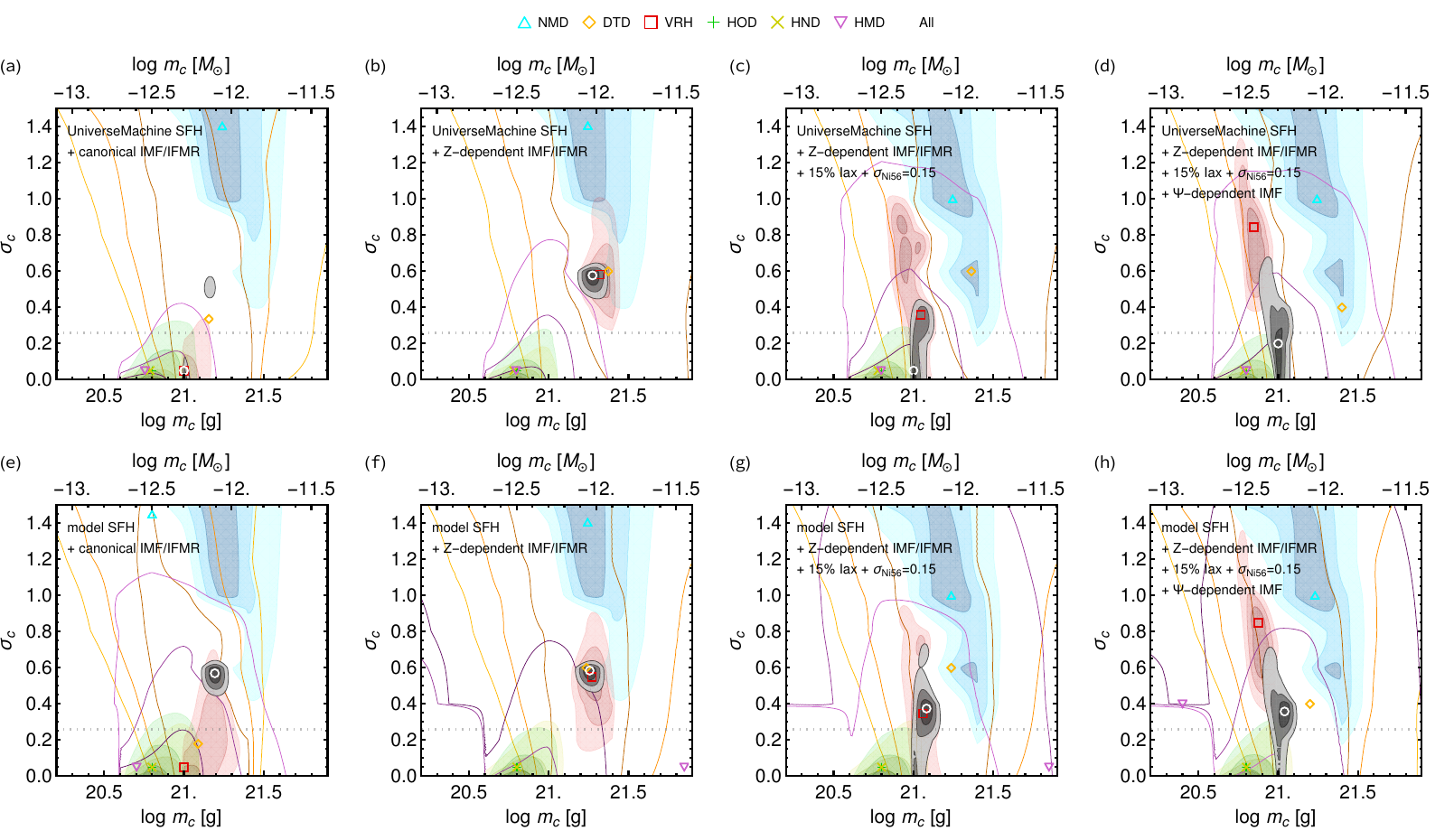}
    \caption{Same as Fig.~\ref{fig:2} but for the restricted data sample.}
    \label{fig:2-restricted}
\end{figure*}

\clearpage
\newpage

\begin{table}[tb]
    \caption{$\chi^2$ results for the complete data sample.}
    \label{tab:chi2-complete}
    \begin{ruledtabular}
    \begin{tabular}{lrrrrrrr}
    Data & NMD & DTD & VRH & HOD & HND & HMD & All \\
    $\#$Bins & 10 & 3 & 11 & 12 & 9 & 13 & 58 \\
    \hline\\[-3mm]
    \input{tables/TeXchi2-complete.txt}\\[-3mm]
    \input{tables/TeXchi2DD-complete.txt}\\[-4mm]
    \end{tabular}
    \end{ruledtabular}
\end{table}

\begin{table}[t!]
    \caption{$\chi^2$ results for the restricted data sample.}
    \label{tab:chi2-restricted}
    \begin{ruledtabular}
    \begin{tabular}{lrrrrrrr}
    Data & NMD & DTD & VRH & HOD & HND & HMD & All \\
    $\#$Bins & 7 & 3 & 11 & 8 & 5 & 13 & 47 \\
    \hline\\[-3mm]
    \input{tables/TeXchi2-restricted.txt}\\[-3mm]
    \input{tables/TeXchi2DD-restricted.txt}\\[-4mm]
    \end{tabular}
    \end{ruledtabular}
\end{table}

\paragraph*{Nickel mass distribution (NMD).}---We use the data of 
\cite{2022MNRAS.509.5275S} 
who determined, for the first time, the intrinsic NMD based on low redshift ($z<0.04$)  SNe Ia
from the Zwicky Transient Facility Bright Transient Survey (ZTF BTS).
The sample used contains 1519 SNe Ia, of which 1371 are used, including sub-classes `normal', 91T-like, 91bg-like, and Ia-pec, but excludes other sub-classes such as Iax.
The data from LOSS 
\cite{2014MNRAS.438.3456P}, 
as reanalyzed by 
\cite{2022MNRAS.509.5275S} 
is shown for comparison in 
Fig.~\ref{fig:1}a of the main text,
but is not used in the data analysis because several events had no decline-rate measurements and were attributed arbitrarily $m_{\rm Ni}=0.1M_{\odot}$ as their sub-class was identified with 91bg-like (and these have  $0.1M_{\odot}$ on average). As can be seen,  
this led to an artificial `bump' around $m_{\rm Ni}=0.1M_{\odot}$ in the LOSS NMD.
The ZTF data was binned into equally spaced bins with mean ranging from $m_{\rm Ni}=0.05M_{\odot}$ to $m_{\rm Ni}=0.95M_{\odot}$ with $\Delta m_{\rm Ni}=0.1M_{\odot}$ 
\cite{2022MNRAS.509.5275S}.
For the data analysis, we use these bins as such, but normalize the data to a probability distribution function (PDF), because the overall rate information is already contained in the cosmic rate histories. 
In our restricted sample, we exclude the lowest two and the highest data points  ($0.05$, $0.15$, $0.95$) because these lie below the ZTF detection limit 
\cite{2022MNRAS.509.5275S} 
(shown as gray dashed line in 
Fig.~\ref{fig:1}a of the main text).
We define the chi square function
\begin{align}\label{eq:chi2-Ni}
    \chi^2_{\rm NMD}(\vec{\theta}) \equiv \sum_j \frac{1}{\sigma_{{\rm NMD},j}^2}\Big[\mu_{{\rm NMD},j} - \frac{d\rr_{\rm Ia}}{dm_{\rm Ni}}(m_{{\rm Ni},j}|\vec{\theta})\Big]^2\!\!,
\end{align}
where $\{m_{{\rm Ni},j},\mu_{{\rm NMD},j}\pm \sigma_{{\rm NMD},j}\}$ are the ZTF data points
and $d\rr_{\rm Ia}/dm_{\rm Ni}$ is given by eq.~(5)
of the main text.

\paragraph*{Delay time distribution (DTD).}---We use the cluster rates 
from table~6 of 
\cite{2018MNRAS.479.3563F}. 
All 11 data points are included and are binned into 3 equally spaced bins with mean ranging from $\tau = 3.2\;\!$Gyr to $9.7\;\!$Gyr and width $\Delta \tau = 3.3\;\!$Gyr. 
We do not use DTD measurements from the `field' or from the Magellanic Clouds, because in DM-ignition scenarios, the predictions depend on the environment.
We define the chi square function
\begin{align}\label{eq:chi2-DTD}
    \chi^2_{\rm DTD}(\vec{\theta}) \equiv \sum_j \frac{1}{\sigma_{{\rm DTD},j}^2}\Big[\mu_{{\rm DTD},j} - \R_{\rm Ia,m_{\star}}^{\rm cl}(\tau_j|\vec{\theta})\Big]^2\!\!,
\end{align}
where $\{\tau_j,\mu_{{\rm DTD},j}\pm \sigma_{{\rm DTD},j}\}$ are the data points shown in 
Fig.~\ref{fig:1}b 
and $\R_{\rm Ia,m_{\star}}^{\rm cl}$ is given by eq.~\eqref{eq:DTD} 
of the main text, respectively. Note that we label cluster galaxy rates per solar mass of stars formed with the acronym `DTD', even though, strictly speaking, there is a minor difference [see comment after eq.~\eqref{eq:DTD} of the main text]. The label `pseudo-DTD' would be more appropriate, however, in order to conform with the nomenclature used in the literature, we stick to using `DTD' and hope that the resulting ambiguity does not confuse the reader unnecessarily. 

\paragraph*{Volumetric rate history (VRH).}---We use the literature data compilation from table~1 of 
Ref.~\cite{2020ApJ...890..140S}. We thank Louis-Gregory Strolger for providing us this data. 
All data points are included in the analysis and are binned into 11 equally spaced bins with mean ranging from $t=3~$Gyr to $t=13$~Gyr and width $\Delta t=1$~Gyr (gray points in 
Fig.~\ref{fig:1}c of the main text).
We define the following chi square function 
\begin{align}\label{eq:chi2-VRH}
    \chi^2_{\rm VRH}(\vec{\theta}) \equiv \sum_j \frac{1}{\sigma_{{\rm VRH},j}^2}\Big[\mu_{{\rm VRH},j} - \rr_{\rm Ia}(t_j|\vec{\theta})\Big]^2\!\!,
\end{align}
where $\{t_j,\mu_{{\rm VRH},j}\pm \sigma_{{\rm VRH},j}\}$ are the binned data points 
shown in Fig.~\ref{fig:1}c 
and $\rr_{\rm Ia}$ is given by 
eq.~\eqref{eq:VRH}
of the main text, respectively.

\paragraph*{Host offset distribution (HOD).}---We use data from 
ZTF 
\cite{2020ApJ...905...58Dmax10}
with galaxy half-light radii supplied therein. We thank Kishalay De for providing us this data. The sample specifies sub-classes, and we keep only those labelled `Ia' (i.e. normal), `Ia-91T', and `Ia-91bg', excluding all others such as `Ia-02cx' (i.e. Iax), `Ca-rich', etc. We normalize the data to a PDF. 
Also shown in 
Fig.~\ref{fig:1}d--e of the main text are data from 
and Palomer Transient Facility (PTF 
\cite{2017ApJ...836...60Lmax10}). 
These data points are shown for comparison only, and are not used formally because error estimates couldn't be recovered reliably [Ragnhild Lunnan, private communication]. The vertical error bars shown are merely crude estimates based on the survey sample size.  Idem for data points from Sloan Digital Sky Survey (SDSS 
\cite{2018MNRAS.481.2766H}). 
In the restricted sample, we limit to $R\leq 20\;\!$kpc and $\tilde{R}\leq 5 R_{1/2}$, for several reasons: (i) large differences between survey results, (ii) net radial migration of stellar populations and/or contributions from dwarf satellite galaxies become  important beyond these scales and were not taken into account in the predictions yet. We define the chi square function
\begin{align}\label{eq:chi2-HOD}
    \chi^2_{\rm HOD}(\vec{\theta}) \equiv \sum_j \frac{1}{\sigma_{{\rm HOD},j}^2}\Big[\mu_{{\rm HOD},j} - \frac{d\rr_{\rm Ia}}{dR}(R_j|\vec{\theta})\Big]^2\!\!,
\end{align}
where $\{R_j,\mu_{{\rm HOD},j}\pm \sigma_{{\rm HOD},j}\}$ are the ZTF data points shown in Fig.~\ref{fig:1}d 
and $d\rr_{\rm Ia}/dR$ is given by 
eq.~(10) of the main text.

\paragraph*{Host-normalized offset distribution (HND).}---Same comments as for the HOD, and we define the chi square function 
\begin{align}\label{eq:chi2-HND}
    \chi^2_{\rm HND}(\vec{\theta}) \equiv \sum_j \frac{1}{\sigma_{{\rm HND},j}^2}\Big[\mu_{{\rm HND},j} - \frac{d\rr_{\rm Ia}}{d\tilde{R}}(\tilde{R}_j|\vec{\theta})\Big]^2\!\!,
\end{align}
where $\{\tilde{R}_j,\mu_{{\rm HND},j}\pm \sigma_{{\rm HND},j}\}$ are the the ZTF data points shown in Fig.~\ref{fig:1}e and $d\rr_{\rm Ia}/d\tilde{R}$ is given by 
eq.~(11) of the main text, respectively.

\paragraph*{Host mass distribution (HMD).}---We use 
data from Dark Energy Survey (DES), PTF, 
 and PanSTARRS (PS1) at $z<0.2$, taken from fig.~14 of 
\cite{2020MNRAS.495.4040Wmax10}. 
We find the data adequately binned. No cuts are applied. We define the chi square function
\begin{align}\label{eq:chi2-HMD}
    \chi^2_{\rm HMD}(\vec{\theta}) \equiv \sum_j \frac{1}{\sigma_{{\rm HMD},j}^2}\Big[\mu_{{\rm HMD},j} - \frac{d\rr_{\rm Ia}}{d\log M_{\star}}(\log M_{\star,j}|\vec{\theta})\Big]^2\!\!,
\end{align}
where $\{\log M_{\star,j},\mu_{{\rm HMD},j}\pm \sigma_{{\rm HMD},i}\}$ where calculated from the mean of the data points shown in 
Fig.~\ref{fig:1}f of the main text and $d\rr_{\rm Ia}/d\log M_{\star}$ is given by 
eq.~(12) there.

Finally, we define the total chi square function
\begin{align}\label{eq:chi2-All}
    \chi_{\rm All}^2 \equiv \chi_{\rm NMD}^2 \!+\! \chi_{\rm DTD}^2 \!+\! \chi_{\rm VRH}^2 \!+\! \chi_{\rm HOD}^2 \!+\! \chi_{\rm HND}^2 \!+\! \chi_{\rm HMD}^2\,.
\end{align}
We minimize this function for both data compilation selections and all eight modeling assumptions to obtain the best-fit PBH models. Confidence intervals on the log-normal parameters from individual and combined rate distributions are shown in Fig.~\ref{fig:2} and Fig.~\ref{fig:2-restricted}, for the complete and restricted data compilation selections, respectively.

The $\chi^2$ values distribution-by-distribution for both best-fit PBH and DD scenarios are reproduced in table~\ref{tab:chi2-complete} and table~\ref{tab:chi2-restricted}, for the complete and restricted data compilation selections, respectively.

The AIC differences for all modeling assumptions and data compilation selections are shown in table~\ref{tab:sigmalikeli}.
\begin{table}[htb]
    \caption{Likelihood differences in standard deviations ($\sigma$) in favor of PBH ignition over DD ignition, for different data samples and modeling assumptions.}
    \label{tab:sigmalikeli}
    \begin{ruledtabular}
    \begin{tabular}{lcccccccc}
    Sample $\diagdown$ Model & 1a & 1b & 1c & 1d & 2a & 2b & 2c & 2d \\
    \hline\\[-3mm]
    $\text{complete}$ & $30.4$ & $32.6$ & $30.3$ & $30.2$ & $37.5$ & $37.5$ & $37.4$ & $36.2$ \\
\\[-4mm]
    $\text{restricted}$ & $21.7$ & $24.6$ & $24.5$ & $23.0$ & $30.8$ & $32.3$ & $32.2$ & $31.5$ 
\\[-0.5mm]
    \end{tabular}
    \end{ruledtabular}
\end{table}

\begin{table}[htb]
    \caption{Same as table~\ref{tab:sigmalikeli}, but for ignition efficiency $\alpha=10$.}
    \label{tab:sigmalikeli-alpha-10}
    \begin{ruledtabular}
    \begin{tabular}{lcccccccc}
    Sample $\diagdown$ Model & 1a & 1b & 1c & 1d & 2a & 2b & 2c & 2d \\
    \hline\\[-3mm]
    $\text{complete}$ & $32.1$ & $34.8$ & $31.0$ & $30.9$ & $37.7$ & $37.7$ & $37.7$ & $36.5$ \\
\\[-4mm]
    $\text{restricted}$ & $23.7$ & $27.3$ & $25.2$ & $23.9$ & $32.2$ & $33.8$ & $32.5$ & $31.6$ 
\\[-0.5mm]
    \end{tabular}
    \end{ruledtabular}
\end{table}

\begin{table}[htb]
    \caption{Same as table~\ref{tab:sigmalikeli}, but for ignition efficiency $\alpha=1$.}
    \label{tab:sigmalikeli-alpha-1}
    \begin{ruledtabular}
    \begin{tabular}{lcccccccc}
    Sample $\diagdown$ Model & 1a & 1b & 1c & 1d & 2a & 2b & 2c & 2d \\
    \hline\\[-3mm]
    $\text{complete}$ & $12.2$ & $14.0$ & $18.6$ & $20.7$ & $25.4$ & $26.7$ & $29.5$ & $29.4$ \\
\\[-4mm]
    $\text{restricted}$ & $17.2$ & $15.4$ & $5.4$ & $7.0$ & $13.9$ & $16.1$ & $22.6$ & $23.2$
\\[-0.5mm]
    \end{tabular}
    \end{ruledtabular}
\end{table}

\begin{figure*}[htb!]
    \centering
    \includegraphics[width=0.952\linewidth]{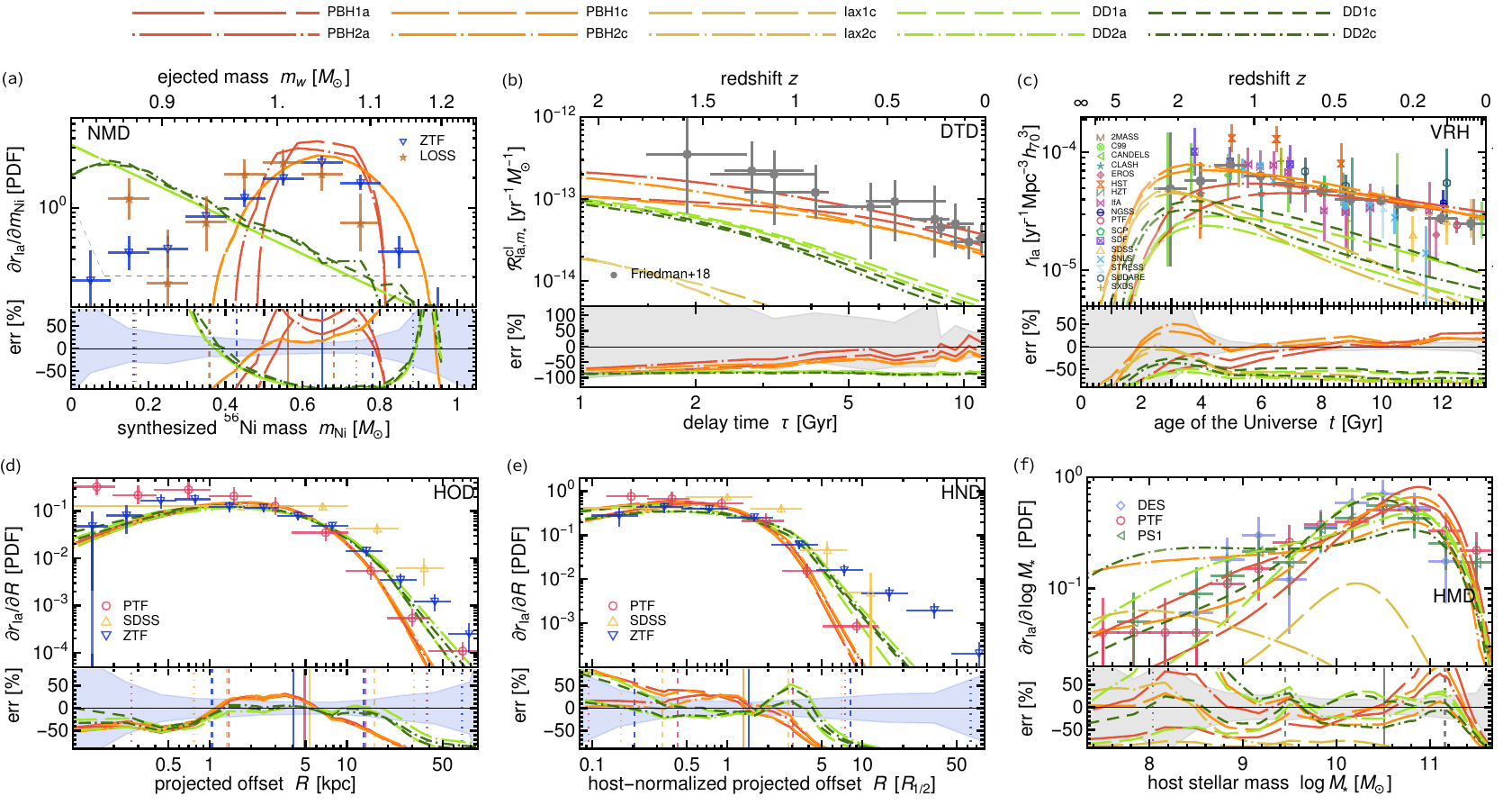}
    \setlength{\belowcaptionskip}{45pt}
    \caption{Same as Fig.~\ref{fig:1} of the main text
    but for ignition efficiency $\alpha = 10$. 
    }
    \label{fig:1-alpha-10}
\end{figure*}

\begin{figure*}[htb!]
    \centering
    \includegraphics[width=0.952\linewidth]{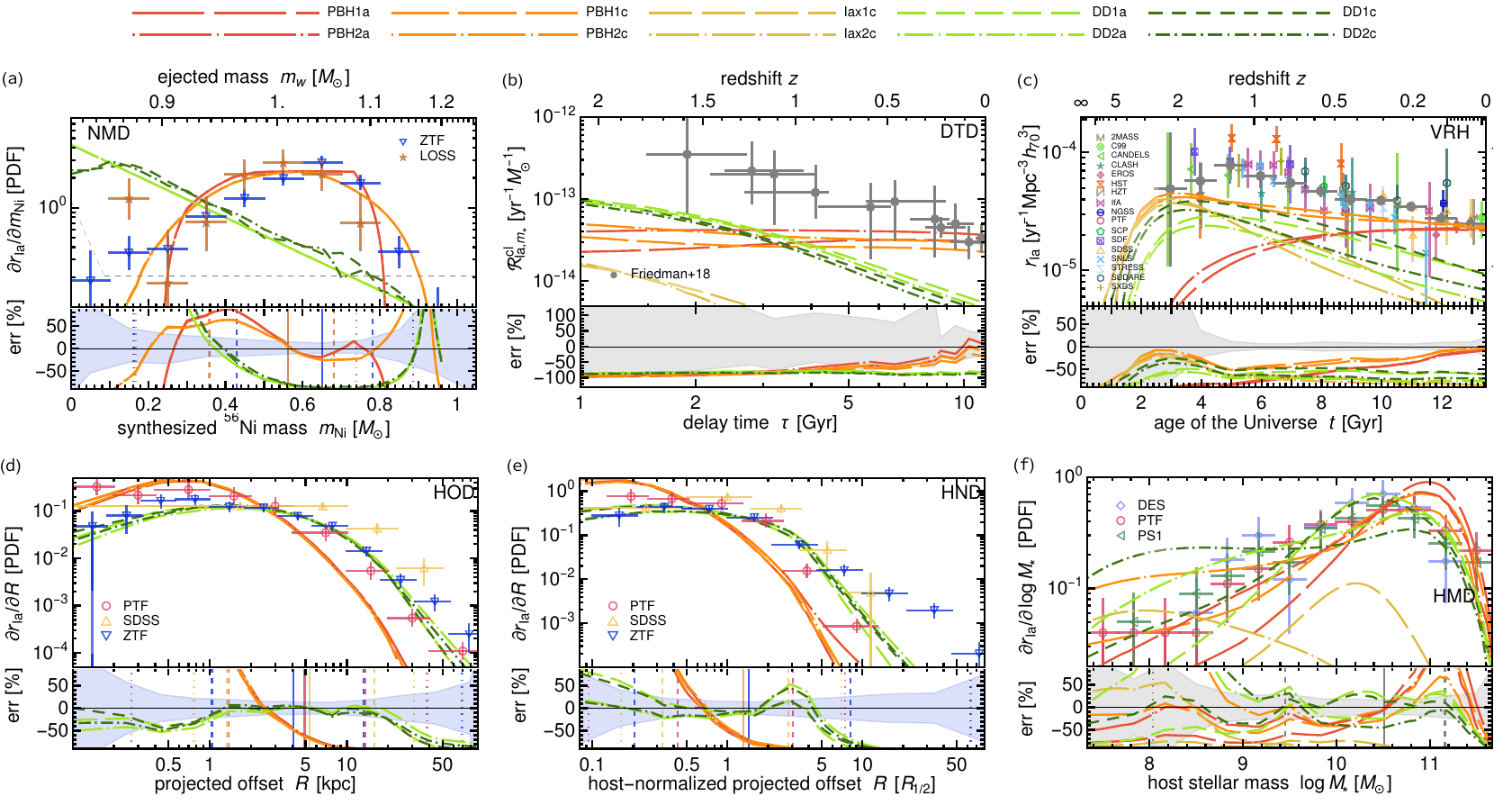}
    \setlength{\belowcaptionskip}{44.5pt}
    \caption{Same as Fig.~\ref{fig:1} of the main text
    but for ignition efficiency $\alpha = 1$. 
    }
    \label{fig:1-alpha-1}
\end{figure*}


\begin{figure*}[htb!]
    \centering
    \includegraphics[width=0.952\linewidth]{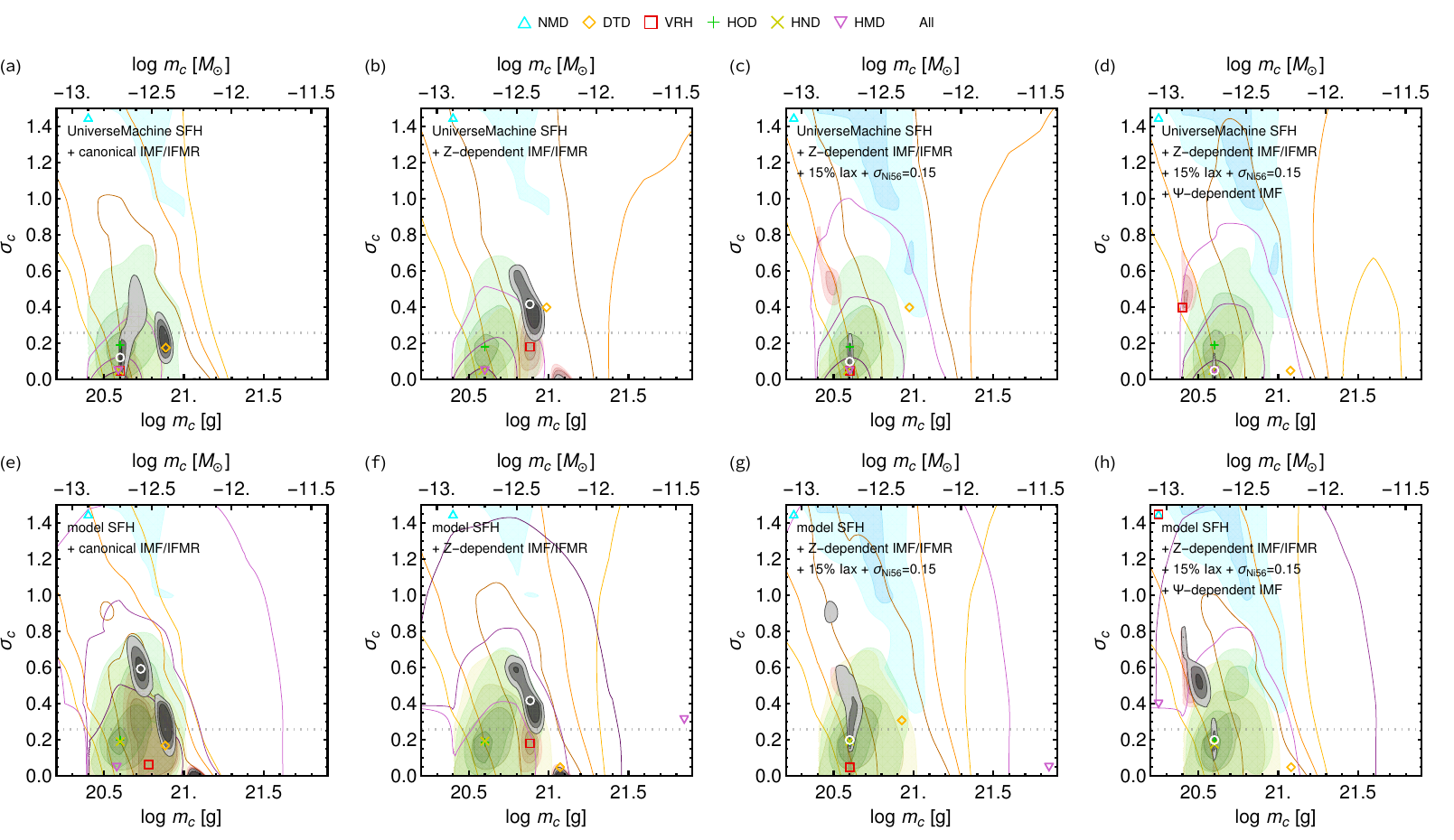}
    \setlength{\belowcaptionskip}{23.5pt}
    \caption{Same as Fig.~\ref{fig:2} but for ignition efficiency $\alpha = 10$. 
    }
    \label{fig:2-alpha-10}
\end{figure*}

\begin{figure*}[htb!]
    \centering
    \includegraphics[width=0.952\linewidth]{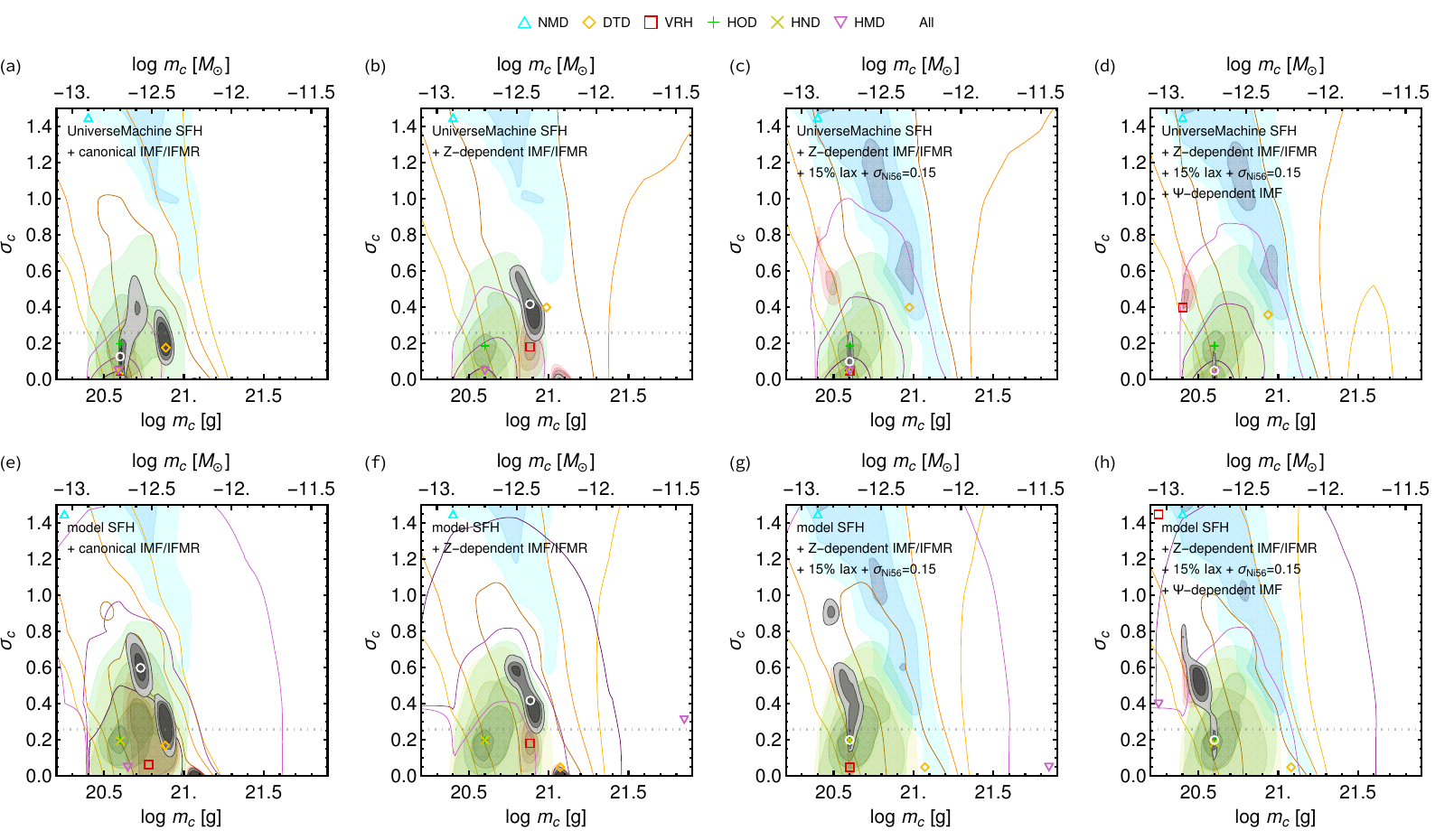}
    \setlength{\belowcaptionskip}{23pt}
    \caption{Same as Fig.~\ref{fig:2-restricted} but for $\alpha = 10$. 
    }
    \label{fig:2-alpha-10-restricted}
\end{figure*}


\begin{figure*}[htb!]
    \centering
    \includegraphics[width=0.952\linewidth]{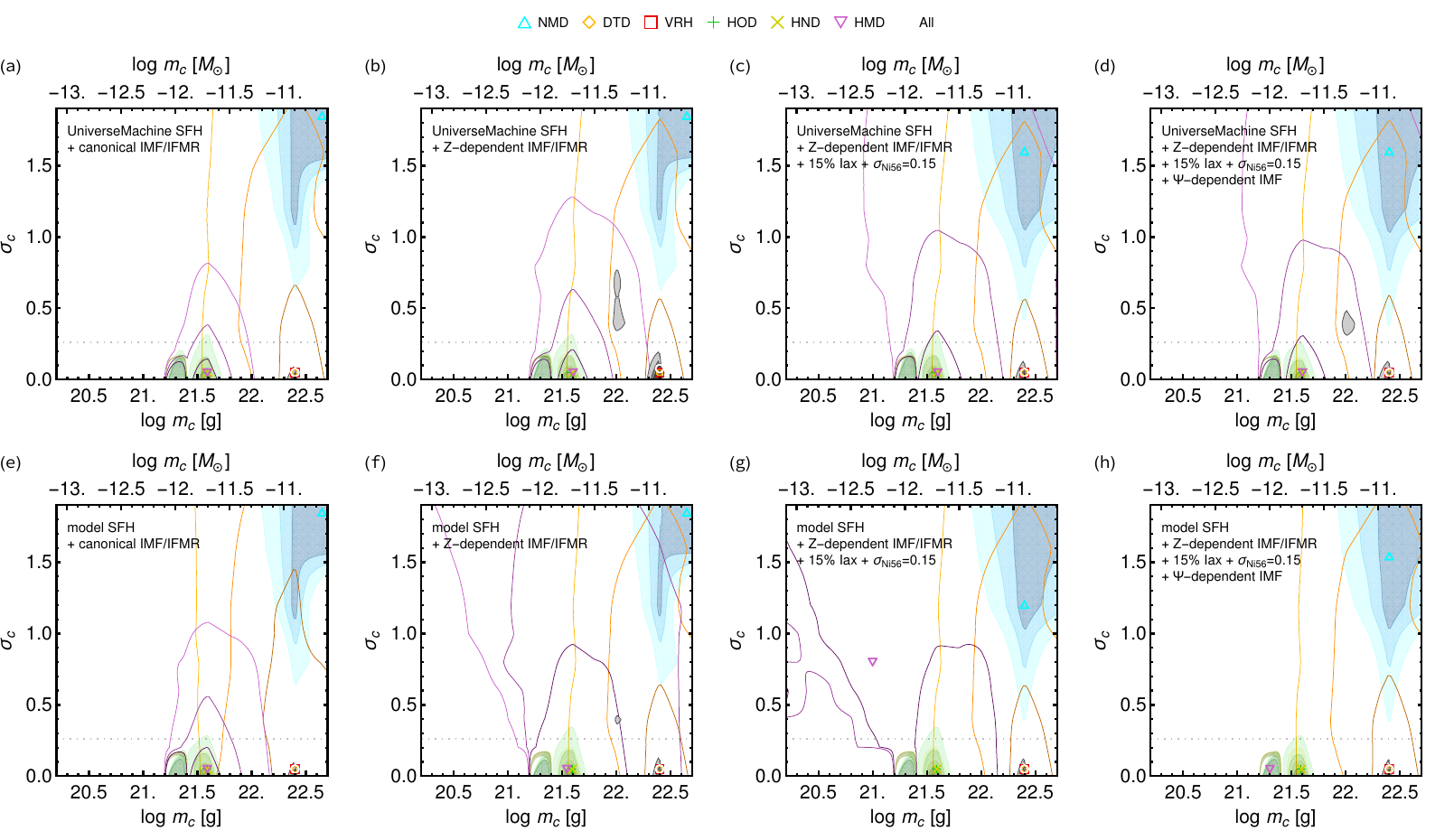}
    \setlength{\belowcaptionskip}{23.5pt}
    \caption{Same as Fig.~\ref{fig:2} but for ignition efficiency $\alpha = 1$.}
    \label{fig:2-alpha-1}
\end{figure*}

\begin{figure*}[htb!]
    \centering
    \includegraphics[width=0.952\linewidth]{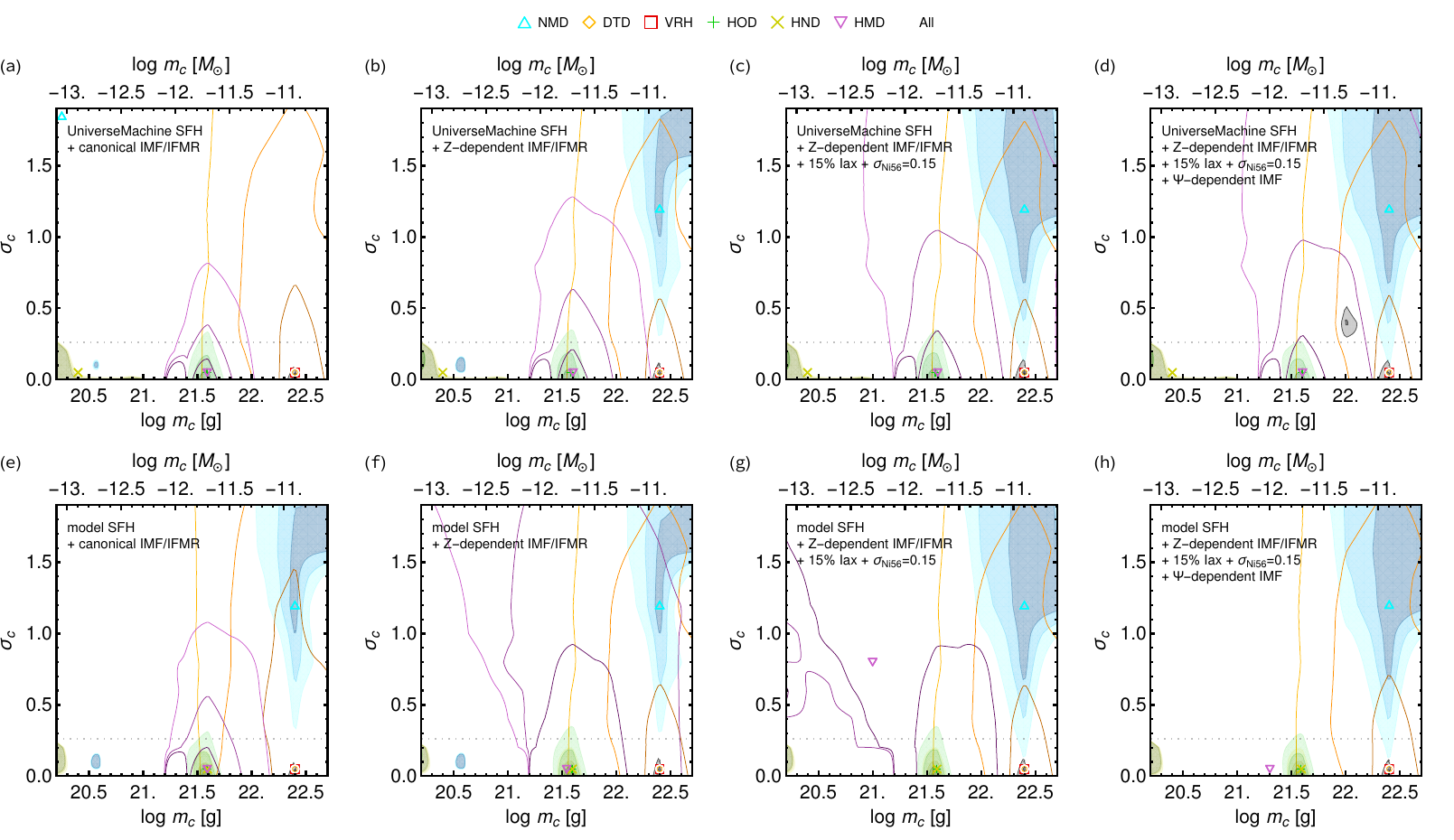}
    \setlength{\belowcaptionskip}{23pt}
    \caption{Same as Fig.~\ref{fig:2-restricted} but for ignition efficiency $\alpha = 1$.}
    \label{fig:2-alpha-1-restricted}
\end{figure*}


\begin{figure*}[htb!]
    \centering
    \includegraphics[width=0.952\linewidth]{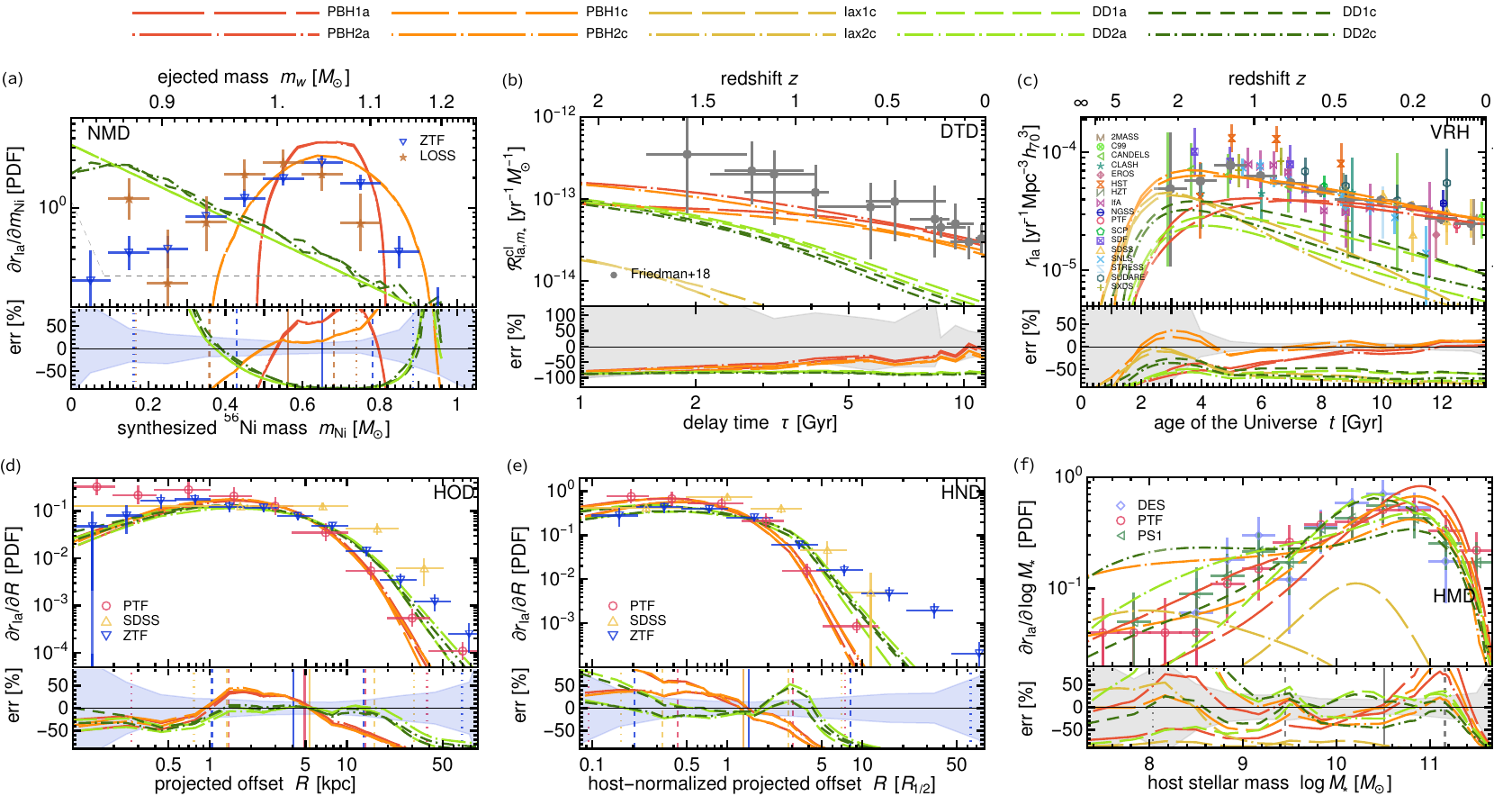}
    \setlength{\belowcaptionskip}{44.5pt}
    \caption{Same as Fig.~\ref{fig:1} of the main text
    but for DM density increased by 100\% or, equivalently, $f_{\rm PBH} = 2$. 
    }
    \label{fig:1-logfPBH-0.3}
\end{figure*}

\begin{figure*}[htb!]
    \centering
    \includegraphics[width=0.952\linewidth]{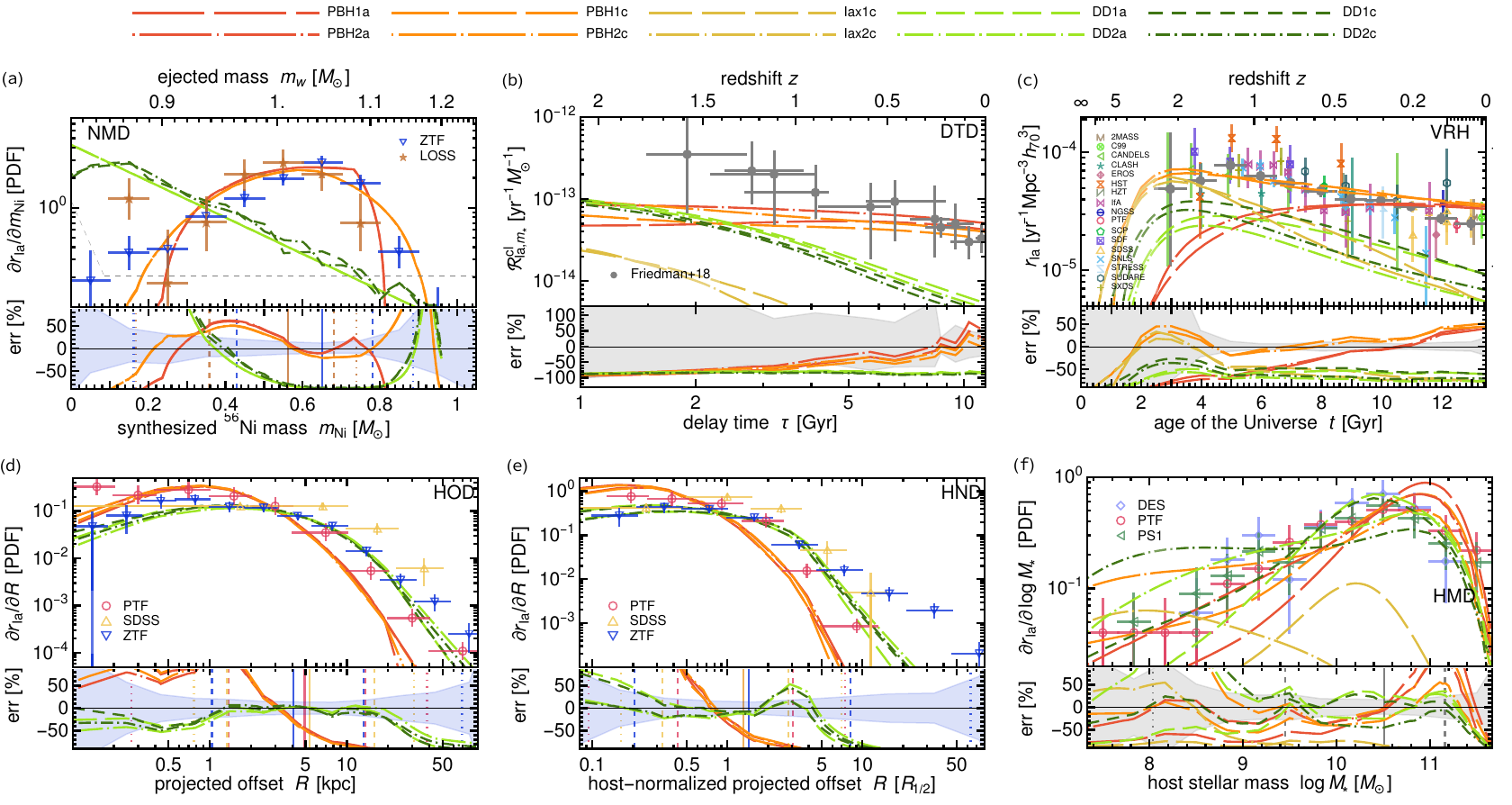}
    \setlength{\belowcaptionskip}{44pt}
    \caption{Same as Fig.~\ref{fig:1} of the main text
    but for DM density reduced by 50\% or, equivalently, $f_{\rm PBH} = 0.5$. 
    }
    \label{fig:1-logfPBH--0.3}
\end{figure*}





%% file: tables/TeXchi2-complete.txt
$\text{PBH1a}$ & $26.1$ & $3.5$ & $101.7$ & $49.4$ & $67.9$ & $78.8$ & $484.9$ \\
$\text{PBH1b}$ & $26.1$ & $4.3$ & $140.7$ & $51.7$ & $67.5$ & $54.9$ & $473.4$ \\
$\text{PBH1c}$ & $15.8$ & $3.8$ & $14.0$ & $51.7$ & $67.5$ & $32.7$ & $278.5$ \\
$\text{PBH1d}$ & $15.9$ & $3.6$ & $8.8$ & $51.1$ & $67.3$ & $39.2$ & $305.0$ \\
$\text{PBH2a}$ & $25.7$ & $2.0$ & $78.5$ & $48.0$ & $48.5$ & $21.1$ & $337.1$ \\
$\text{PBH2b}$ & $26.0$ & $3.0$ & $113.5$ & $53.9$ & $49.9$ & $29.9$ & $375.5$ \\
$\text{PBH2c}$ & $15.7$ & $3.$ & $3.9$ & $53.9$ & $49.9$ & $45.4$ & $285.1$ \\
$\text{PBH2d}$ & $15.6$ & $2.5$ & $5.5$ & $52.9$ & $51.3$ & $43.6$ & $277.0$ \\
\hline

%% file: tables/TeXchi2DD-complete.txt
$\text{DD1a}$ & $772.9$ & $37.1$ & $540.8$ & $16.2$ & $33.1$ & $20.0$ & $1420.2$ \\
$\text{DD1b}$ & $772.9$ & $41.5$ & $679.1$ & $13.2$ & $33.1$ & $10.1$ & $1549.9$ \\
$\text{DD1c}$ & $656.3$ & $41.0$ & $448.8$ & $17.1$ & $34.0$ & $10.2$ & $1207.4$ \\
$\text{DD1d}$ & $704.3$ & $39.9$ & $421.4$ & $16.8$ & $33.8$ & $12.9$ & $1229.1$ \\
$\text{DD2a}$ & $772.9$ & $39.4$ & $864.5$ & $13.0$ & $37.3$ & $29.1$ & $1756.3$ \\
$\text{DD2b}$ & $772.9$ & $43.2$ & $924.3$ & $15.0$ & $33.6$ & $101.2$ & $1890.3$ \\
$\text{DD2c}$ & $731.4$ & $43.0$ & $742.5$ & $16.0$ & $35.4$ & $124.8$ & $1693.0$ \\
$\text{DD2d}$ & $661.4$ & $41.8$ & $752.7$ & $16.5$ & $34.1$ & $95.6$ & $1602.1$ \\

%% file: tables/TeXchi2-restricted.txt
$\text{PBH1a}$ & $17.6$ & $3.5$ & $101.7$ & $28.1$ & $31.4$ & $78.8$ & $398.4$ \\
$\text{PBH1b}$ & $17.5$ & $4.3$ & $140.7$ & $30.0$ & $30.0$ & $54.9$ & $397.7$ \\
$\text{PBH1c}$ & $10.5$ & $3.8$ & $14.0$ & $30.2$ & $31.0$ & $32.7$ & $201.8$ \\
$\text{PBH1d}$ & $10.5$ & $3.6$ & $8.8$ & $29.4$ & $30.7$ & $39.2$ & $231.9$ \\
$\text{PBH2a}$ & $17.5$ & $2.0$ & $78.5$ & $28.1$ & $12.6$ & $21.1$ & $263.5$ \\
$\text{PBH2b}$ & $17.5$ & $3.0$ & $113.5$ & $33.4$ & $13.8$ & $29.9$ & $299.9$ \\
$\text{PBH2c}$ & $10.5$ & $2.8$ & $10.9$ & $33.4$ & $13.8$ & $45.4$ & $207.5$ \\
$\text{PBH2d}$ & $10.4$ & $2.5$ & $5.5$ & $32.2$ & $15.1$ & $43.6$ & $204.1$ \\
\hline

%% file: tables/TeXchi2DD-restricted.txt
$\text{DD1a}$ & $269.9$ & $37.1$ & $540.8$ & $6.3$ & $4.6$ & $20.0$ & $878.7$ \\
$\text{DD1b}$ & $269.9$ & $41.5$ & $679.1$ & $6.9$ & $5.2$ & $10.1$ & $1012.7$ \\
$\text{DD1c}$ & $301.1$ & $41.0$ & $448.8$ & $5.8$ & $4.1$ & $10.2$ & $810.9$ \\
$\text{DD1d}$ & $288.1$ & $39.9$ & $421.4$ & $5.6$ & $4.9$ & $12.9$ & $772.8$ \\
$\text{DD2a}$ & $269.9$ & $39.4$ & $864.5$ & $9.0$ & $12.5$ & $29.1$ & $1224.4$ \\
$\text{DD2b}$ & $269.9$ & $43.2$ & $924.3$ & $7.6$ & $9.0$ & $101.2$ & $1355.2$ \\
$\text{DD2c}$ & $331.9$ & $43.0$ & $742.5$ & $7.4$ & $8.5$ & $124.8$ & $1258.0$ \\
$\text{DD2d}$ & $301.6$ & $41.8$ & $752.7$ & $7.5$ & $5.4$ & $95.6$ & $1204.6$ \\

%% file: main.bbl
\begin{thebibliography}{72}%
\makeatletter
\providecommand \@ifxundefined [1]{%
 \@ifx{#1\undefined}
}%
\providecommand \@ifnum [1]{%
 \ifnum #1\expandafter \@firstoftwo
 \else \expandafter \@secondoftwo
 \fi
}%
\providecommand \@ifx [1]{%
 \ifx #1\expandafter \@firstoftwo
 \else \expandafter \@secondoftwo
 \fi
}%
\providecommand \natexlab [1]{#1}%
\providecommand \enquote  [1]{``#1''}%
\providecommand \bibnamefont  [1]{#1}%
\providecommand \bibfnamefont [1]{#1}%
\providecommand \citenamefont [1]{#1}%
\providecommand \href@noop [0]{\@secondoftwo}%
\providecommand \href [0]{\begingroup \@sanitize@url \@href}%
\providecommand \@href[1]{\@@startlink{#1}\@@href}%
\providecommand \@@href[1]{\endgroup#1\@@endlink}%
\providecommand \@sanitize@url [0]{\catcode `\\12\catcode `\$12\catcode `\&12\catcode `\#12\catcode `\^12\catcode `\_12\catcode `\%12\relax}%
\providecommand \@@startlink[1]{}%
\providecommand \@@endlink[0]{}%
\providecommand \url  [0]{\begingroup\@sanitize@url \@url }%
\providecommand \@url [1]{\endgroup\@href {#1}{\urlprefix }}%
\providecommand \urlprefix  [0]{URL }%
\providecommand \Eprint [0]{\href }%
\providecommand \doibase [0]{https://doi.org/}%
\providecommand \selectlanguage [0]{\@gobble}%
\providecommand \bibinfo  [0]{\@secondoftwo}%
\providecommand \bibfield  [0]{\@secondoftwo}%
\providecommand \translation [1]{[#1]}%
\providecommand \BibitemOpen [0]{}%
\providecommand \bibitemStop [0]{}%
\providecommand \bibitemNoStop [0]{.\EOS\space}%
\providecommand \EOS [0]{\spacefactor3000\relax}%
\providecommand \BibitemShut  [1]{\csname bibitem#1\endcsname}%
\let\auto@bib@innerbib\@empty
\bibitem [{\citenamefont {{Hoyle}}\ and\ \citenamefont {{Fowler}}(1960)}]{1960ApJ...132..565H}%
  \BibitemOpen
  \bibfield  {author} {\bibinfo {author} {\bibfnamefont {F.}~\bibnamefont {{Hoyle}}}\ and\ \bibinfo {author} {\bibfnamefont {W.~A.}\ \bibnamefont {{Fowler}}},\ }\bibfield  {title} {\bibinfo {title} {{Nucleosynthesis in Supernovae.}},\ }\href {https://doi.org/10.1086/146963} {\bibfield  {journal} {\bibinfo  {journal} {\apj}\ }\textbf {\bibinfo {volume} {132}},\ \bibinfo {pages} {565} (\bibinfo {year} {1960})}\BibitemShut {NoStop}%
\bibitem [{\citenamefont {{Arnett}}(1969)}]{1969Ap&SS...5..180A}%
  \BibitemOpen
  \bibfield  {author} {\bibinfo {author} {\bibfnamefont {W.~D.}\ \bibnamefont {{Arnett}}},\ }\bibfield  {title} {\bibinfo {title} {{A Possible Model of Supernovae: Detonation of $^{12}$C}},\ }\href {https://doi.org/10.1007/BF00650291} {\bibfield  {journal} {\bibinfo  {journal} {\apss}\ }\textbf {\bibinfo {volume} {5}},\ \bibinfo {pages} {180} (\bibinfo {year} {1969})}\BibitemShut {NoStop}%
\bibitem [{\citenamefont {{Pskovskii}}(1977)}]{1977SvA....21..675P}%
  \BibitemOpen
  \bibfield  {author} {\bibinfo {author} {\bibfnamefont {I.~P.}\ \bibnamefont {{Pskovskii}}},\ }\bibfield  {title} {\bibinfo {title} {{Light curves, color curves, and expansion velocity of type I supernovae as functions of the rate of brightness decline}},\ }\href@noop {} {\bibfield  {journal} {\bibinfo  {journal} {\sovast}\ }\textbf {\bibinfo {volume} {21}},\ \bibinfo {pages} {675} (\bibinfo {year} {1977})}\BibitemShut {NoStop}%
\bibitem [{\citenamefont {{Phillips}}(1993)}]{1993ApJ...413L.105P}%
  \BibitemOpen
  \bibfield  {author} {\bibinfo {author} {\bibfnamefont {M.~M.}\ \bibnamefont {{Phillips}}},\ }\bibfield  {title} {\bibinfo {title} {{The Absolute Magnitudes of Type IA Supernovae}},\ }\href {https://doi.org/10.1086/186970} {\bibfield  {journal} {\bibinfo  {journal} {\apjl}\ }\textbf {\bibinfo {volume} {413}},\ \bibinfo {pages} {L105} (\bibinfo {year} {1993})}\BibitemShut {NoStop}%
\bibitem [{\citenamefont {{Riess}}\ \emph {et~al.}(1998)\citenamefont {{Riess}}, \citenamefont {{Filippenko}}, \citenamefont {{Challis}}, \citenamefont {{Clocchiatti}}, \citenamefont {{Diercks}}, \citenamefont {{Garnavich}}, \citenamefont {{Gilliland}}, \citenamefont {{Hogan}}, \citenamefont {{Jha}}, \citenamefont {{Kirshner}} \emph {et~al.}}]{1998AJ....116.1009Rmax10}%
  \BibitemOpen
  \bibfield  {author} {\bibinfo {author} {\bibfnamefont {A.~G.}\ \bibnamefont {{Riess}}}, \bibinfo {author} {\bibfnamefont {A.~V.}\ \bibnamefont {{Filippenko}}}, \bibinfo {author} {\bibfnamefont {P.}~\bibnamefont {{Challis}}}, \bibinfo {author} {\bibfnamefont {A.}~\bibnamefont {{Clocchiatti}}}, \bibinfo {author} {\bibfnamefont {A.}~\bibnamefont {{Diercks}}}, \bibinfo {author} {\bibfnamefont {P.~M.}\ \bibnamefont {{Garnavich}}}, \bibinfo {author} {\bibfnamefont {R.~L.}\ \bibnamefont {{Gilliland}}}, \bibinfo {author} {\bibfnamefont {C.~J.}\ \bibnamefont {{Hogan}}}, \bibinfo {author} {\bibfnamefont {S.}~\bibnamefont {{Jha}}}, \bibinfo {author} {\bibfnamefont {R.~P.}\ \bibnamefont {{Kirshner}}}, \emph {et~al.},\ }\bibfield  {title} {\bibinfo {title} {{Observational Evidence from Supernovae for an Accelerating Universe and a Cosmological Constant}},\ }\href {https://doi.org/10.1086/300499} {\bibfield  {journal} {\bibinfo  {journal} {\aj}\ }\textbf {\bibinfo {volume} {116}},\ \bibinfo {pages} {1009} (\bibinfo
  {year} {1998})},\ \Eprint {https://arxiv.org/abs/astro-ph/9805201} {arXiv:astro-ph/9805201 [astro-ph]} \BibitemShut {NoStop}%
\bibitem [{\citenamefont {{Perlmutter}}\ \emph {et~al.}(1999)\citenamefont {{Perlmutter}}, \citenamefont {{Aldering}}, \citenamefont {{Goldhaber}}, \citenamefont {{Knop}}, \citenamefont {{Nugent}}, \citenamefont {{Castro}}, \citenamefont {{Deustua}}, \citenamefont {{Fabbro}}, \citenamefont {{Goobar}}, \citenamefont {{Groom}} \emph {et~al.}}]{1999ApJ...517..565Pmax10}%
  \BibitemOpen
  \bibfield  {author} {\bibinfo {author} {\bibfnamefont {S.}~\bibnamefont {{Perlmutter}}}, \bibinfo {author} {\bibfnamefont {G.}~\bibnamefont {{Aldering}}}, \bibinfo {author} {\bibfnamefont {G.}~\bibnamefont {{Goldhaber}}}, \bibinfo {author} {\bibfnamefont {R.~A.}\ \bibnamefont {{Knop}}}, \bibinfo {author} {\bibfnamefont {P.}~\bibnamefont {{Nugent}}}, \bibinfo {author} {\bibfnamefont {P.~G.}\ \bibnamefont {{Castro}}}, \bibinfo {author} {\bibfnamefont {S.}~\bibnamefont {{Deustua}}}, \bibinfo {author} {\bibfnamefont {S.}~\bibnamefont {{Fabbro}}}, \bibinfo {author} {\bibfnamefont {A.}~\bibnamefont {{Goobar}}}, \bibinfo {author} {\bibfnamefont {D.~E.}\ \bibnamefont {{Groom}}}, \emph {et~al.},\ }\bibfield  {title} {\bibinfo {title} {{Measurements of {\ensuremath{\Omega}} and {\ensuremath{\Lambda}} from 42 High-Redshift Supernovae}},\ }\href {https://doi.org/10.1086/307221} {\bibfield  {journal} {\bibinfo  {journal} {\apj}\ }\textbf {\bibinfo {volume} {517}},\ \bibinfo {pages} {565} (\bibinfo {year} {1999})},\
  \Eprint {https://arxiv.org/abs/astro-ph/9812133} {arXiv:astro-ph/9812133 [astro-ph]} \BibitemShut {NoStop}%
\bibitem [{\citenamefont {{Liu}}\ \emph {et~al.}(2023)\citenamefont {{Liu}}, \citenamefont {{R{\"o}pke}},\ and\ \citenamefont {{Han}}}]{2023RAA....23h2001L}%
  \BibitemOpen
  \bibfield  {author} {\bibinfo {author} {\bibfnamefont {Z.-W.}\ \bibnamefont {{Liu}}}, \bibinfo {author} {\bibfnamefont {F.~K.}\ \bibnamefont {{R{\"o}pke}}},\ and\ \bibinfo {author} {\bibfnamefont {Z.}~\bibnamefont {{Han}}},\ }\bibfield  {title} {\bibinfo {title} {{Type Ia Supernova Explosions in Binary Systems: A Review}},\ }\href {https://doi.org/10.1088/1674-4527/acd89e} {\bibfield  {journal} {\bibinfo  {journal} {Research in Astronomy and Astrophysics}\ }\textbf {\bibinfo {volume} {23}},\ \bibinfo {eid} {082001} (\bibinfo {year} {2023})},\ \Eprint {https://arxiv.org/abs/2305.13305} {arXiv:2305.13305 [astro-ph.HE]} \BibitemShut {NoStop}%
\bibitem [{\citenamefont {{Mannucci}}\ \emph {et~al.}(2008)\citenamefont {{Mannucci}}, \citenamefont {{Maoz}}, \citenamefont {{Sharon}}, \citenamefont {{Botticella}}, \citenamefont {{Della Valle}}, \citenamefont {{Gal-Yam}},\ and\ \citenamefont {{Panagia}}}]{2008MNRAS.383.1121M}%
  \BibitemOpen
  \bibfield  {author} {\bibinfo {author} {\bibfnamefont {F.}~\bibnamefont {{Mannucci}}}, \bibinfo {author} {\bibfnamefont {D.}~\bibnamefont {{Maoz}}}, \bibinfo {author} {\bibfnamefont {K.}~\bibnamefont {{Sharon}}}, \bibinfo {author} {\bibfnamefont {M.~T.}\ \bibnamefont {{Botticella}}}, \bibinfo {author} {\bibfnamefont {M.}~\bibnamefont {{Della Valle}}}, \bibinfo {author} {\bibfnamefont {A.}~\bibnamefont {{Gal-Yam}}},\ and\ \bibinfo {author} {\bibfnamefont {N.}~\bibnamefont {{Panagia}}},\ }\bibfield  {title} {\bibinfo {title} {{The supernova rate in local galaxy clusters}},\ }\href {https://doi.org/10.1111/j.1365-2966.2007.12603.x} {\bibfield  {journal} {\bibinfo  {journal} {\mnras}\ }\textbf {\bibinfo {volume} {383}},\ \bibinfo {pages} {1121} (\bibinfo {year} {2008})},\ \Eprint {https://arxiv.org/abs/0710.1094} {arXiv:0710.1094 [astro-ph]} \BibitemShut {NoStop}%
\bibitem [{\citenamefont {{Friedmann}}\ and\ \citenamefont {{Maoz}}(2018)}]{2018MNRAS.479.3563F}%
  \BibitemOpen
  \bibfield  {author} {\bibinfo {author} {\bibfnamefont {M.}~\bibnamefont {{Friedmann}}}\ and\ \bibinfo {author} {\bibfnamefont {D.}~\bibnamefont {{Maoz}}},\ }\bibfield  {title} {\bibinfo {title} {{The rate of Type-Ia supernovae in galaxy clusters and the delay-time distribution out to redshift 1.75}},\ }\href {https://doi.org/10.1093/mnras/sty1664} {\bibfield  {journal} {\bibinfo  {journal} {\mnras}\ }\textbf {\bibinfo {volume} {479}},\ \bibinfo {pages} {3563} (\bibinfo {year} {2018})},\ \Eprint {https://arxiv.org/abs/1803.04421} {arXiv:1803.04421 [astro-ph.GA]} \BibitemShut {NoStop}%
\bibitem [{\citenamefont {{Greiner}}\ \emph {et~al.}(2023)\citenamefont {{Greiner}}, \citenamefont {{Maitra}}, \citenamefont {{Haberl}}, \citenamefont {{Willer}}, \citenamefont {{Burgess}}, \citenamefont {{Langer}}, \citenamefont {{Bodensteiner}}, \citenamefont {{Buckley}}, \citenamefont {{Monageng}}, \citenamefont {{Udalski}}, \citenamefont {{Ritter}}, \citenamefont {{Werner}}, \citenamefont {{Maggi}}, \citenamefont {{Jayaraman}},\ and\ \citenamefont {{Vanderspek}}}]{2023Natur.615..605G}%
  \BibitemOpen
  \bibfield  {author} {\bibinfo {author} {\bibfnamefont {J.}~\bibnamefont {{Greiner}}}, \bibinfo {author} {\bibfnamefont {C.}~\bibnamefont {{Maitra}}}, \bibinfo {author} {\bibfnamefont {F.}~\bibnamefont {{Haberl}}}, \bibinfo {author} {\bibfnamefont {R.}~\bibnamefont {{Willer}}}, \bibinfo {author} {\bibfnamefont {J.~M.}\ \bibnamefont {{Burgess}}}, \bibinfo {author} {\bibfnamefont {N.}~\bibnamefont {{Langer}}}, \bibinfo {author} {\bibfnamefont {J.}~\bibnamefont {{Bodensteiner}}}, \bibinfo {author} {\bibfnamefont {D.~A.~H.}\ \bibnamefont {{Buckley}}}, \bibinfo {author} {\bibfnamefont {I.~M.}\ \bibnamefont {{Monageng}}}, \bibinfo {author} {\bibfnamefont {A.}~\bibnamefont {{Udalski}}}, \bibinfo {author} {\bibfnamefont {H.}~\bibnamefont {{Ritter}}}, \bibinfo {author} {\bibfnamefont {K.}~\bibnamefont {{Werner}}}, \bibinfo {author} {\bibfnamefont {P.}~\bibnamefont {{Maggi}}}, \bibinfo {author} {\bibfnamefont {R.}~\bibnamefont {{Jayaraman}}},\ and\ \bibinfo {author} {\bibfnamefont {R.}~\bibnamefont {{Vanderspek}}},\
  }\bibfield  {title} {\bibinfo {title} {{A helium-burning white dwarf binary as a supersoft X-ray source}},\ }\href {https://doi.org/10.1038/s41586-023-05714-4} {\bibfield  {journal} {\bibinfo  {journal} {\nat}\ }\textbf {\bibinfo {volume} {615}},\ \bibinfo {pages} {605} (\bibinfo {year} {2023})},\ \Eprint {https://arxiv.org/abs/2303.13338} {arXiv:2303.13338 [astro-ph.HE]} \BibitemShut {NoStop}%
\bibitem [{\citenamefont {{Ruiter}}\ \emph {et~al.}(2011)\citenamefont {{Ruiter}}, \citenamefont {{Belczynski}}, \citenamefont {{Sim}}, \citenamefont {{Hillebrandt}}, \citenamefont {{Fryer}}, \citenamefont {{Fink}},\ and\ \citenamefont {{Kromer}}}]{2011MNRAS.417..408R}%
  \BibitemOpen
  \bibfield  {author} {\bibinfo {author} {\bibfnamefont {A.~J.}\ \bibnamefont {{Ruiter}}}, \bibinfo {author} {\bibfnamefont {K.}~\bibnamefont {{Belczynski}}}, \bibinfo {author} {\bibfnamefont {S.~A.}\ \bibnamefont {{Sim}}}, \bibinfo {author} {\bibfnamefont {W.}~\bibnamefont {{Hillebrandt}}}, \bibinfo {author} {\bibfnamefont {C.~L.}\ \bibnamefont {{Fryer}}}, \bibinfo {author} {\bibfnamefont {M.}~\bibnamefont {{Fink}}},\ and\ \bibinfo {author} {\bibfnamefont {M.}~\bibnamefont {{Kromer}}},\ }\bibfield  {title} {\bibinfo {title} {{Delay times and rates for Type Ia supernovae and thermonuclear explosions from double-detonation sub-Chandrasekhar mass models}},\ }\href {https://doi.org/10.1111/j.1365-2966.2011.19276.x} {\bibfield  {journal} {\bibinfo  {journal} {\mnras}\ }\textbf {\bibinfo {volume} {417}},\ \bibinfo {pages} {408} (\bibinfo {year} {2011})},\ \Eprint {https://arxiv.org/abs/1011.1407} {arXiv:1011.1407 [astro-ph.SR]} \BibitemShut {NoStop}%
\bibitem [{\citenamefont {{Toonen}}\ \emph {et~al.}(2012)\citenamefont {{Toonen}}, \citenamefont {{Nelemans}},\ and\ \citenamefont {{Portegies Zwart}}}]{2012A&A...546A..70T}%
  \BibitemOpen
  \bibfield  {author} {\bibinfo {author} {\bibfnamefont {S.}~\bibnamefont {{Toonen}}}, \bibinfo {author} {\bibfnamefont {G.}~\bibnamefont {{Nelemans}}},\ and\ \bibinfo {author} {\bibfnamefont {S.}~\bibnamefont {{Portegies Zwart}}},\ }\bibfield  {title} {\bibinfo {title} {{Supernova Type Ia progenitors from merging double white dwarfs. Using a new population synthesis model}},\ }\href {https://doi.org/10.1051/0004-6361/201218966} {\bibfield  {journal} {\bibinfo  {journal} {\aap}\ }\textbf {\bibinfo {volume} {546}},\ \bibinfo {eid} {A70} (\bibinfo {year} {2012})},\ \Eprint {https://arxiv.org/abs/1208.6446} {arXiv:1208.6446 [astro-ph.HE]} \BibitemShut {NoStop}%
\bibitem [{\citenamefont {{Sharon}}\ and\ \citenamefont {{Kushnir}}(2022)}]{2022MNRAS.509.5275S}%
  \BibitemOpen
  \bibfield  {author} {\bibinfo {author} {\bibfnamefont {A.}~\bibnamefont {{Sharon}}}\ and\ \bibinfo {author} {\bibfnamefont {D.}~\bibnamefont {{Kushnir}}},\ }\bibfield  {title} {\bibinfo {title} {{The ZTF-BTS Type Ia supernovae luminosity function is consistent with a single progenitor channel for the explosions}},\ }\href {https://doi.org/10.1093/mnras/stab3380} {\bibfield  {journal} {\bibinfo  {journal} {\mnras}\ }\textbf {\bibinfo {volume} {509}},\ \bibinfo {pages} {5275} (\bibinfo {year} {2022})},\ \Eprint {https://arxiv.org/abs/2109.06219} {arXiv:2109.06219 [astro-ph.HE]} \BibitemShut {NoStop}%
\bibitem [{\citenamefont {{Piro}}\ \emph {et~al.}(2014)\citenamefont {{Piro}}, \citenamefont {{Thompson}},\ and\ \citenamefont {{Kochanek}}}]{2014MNRAS.438.3456P}%
  \BibitemOpen
  \bibfield  {author} {\bibinfo {author} {\bibfnamefont {A.~L.}\ \bibnamefont {{Piro}}}, \bibinfo {author} {\bibfnamefont {T.~A.}\ \bibnamefont {{Thompson}}},\ and\ \bibinfo {author} {\bibfnamefont {C.~S.}\ \bibnamefont {{Kochanek}}},\ }\bibfield  {title} {\bibinfo {title} {{Reconciling $^{56}$Ni production in Type Ia supernovae with double degenerate scenarios}},\ }\href {https://doi.org/10.1093/mnras/stt2451} {\bibfield  {journal} {\bibinfo  {journal} {\mnras}\ }\textbf {\bibinfo {volume} {438}},\ \bibinfo {pages} {3456} (\bibinfo {year} {2014})},\ \Eprint {https://arxiv.org/abs/1308.0334} {arXiv:1308.0334 [astro-ph.HE]} \BibitemShut {NoStop}%
\bibitem [{\citenamefont {{Ghosh}}\ and\ \citenamefont {{Kushnir}}(2022)}]{2022MNRAS.515..286G}%
  \BibitemOpen
  \bibfield  {author} {\bibinfo {author} {\bibfnamefont {A.}~\bibnamefont {{Ghosh}}}\ and\ \bibinfo {author} {\bibfnamefont {D.}~\bibnamefont {{Kushnir}}},\ }\bibfield  {title} {\bibinfo {title} {{Confronting double-detonation sub-Chandrasekhar models with the low-luminosity suppression of Type Ia supernovae}},\ }\href {https://doi.org/10.1093/mnras/stac1846} {\bibfield  {journal} {\bibinfo  {journal} {\mnras}\ }\textbf {\bibinfo {volume} {515}},\ \bibinfo {pages} {286} (\bibinfo {year} {2022})},\ \Eprint {https://arxiv.org/abs/2110.13169} {arXiv:2110.13169 [astro-ph.HE]} \BibitemShut {NoStop}%
\bibitem [{\citenamefont {{Graham}}\ \emph {et~al.}(2015)\citenamefont {{Graham}}, \citenamefont {{Rajendran}},\ and\ \citenamefont {{Varela}}}]{2015PhRvD..92f3007G}%
  \BibitemOpen
  \bibfield  {author} {\bibinfo {author} {\bibfnamefont {P.~W.}\ \bibnamefont {{Graham}}}, \bibinfo {author} {\bibfnamefont {S.}~\bibnamefont {{Rajendran}}},\ and\ \bibinfo {author} {\bibfnamefont {J.}~\bibnamefont {{Varela}}},\ }\bibfield  {title} {\bibinfo {title} {{Dark matter triggers of supernovae}},\ }\href {https://doi.org/10.1103/PhysRevD.92.063007} {\bibfield  {journal} {\bibinfo  {journal} {\prd}\ }\textbf {\bibinfo {volume} {92}},\ \bibinfo {eid} {063007} (\bibinfo {year} {2015})},\ \Eprint {https://arxiv.org/abs/1505.04444} {arXiv:1505.04444 [hep-ph]} \BibitemShut {NoStop}%
\bibitem [{\citenamefont {{Montero-Camacho}}\ \emph {et~al.}(2019)\citenamefont {{Montero-Camacho}}, \citenamefont {{Fang}}, \citenamefont {{Vasquez}}, \citenamefont {{Silva}},\ and\ \citenamefont {{Hirata}}}]{2019JCAP...08..031M}%
  \BibitemOpen
  \bibfield  {author} {\bibinfo {author} {\bibfnamefont {P.}~\bibnamefont {{Montero-Camacho}}}, \bibinfo {author} {\bibfnamefont {X.}~\bibnamefont {{Fang}}}, \bibinfo {author} {\bibfnamefont {G.}~\bibnamefont {{Vasquez}}}, \bibinfo {author} {\bibfnamefont {M.}~\bibnamefont {{Silva}}},\ and\ \bibinfo {author} {\bibfnamefont {C.~M.}\ \bibnamefont {{Hirata}}},\ }\bibfield  {title} {\bibinfo {title} {{Revisiting constraints on asteroid-mass primordial black holes as dark matter candidates}},\ }\href {https://doi.org/10.1088/1475-7516/2019/08/031} {\bibfield  {journal} {\bibinfo  {journal} {\jcap}\ }\textbf {\bibinfo {volume} {2019}},\ \bibinfo {eid} {031} (\bibinfo {year} {2019})},\ \Eprint {https://arxiv.org/abs/1906.05950} {arXiv:1906.05950 [astro-ph.CO]} \BibitemShut {NoStop}%
\bibitem [{\citenamefont {{Steigerwald}}\ and\ \citenamefont {{Tejeda}}(2021)}]{2021PhRvL.127a1101S}%
  \BibitemOpen
  \bibfield  {author} {\bibinfo {author} {\bibfnamefont {H.}~\bibnamefont {{Steigerwald}}}\ and\ \bibinfo {author} {\bibfnamefont {E.}~\bibnamefont {{Tejeda}}},\ }\bibfield  {title} {\bibinfo {title} {{Bondi-Hoyle-Lyttleton Accretion in a Reactive Medium: Detonation Ignition and a Mechanism for Type Ia Supernovae}},\ }\href {https://doi.org/10.1103/PhysRevLett.127.011101} {\bibfield  {journal} {\bibinfo  {journal} {\prl}\ }\textbf {\bibinfo {volume} {127}},\ \bibinfo {eid} {011101} (\bibinfo {year} {2021})},\ \Eprint {https://arxiv.org/abs/2104.07066} {arXiv:2104.07066 [astro-ph.HE]} \BibitemShut {NoStop}%
\bibitem [{\citenamefont {{Olling}}\ \emph {et~al.}(2015)\citenamefont {{Olling}}, \citenamefont {{Mushotzky}}, \citenamefont {{Shaya}}, \citenamefont {{Rest}}, \citenamefont {{Garnavich}}, \citenamefont {{Tucker}}, \citenamefont {{Kasen}}, \citenamefont {{Margheim}},\ and\ \citenamefont {{Filippenko}}}]{2015Natur.521..332O}%
  \BibitemOpen
  \bibfield  {author} {\bibinfo {author} {\bibfnamefont {R.~P.}\ \bibnamefont {{Olling}}}, \bibinfo {author} {\bibfnamefont {R.}~\bibnamefont {{Mushotzky}}}, \bibinfo {author} {\bibfnamefont {E.~J.}\ \bibnamefont {{Shaya}}}, \bibinfo {author} {\bibfnamefont {A.}~\bibnamefont {{Rest}}}, \bibinfo {author} {\bibfnamefont {P.~M.}\ \bibnamefont {{Garnavich}}}, \bibinfo {author} {\bibfnamefont {B.~E.}\ \bibnamefont {{Tucker}}}, \bibinfo {author} {\bibfnamefont {D.}~\bibnamefont {{Kasen}}}, \bibinfo {author} {\bibfnamefont {S.}~\bibnamefont {{Margheim}}},\ and\ \bibinfo {author} {\bibfnamefont {A.~V.}\ \bibnamefont {{Filippenko}}},\ }\bibfield  {title} {\bibinfo {title} {{No signature of ejecta interaction with a stellar companion in three type Ia supernovae}},\ }\href {https://doi.org/10.1038/nature14455} {\bibfield  {journal} {\bibinfo  {journal} {\nat}\ }\textbf {\bibinfo {volume} {521}},\ \bibinfo {pages} {332} (\bibinfo {year} {2015})}\BibitemShut {NoStop}%
\bibitem [{\citenamefont {{Shields}}\ \emph {et~al.}(2022)\citenamefont {{Shields}}, \citenamefont {{Kerzendorf}}, \citenamefont {{Hosek}}, \citenamefont {{Shen}}, \citenamefont {{Rest}}, \citenamefont {{Do}}, \citenamefont {{Lu}}, \citenamefont {{Fullard}}, \citenamefont {{Strampelli}},\ and\ \citenamefont {{Zenteno}}}]{2022ApJ...933L..31S}%
  \BibitemOpen
  \bibfield  {author} {\bibinfo {author} {\bibfnamefont {J.~V.}\ \bibnamefont {{Shields}}}, \bibinfo {author} {\bibfnamefont {W.}~\bibnamefont {{Kerzendorf}}}, \bibinfo {author} {\bibfnamefont {M.~W.}\ \bibnamefont {{Hosek}}}, \bibinfo {author} {\bibfnamefont {K.~J.}\ \bibnamefont {{Shen}}}, \bibinfo {author} {\bibfnamefont {A.}~\bibnamefont {{Rest}}}, \bibinfo {author} {\bibfnamefont {T.}~\bibnamefont {{Do}}}, \bibinfo {author} {\bibfnamefont {J.~R.}\ \bibnamefont {{Lu}}}, \bibinfo {author} {\bibfnamefont {A.~G.}\ \bibnamefont {{Fullard}}}, \bibinfo {author} {\bibfnamefont {G.}~\bibnamefont {{Strampelli}}},\ and\ \bibinfo {author} {\bibfnamefont {A.}~\bibnamefont {{Zenteno}}},\ }\bibfield  {title} {\bibinfo {title} {{Searching for a Hypervelocity White Dwarf SN Ia Companion: A Proper-motion Survey of SN 1006}},\ }\href {https://doi.org/10.3847/2041-8213/ac7950} {\bibfield  {journal} {\bibinfo  {journal} {\apjl}\ }\textbf {\bibinfo {volume} {933}},\ \bibinfo {eid} {L31} (\bibinfo {year} {2022})},\ \Eprint
  {https://arxiv.org/abs/2206.04095} {arXiv:2206.04095 [astro-ph.SR]} \BibitemShut {NoStop}%
\bibitem [{\citenamefont {{Sim}}\ \emph {et~al.}(2010)\citenamefont {{Sim}}, \citenamefont {{R{\"o}pke}}, \citenamefont {{Hillebrandt}}, \citenamefont {{Kromer}}, \citenamefont {{Pakmor}}, \citenamefont {{Fink}}, \citenamefont {{Ruiter}},\ and\ \citenamefont {{Seitenzahl}}}]{2010ApJ...714L..52S}%
  \BibitemOpen
  \bibfield  {author} {\bibinfo {author} {\bibfnamefont {S.~A.}\ \bibnamefont {{Sim}}}, \bibinfo {author} {\bibfnamefont {F.~K.}\ \bibnamefont {{R{\"o}pke}}}, \bibinfo {author} {\bibfnamefont {W.}~\bibnamefont {{Hillebrandt}}}, \bibinfo {author} {\bibfnamefont {M.}~\bibnamefont {{Kromer}}}, \bibinfo {author} {\bibfnamefont {R.}~\bibnamefont {{Pakmor}}}, \bibinfo {author} {\bibfnamefont {M.}~\bibnamefont {{Fink}}}, \bibinfo {author} {\bibfnamefont {A.~J.}\ \bibnamefont {{Ruiter}}},\ and\ \bibinfo {author} {\bibfnamefont {I.~R.}\ \bibnamefont {{Seitenzahl}}},\ }\bibfield  {title} {\bibinfo {title} {{Detonations in Sub-Chandrasekhar-mass C+O White Dwarfs}},\ }\href {https://doi.org/10.1088/2041-8205/714/1/L52} {\bibfield  {journal} {\bibinfo  {journal} {\apjl}\ }\textbf {\bibinfo {volume} {714}},\ \bibinfo {pages} {L52} (\bibinfo {year} {2010})},\ \Eprint {https://arxiv.org/abs/1003.2917} {arXiv:1003.2917 [astro-ph.HE]} \BibitemShut {NoStop}%
\bibitem [{\citenamefont {{Blondin}}\ \emph {et~al.}(2017)\citenamefont {{Blondin}}, \citenamefont {{Dessart}}, \citenamefont {{Hillier}},\ and\ \citenamefont {{Khokhlov}}}]{2017MNRAS.470..157B}%
  \BibitemOpen
  \bibfield  {author} {\bibinfo {author} {\bibfnamefont {S.}~\bibnamefont {{Blondin}}}, \bibinfo {author} {\bibfnamefont {L.}~\bibnamefont {{Dessart}}}, \bibinfo {author} {\bibfnamefont {D.~J.}\ \bibnamefont {{Hillier}}},\ and\ \bibinfo {author} {\bibfnamefont {A.~M.}\ \bibnamefont {{Khokhlov}}},\ }\bibfield  {title} {\bibinfo {title} {{Evidence for sub-Chandrasekhar-mass progenitors of Type Ia supernovae at the faint end of the width-luminosity relation}},\ }\href {https://doi.org/10.1093/mnras/stw2492} {\bibfield  {journal} {\bibinfo  {journal} {\mnras}\ }\textbf {\bibinfo {volume} {470}},\ \bibinfo {pages} {157} (\bibinfo {year} {2017})},\ \Eprint {https://arxiv.org/abs/1706.01901} {arXiv:1706.01901 [astro-ph.SR]} \BibitemShut {NoStop}%
\bibitem [{\citenamefont {{Shen}}\ \emph {et~al.}(2018)\citenamefont {{Shen}}, \citenamefont {{Kasen}}, \citenamefont {{Miles}},\ and\ \citenamefont {{Townsley}}}]{2018ApJ...854...52S}%
  \BibitemOpen
  \bibfield  {author} {\bibinfo {author} {\bibfnamefont {K.~J.}\ \bibnamefont {{Shen}}}, \bibinfo {author} {\bibfnamefont {D.}~\bibnamefont {{Kasen}}}, \bibinfo {author} {\bibfnamefont {B.~J.}\ \bibnamefont {{Miles}}},\ and\ \bibinfo {author} {\bibfnamefont {D.~M.}\ \bibnamefont {{Townsley}}},\ }\bibfield  {title} {\bibinfo {title} {{Sub-Chandrasekhar-mass White Dwarf Detonations Revisited}},\ }\href {https://doi.org/10.3847/1538-4357/aaa8de} {\bibfield  {journal} {\bibinfo  {journal} {\apj}\ }\textbf {\bibinfo {volume} {854}},\ \bibinfo {eid} {52} (\bibinfo {year} {2018})},\ \Eprint {https://arxiv.org/abs/1706.01898} {arXiv:1706.01898 [astro-ph.HE]} \BibitemShut {NoStop}%
\bibitem [{\citenamefont {{Shen}}\ \emph {et~al.}(2021)\citenamefont {{Shen}}, \citenamefont {{Blondin}}, \citenamefont {{Kasen}}, \citenamefont {{Dessart}}, \citenamefont {{Townsley}}, \citenamefont {{Boos}},\ and\ \citenamefont {{Hillier}}}]{2021ApJ...909L..18S}%
  \BibitemOpen
  \bibfield  {author} {\bibinfo {author} {\bibfnamefont {K.~J.}\ \bibnamefont {{Shen}}}, \bibinfo {author} {\bibfnamefont {S.}~\bibnamefont {{Blondin}}}, \bibinfo {author} {\bibfnamefont {D.}~\bibnamefont {{Kasen}}}, \bibinfo {author} {\bibfnamefont {L.}~\bibnamefont {{Dessart}}}, \bibinfo {author} {\bibfnamefont {D.~M.}\ \bibnamefont {{Townsley}}}, \bibinfo {author} {\bibfnamefont {S.}~\bibnamefont {{Boos}}},\ and\ \bibinfo {author} {\bibfnamefont {D.~J.}\ \bibnamefont {{Hillier}}},\ }\bibfield  {title} {\bibinfo {title} {{Non-local Thermodynamic Equilibrium Radiative Transfer Simulations of Sub-Chandrasekhar-mass White Dwarf Detonations}},\ }\href {https://doi.org/10.3847/2041-8213/abe69b} {\bibfield  {journal} {\bibinfo  {journal} {\apjl}\ }\textbf {\bibinfo {volume} {909}},\ \bibinfo {eid} {L18} (\bibinfo {year} {2021})},\ \Eprint {https://arxiv.org/abs/2102.08238} {arXiv:2102.08238 [astro-ph.HE]} \BibitemShut {NoStop}%
\bibitem [{\citenamefont {{Livneh}}\ and\ \citenamefont {{Katz}}(2022)}]{2022MNRAS.511.2994L}%
  \BibitemOpen
  \bibfield  {author} {\bibinfo {author} {\bibfnamefont {R.}~\bibnamefont {{Livneh}}}\ and\ \bibinfo {author} {\bibfnamefont {B.}~\bibnamefont {{Katz}}},\ }\bibfield  {title} {\bibinfo {title} {{Polarization signatures of the head-on collision model for Type Ia supernovae: how much asymmetry is too much?}},\ }\href {https://doi.org/10.1093/mnras/stab3787} {\bibfield  {journal} {\bibinfo  {journal} {\mnras}\ }\textbf {\bibinfo {volume} {511}},\ \bibinfo {pages} {2994} (\bibinfo {year} {2022})},\ \Eprint {https://arxiv.org/abs/2109.10371} {arXiv:2109.10371 [astro-ph.HE]} \BibitemShut {NoStop}%
\bibitem [{\citenamefont {{Kushnir}}\ \emph {et~al.}(2020)\citenamefont {{Kushnir}}, \citenamefont {{Wygoda}},\ and\ \citenamefont {{Sharon}}}]{2020MNRAS.499.4725K}%
  \BibitemOpen
  \bibfield  {author} {\bibinfo {author} {\bibfnamefont {D.}~\bibnamefont {{Kushnir}}}, \bibinfo {author} {\bibfnamefont {N.}~\bibnamefont {{Wygoda}}},\ and\ \bibinfo {author} {\bibfnamefont {A.}~\bibnamefont {{Sharon}}},\ }\bibfield  {title} {\bibinfo {title} {{Sub-Chandrasekhar-mass detonations are in tension with the observed t$_{0}$-M$_{Ni56}$ relation of type Ia supernovae}},\ }\href {https://doi.org/10.1093/mnras/staa3017} {\bibfield  {journal} {\bibinfo  {journal} {\mnras}\ }\textbf {\bibinfo {volume} {499}},\ \bibinfo {pages} {4725} (\bibinfo {year} {2020})},\ \Eprint {https://arxiv.org/abs/2008.08592} {arXiv:2008.08592 [astro-ph.HE]} \BibitemShut {NoStop}%
\bibitem [{\citenamefont {{Carr}}\ and\ \citenamefont {{K{\"u}hnel}}(2020)}]{2020ARNPS..70..355C}%
  \BibitemOpen
  \bibfield  {author} {\bibinfo {author} {\bibfnamefont {B.}~\bibnamefont {{Carr}}}\ and\ \bibinfo {author} {\bibfnamefont {F.}~\bibnamefont {{K{\"u}hnel}}},\ }\bibfield  {title} {\bibinfo {title} {{Primordial Black Holes as Dark Matter: Recent Developments}},\ }\href {https://doi.org/10.1146/annurev-nucl-050520-125911} {\bibfield  {journal} {\bibinfo  {journal} {Annual Review of Nuclear and Particle Science}\ }\textbf {\bibinfo {volume} {70}},\ \bibinfo {pages} {355} (\bibinfo {year} {2020})},\ \Eprint {https://arxiv.org/abs/2006.02838} {arXiv:2006.02838 [astro-ph.CO]} \BibitemShut {NoStop}%
\bibitem [{\citenamefont {{Carr}}\ \emph {et~al.}(2021)\citenamefont {{Carr}}, \citenamefont {{Kohri}}, \citenamefont {{Sendouda}},\ and\ \citenamefont {{Yokoyama}}}]{2021RPPh...84k6902C}%
  \BibitemOpen
  \bibfield  {author} {\bibinfo {author} {\bibfnamefont {B.}~\bibnamefont {{Carr}}}, \bibinfo {author} {\bibfnamefont {K.}~\bibnamefont {{Kohri}}}, \bibinfo {author} {\bibfnamefont {Y.}~\bibnamefont {{Sendouda}}},\ and\ \bibinfo {author} {\bibfnamefont {J.}~\bibnamefont {{Yokoyama}}},\ }\bibfield  {title} {\bibinfo {title} {{Constraints on primordial black holes}},\ }\href {https://doi.org/10.1088/1361-6633/ac1e31} {\bibfield  {journal} {\bibinfo  {journal} {Reports on Progress in Physics}\ }\textbf {\bibinfo {volume} {84}},\ \bibinfo {eid} {116902} (\bibinfo {year} {2021})},\ \Eprint {https://arxiv.org/abs/2002.12778} {arXiv:2002.12778 [astro-ph.CO]} \BibitemShut {NoStop}%
\bibitem [{\citenamefont {{Smyth}}\ \emph {et~al.}(2020)\citenamefont {{Smyth}}, \citenamefont {{Profumo}}, \citenamefont {{English}}, \citenamefont {{Jeltema}}, \citenamefont {{McKinnon}},\ and\ \citenamefont {{Guhathakurta}}}]{2020PhRvD.101f3005S}%
  \BibitemOpen
  \bibfield  {author} {\bibinfo {author} {\bibfnamefont {N.}~\bibnamefont {{Smyth}}}, \bibinfo {author} {\bibfnamefont {S.}~\bibnamefont {{Profumo}}}, \bibinfo {author} {\bibfnamefont {S.}~\bibnamefont {{English}}}, \bibinfo {author} {\bibfnamefont {T.}~\bibnamefont {{Jeltema}}}, \bibinfo {author} {\bibfnamefont {K.}~\bibnamefont {{McKinnon}}},\ and\ \bibinfo {author} {\bibfnamefont {P.}~\bibnamefont {{Guhathakurta}}},\ }\bibfield  {title} {\bibinfo {title} {{Updated constraints on asteroid-mass primordial black holes as dark matter}},\ }\href {https://doi.org/10.1103/PhysRevD.101.063005} {\bibfield  {journal} {\bibinfo  {journal} {\prd}\ }\textbf {\bibinfo {volume} {101}},\ \bibinfo {eid} {063005} (\bibinfo {year} {2020})},\ \Eprint {https://arxiv.org/abs/1910.01285} {arXiv:1910.01285 [astro-ph.CO]} \BibitemShut {NoStop}%
\bibitem [{\citenamefont {{Korwar}}\ and\ \citenamefont {{Profumo}}(2023)}]{2023JCAP...05..054K}%
  \BibitemOpen
  \bibfield  {author} {\bibinfo {author} {\bibfnamefont {M.}~\bibnamefont {{Korwar}}}\ and\ \bibinfo {author} {\bibfnamefont {S.}~\bibnamefont {{Profumo}}},\ }\bibfield  {title} {\bibinfo {title} {{Updated constraints on primordial black hole evaporation}},\ }\href {https://doi.org/10.1088/1475-7516/2023/05/054} {\bibfield  {journal} {\bibinfo  {journal} {\jcap}\ }\textbf {\bibinfo {volume} {2023}},\ \bibinfo {eid} {054} (\bibinfo {year} {2023})},\ \Eprint {https://arxiv.org/abs/2302.04408} {arXiv:2302.04408 [hep-ph]} \BibitemShut {NoStop}%
\bibitem [{\citenamefont {{Steigerwald}}(2025)}]{2025arXiv250521260S}%
  \BibitemOpen
  \bibfield  {author} {\bibinfo {author} {\bibfnamefont {H.}~\bibnamefont {{Steigerwald}}},\ }\bibfield  {title} {\bibinfo {title} {{Dark-matter-induced transients over cosmic time: The role of star formation history profiles}},\ }\href {https://doi.org/10.1103/PhysRevD.111.123543} {\bibfield  {journal} {\bibinfo  {journal} {Phys. Rev. D}\ }\textbf {\bibinfo {volume} {111}},\ \bibinfo {pages} {123543} (\bibinfo {year} {2025})},\ \Eprint {https://arxiv.org/abs/2505.21260} {arXiv:2505.21260 [astro-ph.GA]} \BibitemShut {NoStop}%
\bibitem [{\citenamefont {{Steigerwald}}\ \emph {et~al.}(2022)\citenamefont {{Steigerwald}}, \citenamefont {{Rodrigues}}, \citenamefont {{Profumo}},\ and\ \citenamefont {{Marra}}}]{2022MNRAS.510.4779S}%
  \BibitemOpen
  \bibfield  {author} {\bibinfo {author} {\bibfnamefont {H.}~\bibnamefont {{Steigerwald}}}, \bibinfo {author} {\bibfnamefont {D.}~\bibnamefont {{Rodrigues}}}, \bibinfo {author} {\bibfnamefont {S.}~\bibnamefont {{Profumo}}},\ and\ \bibinfo {author} {\bibfnamefont {V.}~\bibnamefont {{Marra}}},\ }\bibfield  {title} {\bibinfo {title} {{Type Ia supernova magnitude step from the local dark matter environment}},\ }\href {https://doi.org/10.1093/mnras/stab3747} {\bibfield  {journal} {\bibinfo  {journal} {\mnras}\ }\textbf {\bibinfo {volume} {510}},\ \bibinfo {pages} {4779} (\bibinfo {year} {2022})},\ \Eprint {https://arxiv.org/abs/2112.09739} {arXiv:2112.09739 [astro-ph.CO]} \BibitemShut {NoStop}%
\bibitem [{\citenamefont {{Moster}}\ \emph {et~al.}(2010)\citenamefont {{Moster}}, \citenamefont {{Somerville}}, \citenamefont {{Maulbetsch}}, \citenamefont {{van den Bosch}}, \citenamefont {{Macci{\`o}}}, \citenamefont {{Naab}},\ and\ \citenamefont {{Oser}}}]{2010ApJ...710..903M}%
  \BibitemOpen
  \bibfield  {author} {\bibinfo {author} {\bibfnamefont {B.~P.}\ \bibnamefont {{Moster}}}, \bibinfo {author} {\bibfnamefont {R.~S.}\ \bibnamefont {{Somerville}}}, \bibinfo {author} {\bibfnamefont {C.}~\bibnamefont {{Maulbetsch}}}, \bibinfo {author} {\bibfnamefont {F.~C.}\ \bibnamefont {{van den Bosch}}}, \bibinfo {author} {\bibfnamefont {A.~V.}\ \bibnamefont {{Macci{\`o}}}}, \bibinfo {author} {\bibfnamefont {T.}~\bibnamefont {{Naab}}},\ and\ \bibinfo {author} {\bibfnamefont {L.}~\bibnamefont {{Oser}}},\ }\bibfield  {title} {\bibinfo {title} {{Constraints on the Relationship between Stellar Mass and Halo Mass at Low and High Redshift}},\ }\href {https://doi.org/10.1088/0004-637X/710/2/903} {\bibfield  {journal} {\bibinfo  {journal} {\apj}\ }\textbf {\bibinfo {volume} {710}},\ \bibinfo {pages} {903} (\bibinfo {year} {2010})},\ \Eprint {https://arxiv.org/abs/0903.4682} {arXiv:0903.4682 [astro-ph.CO]} \BibitemShut {NoStop}%
\bibitem [{\citenamefont {{Girelli}}\ \emph {et~al.}(2020)\citenamefont {{Girelli}}, \citenamefont {{Pozzetti}}, \citenamefont {{Bolzonella}}, \citenamefont {{Giocoli}}, \citenamefont {{Marulli}},\ and\ \citenamefont {{Baldi}}}]{2020A&A...634A.135G}%
  \BibitemOpen
  \bibfield  {author} {\bibinfo {author} {\bibfnamefont {G.}~\bibnamefont {{Girelli}}}, \bibinfo {author} {\bibfnamefont {L.}~\bibnamefont {{Pozzetti}}}, \bibinfo {author} {\bibfnamefont {M.}~\bibnamefont {{Bolzonella}}}, \bibinfo {author} {\bibfnamefont {C.}~\bibnamefont {{Giocoli}}}, \bibinfo {author} {\bibfnamefont {F.}~\bibnamefont {{Marulli}}},\ and\ \bibinfo {author} {\bibfnamefont {M.}~\bibnamefont {{Baldi}}},\ }\bibfield  {title} {\bibinfo {title} {{The stellar-to-halo mass relation over the past 12 Gyr. I. Standard {\ensuremath{\Lambda}}CDM model}},\ }\href {https://doi.org/10.1051/0004-6361/201936329} {\bibfield  {journal} {\bibinfo  {journal} {\aap}\ }\textbf {\bibinfo {volume} {634}},\ \bibinfo {eid} {A135} (\bibinfo {year} {2020})},\ \Eprint {https://arxiv.org/abs/2001.02230} {arXiv:2001.02230 [astro-ph.CO]} \BibitemShut {NoStop}%
\bibitem [{Note10()}]{Note10}%
  \BibitemOpen
  \bibinfo {note} {See Supplemental Material (after the bibliography).}\BibitemShut {Stop}%
\bibitem [{\citenamefont {{Shen}}\ and\ \citenamefont {{Bildsten}}(2014)}]{2014ApJ...785...61S}%
  \BibitemOpen
  \bibfield  {author} {\bibinfo {author} {\bibfnamefont {K.~J.}\ \bibnamefont {{Shen}}}\ and\ \bibinfo {author} {\bibfnamefont {L.}~\bibnamefont {{Bildsten}}},\ }\bibfield  {title} {\bibinfo {title} {{The Ignition of Carbon Detonations via Converging Shock Waves in White Dwarfs}},\ }\href {https://doi.org/10.1088/0004-637X/785/1/61} {\bibfield  {journal} {\bibinfo  {journal} {\apj}\ }\textbf {\bibinfo {volume} {785}},\ \bibinfo {eid} {61} (\bibinfo {year} {2014})},\ \Eprint {https://arxiv.org/abs/1305.6925} {arXiv:1305.6925 [astro-ph.HE]} \BibitemShut {NoStop}%
\bibitem [{\citenamefont {{Timmes}}\ and\ \citenamefont {{Woosley}}(1992)}]{1992ApJ...396..649T}%
  \BibitemOpen
  \bibfield  {author} {\bibinfo {author} {\bibfnamefont {F.~X.}\ \bibnamefont {{Timmes}}}\ and\ \bibinfo {author} {\bibfnamefont {S.~E.}\ \bibnamefont {{Woosley}}},\ }\bibfield  {title} {\bibinfo {title} {{The Conductive Propagation of Nuclear Flames. I. Degenerate C + O and O + NE + MG White Dwarfs}},\ }\href {https://doi.org/10.1086/171746} {\bibfield  {journal} {\bibinfo  {journal} {\apj}\ }\textbf {\bibinfo {volume} {396}},\ \bibinfo {pages} {649} (\bibinfo {year} {1992})}\BibitemShut {NoStop}%
\bibitem [{\citenamefont {{Romero}}\ \emph {et~al.}(2013)\citenamefont {{Romero}}, \citenamefont {{Kepler}}, \citenamefont {{C{\'o}rsico}}, \citenamefont {{Althaus}},\ and\ \citenamefont {{Fraga}}}]{2013ApJ...779...58R}%
  \BibitemOpen
  \bibfield  {author} {\bibinfo {author} {\bibfnamefont {A.~D.}\ \bibnamefont {{Romero}}}, \bibinfo {author} {\bibfnamefont {S.~O.}\ \bibnamefont {{Kepler}}}, \bibinfo {author} {\bibfnamefont {A.~H.}\ \bibnamefont {{C{\'o}rsico}}}, \bibinfo {author} {\bibfnamefont {L.~G.}\ \bibnamefont {{Althaus}}},\ and\ \bibinfo {author} {\bibfnamefont {L.}~\bibnamefont {{Fraga}}},\ }\bibfield  {title} {\bibinfo {title} {{Asteroseismological Study of Massive ZZ Ceti Stars with Fully Evolutionary Models}},\ }\href {https://doi.org/10.1088/0004-637X/779/1/58} {\bibfield  {journal} {\bibinfo  {journal} {\apj}\ }\textbf {\bibinfo {volume} {779}},\ \bibinfo {eid} {58} (\bibinfo {year} {2013})},\ \Eprint {https://arxiv.org/abs/1310.4137} {arXiv:1310.4137 [astro-ph.SR]} \BibitemShut {NoStop}%
\bibitem [{\citenamefont {{Lauffer}}\ \emph {et~al.}(2018)\citenamefont {{Lauffer}}, \citenamefont {{Romero}},\ and\ \citenamefont {{Kepler}}}]{2018MNRAS.480.1547L}%
  \BibitemOpen
  \bibfield  {author} {\bibinfo {author} {\bibfnamefont {G.~R.}\ \bibnamefont {{Lauffer}}}, \bibinfo {author} {\bibfnamefont {A.~D.}\ \bibnamefont {{Romero}}},\ and\ \bibinfo {author} {\bibfnamefont {S.~O.}\ \bibnamefont {{Kepler}}},\ }\bibfield  {title} {\bibinfo {title} {{New full evolutionary sequences of H- and He-atmosphere massive white dwarf stars using MESA}},\ }\href {https://doi.org/10.1093/mnras/sty1925} {\bibfield  {journal} {\bibinfo  {journal} {\mnras}\ }\textbf {\bibinfo {volume} {480}},\ \bibinfo {pages} {1547} (\bibinfo {year} {2018})},\ \Eprint {https://arxiv.org/abs/1807.04774} {arXiv:1807.04774 [astro-ph.SR]} \BibitemShut {NoStop}%
\bibitem [{\citenamefont {{Bauer}}(2023)}]{2023ApJ...950..115B}%
  \BibitemOpen
  \bibfield  {author} {\bibinfo {author} {\bibfnamefont {E.~B.}\ \bibnamefont {{Bauer}}},\ }\bibfield  {title} {\bibinfo {title} {{Carbon-Oxygen Phase Separation in Modules for Experiments in Stellar Astrophysics (MESA) White Dwarf Models}},\ }\href {https://doi.org/10.3847/1538-4357/acd057} {\bibfield  {journal} {\bibinfo  {journal} {\apj}\ }\textbf {\bibinfo {volume} {950}},\ \bibinfo {eid} {115} (\bibinfo {year} {2023})},\ \Eprint {https://arxiv.org/abs/2303.10110} {arXiv:2303.10110 [astro-ph.SR]} \BibitemShut {NoStop}%
\bibitem [{\citenamefont {{Dolgov}}\ and\ \citenamefont {{Silk}}(1993)}]{1993PhRvD..47.4244D}%
  \BibitemOpen
  \bibfield  {author} {\bibinfo {author} {\bibfnamefont {A.}~\bibnamefont {{Dolgov}}}\ and\ \bibinfo {author} {\bibfnamefont {J.}~\bibnamefont {{Silk}}},\ }\bibfield  {title} {\bibinfo {title} {{Baryon isocurvature fluctuations at small scales and baryonic dark matter}},\ }\href {https://doi.org/10.1103/PhysRevD.47.4244} {\bibfield  {journal} {\bibinfo  {journal} {\prd}\ }\textbf {\bibinfo {volume} {47}},\ \bibinfo {pages} {4244} (\bibinfo {year} {1993})}\BibitemShut {NoStop}%
\bibitem [{\citenamefont {{Green}}(2016)}]{2016PhRvD..94f3530G}%
  \BibitemOpen
  \bibfield  {author} {\bibinfo {author} {\bibfnamefont {A.~M.}\ \bibnamefont {{Green}}},\ }\bibfield  {title} {\bibinfo {title} {{Microlensing and dynamical constraints on primordial black hole dark matter with an extended mass function}},\ }\href {https://doi.org/10.1103/PhysRevD.94.063530} {\bibfield  {journal} {\bibinfo  {journal} {\prd}\ }\textbf {\bibinfo {volume} {94}},\ \bibinfo {eid} {063530} (\bibinfo {year} {2016})},\ \Eprint {https://arxiv.org/abs/1609.01143} {arXiv:1609.01143 [astro-ph.CO]} \BibitemShut {NoStop}%
\bibitem [{\citenamefont {{Kannike}}\ \emph {et~al.}(2017)\citenamefont {{Kannike}}, \citenamefont {{Marzola}}, \citenamefont {{Raidal}},\ and\ \citenamefont {{Veerm{\"a}e}}}]{2017JCAP...09..020K}%
  \BibitemOpen
  \bibfield  {author} {\bibinfo {author} {\bibfnamefont {K.}~\bibnamefont {{Kannike}}}, \bibinfo {author} {\bibfnamefont {L.}~\bibnamefont {{Marzola}}}, \bibinfo {author} {\bibfnamefont {M.}~\bibnamefont {{Raidal}}},\ and\ \bibinfo {author} {\bibfnamefont {H.}~\bibnamefont {{Veerm{\"a}e}}},\ }\bibfield  {title} {\bibinfo {title} {{Single field double inflation and primordial black holes}},\ }\href {https://doi.org/10.1088/1475-7516/2017/09/020} {\bibfield  {journal} {\bibinfo  {journal} {\jcap}\ }\textbf {\bibinfo {volume} {2017}},\ \bibinfo {eid} {020} (\bibinfo {year} {2017})},\ \Eprint {https://arxiv.org/abs/1705.06225} {arXiv:1705.06225 [astro-ph.CO]} \BibitemShut {NoStop}%
\bibitem [{\citenamefont {{Carr}}\ \emph {et~al.}(2017)\citenamefont {{Carr}}, \citenamefont {{Raidal}}, \citenamefont {{Tenkanen}}, \citenamefont {{Vaskonen}},\ and\ \citenamefont {{Veerm{\"a}e}}}]{2017PhRvD..96b3514C}%
  \BibitemOpen
  \bibfield  {author} {\bibinfo {author} {\bibfnamefont {B.}~\bibnamefont {{Carr}}}, \bibinfo {author} {\bibfnamefont {M.}~\bibnamefont {{Raidal}}}, \bibinfo {author} {\bibfnamefont {T.}~\bibnamefont {{Tenkanen}}}, \bibinfo {author} {\bibfnamefont {V.}~\bibnamefont {{Vaskonen}}},\ and\ \bibinfo {author} {\bibfnamefont {H.}~\bibnamefont {{Veerm{\"a}e}}},\ }\bibfield  {title} {\bibinfo {title} {{Primordial black hole constraints for extended mass functions}},\ }\href {https://doi.org/10.1103/PhysRevD.96.023514} {\bibfield  {journal} {\bibinfo  {journal} {\prd}\ }\textbf {\bibinfo {volume} {96}},\ \bibinfo {eid} {023514} (\bibinfo {year} {2017})},\ \Eprint {https://arxiv.org/abs/1705.05567} {arXiv:1705.05567 [astro-ph.CO]} \BibitemShut {NoStop}%
\bibitem [{\citenamefont {{Frankel}}\ \emph {et~al.}(2018)\citenamefont {{Frankel}}, \citenamefont {{Rix}}, \citenamefont {{Ting}}, \citenamefont {{Ness}},\ and\ \citenamefont {{Hogg}}}]{2018ApJ...865...96F}%
  \BibitemOpen
  \bibfield  {author} {\bibinfo {author} {\bibfnamefont {N.}~\bibnamefont {{Frankel}}}, \bibinfo {author} {\bibfnamefont {H.-W.}\ \bibnamefont {{Rix}}}, \bibinfo {author} {\bibfnamefont {Y.-S.}\ \bibnamefont {{Ting}}}, \bibinfo {author} {\bibfnamefont {M.}~\bibnamefont {{Ness}}},\ and\ \bibinfo {author} {\bibfnamefont {D.~W.}\ \bibnamefont {{Hogg}}},\ }\bibfield  {title} {\bibinfo {title} {{Measuring Radial Orbit Migration in the Galactic Disk}},\ }\href {https://doi.org/10.3847/1538-4357/aadba5} {\bibfield  {journal} {\bibinfo  {journal} {\apj}\ }\textbf {\bibinfo {volume} {865}},\ \bibinfo {eid} {96} (\bibinfo {year} {2018})},\ \Eprint {https://arxiv.org/abs/1805.09198} {arXiv:1805.09198 [astro-ph.GA]} \BibitemShut {NoStop}%
\bibitem [{\citenamefont {{Johnson}}\ \emph {et~al.}(2021)\citenamefont {{Johnson}}, \citenamefont {{Weinberg}}, \citenamefont {{Vincenzo}}, \citenamefont {{Bird}}, \citenamefont {{Loebman}}, \citenamefont {{Brooks}}, \citenamefont {{Quinn}}, \citenamefont {{Christensen}},\ and\ \citenamefont {{Griffith}}}]{2021MNRAS.508.4484J}%
  \BibitemOpen
  \bibfield  {author} {\bibinfo {author} {\bibfnamefont {J.~W.}\ \bibnamefont {{Johnson}}}, \bibinfo {author} {\bibfnamefont {D.~H.}\ \bibnamefont {{Weinberg}}}, \bibinfo {author} {\bibfnamefont {F.}~\bibnamefont {{Vincenzo}}}, \bibinfo {author} {\bibfnamefont {J.~C.}\ \bibnamefont {{Bird}}}, \bibinfo {author} {\bibfnamefont {S.~R.}\ \bibnamefont {{Loebman}}}, \bibinfo {author} {\bibfnamefont {A.~M.}\ \bibnamefont {{Brooks}}}, \bibinfo {author} {\bibfnamefont {T.~R.}\ \bibnamefont {{Quinn}}}, \bibinfo {author} {\bibfnamefont {C.~R.}\ \bibnamefont {{Christensen}}},\ and\ \bibinfo {author} {\bibfnamefont {E.~J.}\ \bibnamefont {{Griffith}}},\ }\bibfield  {title} {\bibinfo {title} {{Stellar migration and chemical enrichment in the milky way disc: a hybrid model}},\ }\href {https://doi.org/10.1093/mnras/stab2718} {\bibfield  {journal} {\bibinfo  {journal} {\mnras}\ }\textbf {\bibinfo {volume} {508}},\ \bibinfo {pages} {4484} (\bibinfo {year} {2021})},\ \Eprint {https://arxiv.org/abs/2103.09838} {arXiv:2103.09838
  [astro-ph.GA]} \BibitemShut {NoStop}%
\bibitem [{\citenamefont {{Lunnan}}\ \emph {et~al.}(2017)\citenamefont {{Lunnan}}, \citenamefont {{Kasliwal}}, \citenamefont {{Cao}}, \citenamefont {{Hangard}}, \citenamefont {{Yaron}}, \citenamefont {{Parrent}}, \citenamefont {{McCully}}, \citenamefont {{Gal-Yam}}, \citenamefont {{Mulchaey}}, \citenamefont {{Ben-Ami}} \emph {et~al.}}]{2017ApJ...836...60Lmax10}%
  \BibitemOpen
  \bibfield  {author} {\bibinfo {author} {\bibfnamefont {R.}~\bibnamefont {{Lunnan}}}, \bibinfo {author} {\bibfnamefont {M.~M.}\ \bibnamefont {{Kasliwal}}}, \bibinfo {author} {\bibfnamefont {Y.}~\bibnamefont {{Cao}}}, \bibinfo {author} {\bibfnamefont {L.}~\bibnamefont {{Hangard}}}, \bibinfo {author} {\bibfnamefont {O.}~\bibnamefont {{Yaron}}}, \bibinfo {author} {\bibfnamefont {J.~T.}\ \bibnamefont {{Parrent}}}, \bibinfo {author} {\bibfnamefont {C.}~\bibnamefont {{McCully}}}, \bibinfo {author} {\bibfnamefont {A.}~\bibnamefont {{Gal-Yam}}}, \bibinfo {author} {\bibfnamefont {J.~S.}\ \bibnamefont {{Mulchaey}}}, \bibinfo {author} {\bibfnamefont {S.}~\bibnamefont {{Ben-Ami}}}, \emph {et~al.},\ }\bibfield  {title} {\bibinfo {title} {{Two New Calcium-rich Gap Transients in Group and Cluster Environments}},\ }\href {https://doi.org/10.3847/1538-4357/836/1/6010.48550/arXiv.1612.00454} {\bibfield  {journal} {\bibinfo  {journal} {\apj}\ }\textbf {\bibinfo {volume} {836}},\ \bibinfo {eid} {60} (\bibinfo {year} {2017})},\
  \Eprint {https://arxiv.org/abs/1612.00454} {arXiv:1612.00454 [astro-ph.HE]} \BibitemShut {NoStop}%
\bibitem [{\citenamefont {{Hill}}\ \emph {et~al.}(2018)\citenamefont {{Hill}}, \citenamefont {{Shariff}}, \citenamefont {{Trotta}}, \citenamefont {{Ali-Khan}}, \citenamefont {{Jiao}}, \citenamefont {{Liu}}, \citenamefont {{Moon}}, \citenamefont {{Parker}}, \citenamefont {{Paulus}}, \citenamefont {{van Dyk}},\ and\ \citenamefont {{Lucy}}}]{2018MNRAS.481.2766H}%
  \BibitemOpen
  \bibfield  {author} {\bibinfo {author} {\bibfnamefont {R.}~\bibnamefont {{Hill}}}, \bibinfo {author} {\bibfnamefont {H.}~\bibnamefont {{Shariff}}}, \bibinfo {author} {\bibfnamefont {R.}~\bibnamefont {{Trotta}}}, \bibinfo {author} {\bibfnamefont {S.}~\bibnamefont {{Ali-Khan}}}, \bibinfo {author} {\bibfnamefont {X.}~\bibnamefont {{Jiao}}}, \bibinfo {author} {\bibfnamefont {Y.}~\bibnamefont {{Liu}}}, \bibinfo {author} {\bibfnamefont {S.~K.}\ \bibnamefont {{Moon}}}, \bibinfo {author} {\bibfnamefont {W.}~\bibnamefont {{Parker}}}, \bibinfo {author} {\bibfnamefont {M.}~\bibnamefont {{Paulus}}}, \bibinfo {author} {\bibfnamefont {D.~A.}\ \bibnamefont {{van Dyk}}},\ and\ \bibinfo {author} {\bibfnamefont {L.~B.}\ \bibnamefont {{Lucy}}},\ }\bibfield  {title} {\bibinfo {title} {{Projected distances to host galaxy reduce SNIa dispersion}},\ }\href {https://doi.org/10.1093/mnras/sty2510} {\bibfield  {journal} {\bibinfo  {journal} {\mnras}\ }\textbf {\bibinfo {volume} {481}},\ \bibinfo {pages} {2766} (\bibinfo {year}
  {2018})},\ \Eprint {https://arxiv.org/abs/1612.04417} {arXiv:1612.04417 [astro-ph.GA]} \BibitemShut {NoStop}%
\bibitem [{\citenamefont {{De}}\ \emph {et~al.}(2020)\citenamefont {{De}}, \citenamefont {{Kasliwal}}, \citenamefont {{Tzanidakis}}, \citenamefont {{Fremling}}, \citenamefont {{Adams}}, \citenamefont {{Aloisi}}, \citenamefont {{Andreoni}}, \citenamefont {{Bagdasaryan}}, \citenamefont {{Bellm}}, \citenamefont {{Bildsten}} \emph {et~al.}}]{2020ApJ...905...58Dmax10}%
  \BibitemOpen
  \bibfield  {author} {\bibinfo {author} {\bibfnamefont {K.}~\bibnamefont {{De}}}, \bibinfo {author} {\bibfnamefont {M.~M.}\ \bibnamefont {{Kasliwal}}}, \bibinfo {author} {\bibfnamefont {A.}~\bibnamefont {{Tzanidakis}}}, \bibinfo {author} {\bibfnamefont {U.~C.}\ \bibnamefont {{Fremling}}}, \bibinfo {author} {\bibfnamefont {S.}~\bibnamefont {{Adams}}}, \bibinfo {author} {\bibfnamefont {R.}~\bibnamefont {{Aloisi}}}, \bibinfo {author} {\bibfnamefont {I.}~\bibnamefont {{Andreoni}}}, \bibinfo {author} {\bibfnamefont {A.}~\bibnamefont {{Bagdasaryan}}}, \bibinfo {author} {\bibfnamefont {E.~C.}\ \bibnamefont {{Bellm}}}, \bibinfo {author} {\bibfnamefont {L.}~\bibnamefont {{Bildsten}}}, \emph {et~al.},\ }\bibfield  {title} {\bibinfo {title} {{The Zwicky Transient Facility Census of the Local Universe. I. Systematic Search for Calcium-rich Gap Transients Reveals Three Related Spectroscopic Subclasses}},\ }\href {https://doi.org/10.3847/1538-4357/abb45c} {\bibfield  {journal} {\bibinfo  {journal} {\apj}\ }\textbf
  {\bibinfo {volume} {905}},\ \bibinfo {eid} {58} (\bibinfo {year} {2020})},\ \Eprint {https://arxiv.org/abs/2004.09029} {arXiv:2004.09029 [astro-ph.HE]} \BibitemShut {NoStop}%
\bibitem [{\citenamefont {{Marquardt}}\ \emph {et~al.}(2015)\citenamefont {{Marquardt}}, \citenamefont {{Sim}}, \citenamefont {{Ruiter}}, \citenamefont {{Seitenzahl}}, \citenamefont {{Ohlmann}}, \citenamefont {{Kromer}}, \citenamefont {{Pakmor}},\ and\ \citenamefont {{R{\"o}pke}}}]{2015A&A...580A.118M}%
  \BibitemOpen
  \bibfield  {author} {\bibinfo {author} {\bibfnamefont {K.~S.}\ \bibnamefont {{Marquardt}}}, \bibinfo {author} {\bibfnamefont {S.~A.}\ \bibnamefont {{Sim}}}, \bibinfo {author} {\bibfnamefont {A.~J.}\ \bibnamefont {{Ruiter}}}, \bibinfo {author} {\bibfnamefont {I.~R.}\ \bibnamefont {{Seitenzahl}}}, \bibinfo {author} {\bibfnamefont {S.~T.}\ \bibnamefont {{Ohlmann}}}, \bibinfo {author} {\bibfnamefont {M.}~\bibnamefont {{Kromer}}}, \bibinfo {author} {\bibfnamefont {R.}~\bibnamefont {{Pakmor}}},\ and\ \bibinfo {author} {\bibfnamefont {F.~K.}\ \bibnamefont {{R{\"o}pke}}},\ }\bibfield  {title} {\bibinfo {title} {{Type Ia supernovae from exploding oxygen-neon white dwarfs}},\ }\href {https://doi.org/10.1051/0004-6361/201525761} {\bibfield  {journal} {\bibinfo  {journal} {\aap}\ }\textbf {\bibinfo {volume} {580}},\ \bibinfo {eid} {A118} (\bibinfo {year} {2015})},\ \Eprint {https://arxiv.org/abs/1506.05809} {arXiv:1506.05809 [astro-ph.SR]} \BibitemShut {NoStop}%
\bibitem [{\citenamefont {{Strolger}}\ \emph {et~al.}(2020)\citenamefont {{Strolger}}, \citenamefont {{Rodney}}, \citenamefont {{Pacifici}}, \citenamefont {{Narayan}},\ and\ \citenamefont {{Graur}}}]{2020ApJ...890..140S}%
  \BibitemOpen
  \bibfield  {author} {\bibinfo {author} {\bibfnamefont {L.-G.}\ \bibnamefont {{Strolger}}}, \bibinfo {author} {\bibfnamefont {S.~A.}\ \bibnamefont {{Rodney}}}, \bibinfo {author} {\bibfnamefont {C.}~\bibnamefont {{Pacifici}}}, \bibinfo {author} {\bibfnamefont {G.}~\bibnamefont {{Narayan}}},\ and\ \bibinfo {author} {\bibfnamefont {O.}~\bibnamefont {{Graur}}},\ }\bibfield  {title} {\bibinfo {title} {{Delay Time Distributions of Type Ia Supernovae from Galaxy and Cosmic Star Formation Histories}},\ }\href {https://doi.org/10.3847/1538-4357/ab6a97} {\bibfield  {journal} {\bibinfo  {journal} {\apj}\ }\textbf {\bibinfo {volume} {890}},\ \bibinfo {eid} {140} (\bibinfo {year} {2020})},\ \Eprint {https://arxiv.org/abs/2001.05967} {arXiv:2001.05967 [astro-ph.GA]} \BibitemShut {NoStop}%
\bibitem [{\citenamefont {{Wiseman}}\ \emph {et~al.}(2020)\citenamefont {{Wiseman}}, \citenamefont {{Smith}}, \citenamefont {{Childress}}, \citenamefont {{Kelsey}}, \citenamefont {{M{\"o}ller}}, \citenamefont {{Gupta}}, \citenamefont {{Swann}}, \citenamefont {{Angus}}, \citenamefont {{Brout}}, \citenamefont {{Davis}} \emph {et~al.}}]{2020MNRAS.495.4040Wmax10}%
  \BibitemOpen
  \bibfield  {author} {\bibinfo {author} {\bibfnamefont {P.}~\bibnamefont {{Wiseman}}}, \bibinfo {author} {\bibfnamefont {M.}~\bibnamefont {{Smith}}}, \bibinfo {author} {\bibfnamefont {M.}~\bibnamefont {{Childress}}}, \bibinfo {author} {\bibfnamefont {L.}~\bibnamefont {{Kelsey}}}, \bibinfo {author} {\bibfnamefont {A.}~\bibnamefont {{M{\"o}ller}}}, \bibinfo {author} {\bibfnamefont {R.~R.}\ \bibnamefont {{Gupta}}}, \bibinfo {author} {\bibfnamefont {E.}~\bibnamefont {{Swann}}}, \bibinfo {author} {\bibfnamefont {C.~R.}\ \bibnamefont {{Angus}}}, \bibinfo {author} {\bibfnamefont {D.}~\bibnamefont {{Brout}}}, \bibinfo {author} {\bibfnamefont {T.~M.}\ \bibnamefont {{Davis}}}, \emph {et~al.},\ }\bibfield  {title} {\bibinfo {title} {{Supernova host galaxies in the dark energy survey: I. Deep coadds, photometry, and stellar masses}},\ }\href {https://doi.org/10.1093/mnras/staa1302} {\bibfield  {journal} {\bibinfo  {journal} {\mnras}\ }\textbf {\bibinfo {volume} {495}},\ \bibinfo {pages} {4040} (\bibinfo {year}
  {2020})},\ \Eprint {https://arxiv.org/abs/2001.02640} {arXiv:2001.02640 [astro-ph.GA]} \BibitemShut {NoStop}%
\bibitem [{\citenamefont {{Behroozi}}\ \emph {et~al.}(2013{\natexlab{a}})\citenamefont {{Behroozi}}, \citenamefont {{Wechsler}},\ and\ \citenamefont {{Conroy}}}]{2013ApJ...770...57B}%
  \BibitemOpen
  \bibfield  {author} {\bibinfo {author} {\bibfnamefont {P.~S.}\ \bibnamefont {{Behroozi}}}, \bibinfo {author} {\bibfnamefont {R.~H.}\ \bibnamefont {{Wechsler}}},\ and\ \bibinfo {author} {\bibfnamefont {C.}~\bibnamefont {{Conroy}}},\ }\bibfield  {title} {\bibinfo {title} {{The Average Star Formation Histories of Galaxies in Dark Matter Halos from z = 0-8}},\ }\href {https://doi.org/10.1088/0004-637X/770/1/57} {\bibfield  {journal} {\bibinfo  {journal} {\apj}\ }\textbf {\bibinfo {volume} {770}},\ \bibinfo {eid} {57} (\bibinfo {year} {2013}{\natexlab{a}})}\BibitemShut {NoStop}%
\bibitem [{\citenamefont {{Behroozi}}\ \emph {et~al.}(2013{\natexlab{b}})\citenamefont {{Behroozi}}, \citenamefont {{Wechsler}},\ and\ \citenamefont {{Conroy}}}]{2013ApJ...762L..31B}%
  \BibitemOpen
  \bibfield  {author} {\bibinfo {author} {\bibfnamefont {P.~S.}\ \bibnamefont {{Behroozi}}}, \bibinfo {author} {\bibfnamefont {R.~H.}\ \bibnamefont {{Wechsler}}},\ and\ \bibinfo {author} {\bibfnamefont {C.}~\bibnamefont {{Conroy}}},\ }\bibfield  {title} {\bibinfo {title} {{On the Lack of Evolution in Galaxy Star Formation Efficiency}},\ }\href {https://doi.org/10.1088/2041-8205/762/2/L31} {\bibfield  {journal} {\bibinfo  {journal} {\apjl}\ }\textbf {\bibinfo {volume} {762}},\ \bibinfo {eid} {L31} (\bibinfo {year} {2013}{\natexlab{b}})},\ \Eprint {https://arxiv.org/abs/1209.3013} {arXiv:1209.3013 [astro-ph.CO]} \BibitemShut {NoStop}%
\bibitem [{\citenamefont {{Leitner}}\ and\ \citenamefont {{Kravtsov}}(2011)}]{2011ApJ...734...48L}%
  \BibitemOpen
  \bibfield  {author} {\bibinfo {author} {\bibfnamefont {S.~N.}\ \bibnamefont {{Leitner}}}\ and\ \bibinfo {author} {\bibfnamefont {A.~V.}\ \bibnamefont {{Kravtsov}}},\ }\bibfield  {title} {\bibinfo {title} {{Fuel Efficient Galaxies: Sustaining Star Formation with Stellar Mass Loss}},\ }\href {https://doi.org/10.1088/0004-637X/734/1/48} {\bibfield  {journal} {\bibinfo  {journal} {\apj}\ }\textbf {\bibinfo {volume} {734}},\ \bibinfo {eid} {48} (\bibinfo {year} {2011})},\ \Eprint {https://arxiv.org/abs/1011.1252} {arXiv:1011.1252 [astro-ph.GA]} \BibitemShut {NoStop}%
\bibitem [{\citenamefont {{Childress}}\ \emph {et~al.}(2014)\citenamefont {{Childress}}, \citenamefont {{Wolf}},\ and\ \citenamefont {{Zahid}}}]{2014MNRAS.445.1898C}%
  \BibitemOpen
  \bibfield  {author} {\bibinfo {author} {\bibfnamefont {M.~J.}\ \bibnamefont {{Childress}}}, \bibinfo {author} {\bibfnamefont {C.}~\bibnamefont {{Wolf}}},\ and\ \bibinfo {author} {\bibfnamefont {H.~J.}\ \bibnamefont {{Zahid}}},\ }\bibfield  {title} {\bibinfo {title} {{Ages of Type Ia supernovae over cosmic time}},\ }\href {https://doi.org/10.1093/mnras/stu1892} {\bibfield  {journal} {\bibinfo  {journal} {\mnras}\ }\textbf {\bibinfo {volume} {445}},\ \bibinfo {pages} {1898} (\bibinfo {year} {2014})},\ \Eprint {https://arxiv.org/abs/1409.2951} {arXiv:1409.2951 [astro-ph.CO]} \BibitemShut {NoStop}%
\bibitem [{\citenamefont {{Marks}}\ \emph {et~al.}(2012)\citenamefont {{Marks}}, \citenamefont {{Kroupa}}, \citenamefont {{Dabringhausen}},\ and\ \citenamefont {{Pawlowski}}}]{2012MNRAS.422.2246M}%
  \BibitemOpen
  \bibfield  {author} {\bibinfo {author} {\bibfnamefont {M.}~\bibnamefont {{Marks}}}, \bibinfo {author} {\bibfnamefont {P.}~\bibnamefont {{Kroupa}}}, \bibinfo {author} {\bibfnamefont {J.}~\bibnamefont {{Dabringhausen}}},\ and\ \bibinfo {author} {\bibfnamefont {M.~S.}\ \bibnamefont {{Pawlowski}}},\ }\bibfield  {title} {\bibinfo {title} {{Evidence for top-heavy stellar initial mass functions with increasing density and decreasing metallicity}},\ }\href {https://doi.org/10.1111/j.1365-2966.2012.20767.x} {\bibfield  {journal} {\bibinfo  {journal} {\mnras}\ }\textbf {\bibinfo {volume} {422}},\ \bibinfo {pages} {2246} (\bibinfo {year} {2012})},\ \Eprint {https://arxiv.org/abs/1202.4755} {arXiv:1202.4755 [astro-ph.GA]} \BibitemShut {NoStop}%
\bibitem [{\citenamefont {{Yan}}\ \emph {et~al.}(2020)\citenamefont {{Yan}}, \citenamefont {{Jerabkova}},\ and\ \citenamefont {{Kroupa}}}]{2020A&A...637A..68Y}%
  \BibitemOpen
  \bibfield  {author} {\bibinfo {author} {\bibfnamefont {Z.}~\bibnamefont {{Yan}}}, \bibinfo {author} {\bibfnamefont {T.}~\bibnamefont {{Jerabkova}}},\ and\ \bibinfo {author} {\bibfnamefont {P.}~\bibnamefont {{Kroupa}}},\ }\bibfield  {title} {\bibinfo {title} {{Chemical evolution of ultra-faint dwarf galaxies in the self-consistently calculated integrated galactic IMF theory}},\ }\href {https://doi.org/10.1051/0004-6361/202037567} {\bibfield  {journal} {\bibinfo  {journal} {\aap}\ }\textbf {\bibinfo {volume} {637}},\ \bibinfo {eid} {A68} (\bibinfo {year} {2020})},\ \Eprint {https://arxiv.org/abs/2003.11029} {arXiv:2003.11029 [astro-ph.GA]} \BibitemShut {NoStop}%
\bibitem [{\citenamefont {{Kroupa}}\ and\ \citenamefont {{Weidner}}(2003)}]{2003ApJ...598.1076K}%
  \BibitemOpen
  \bibfield  {author} {\bibinfo {author} {\bibfnamefont {P.}~\bibnamefont {{Kroupa}}}\ and\ \bibinfo {author} {\bibfnamefont {C.}~\bibnamefont {{Weidner}}},\ }\bibfield  {title} {\bibinfo {title} {{Galactic-Field Initial Mass Functions of Massive Stars}},\ }\href {https://doi.org/10.1086/379105} {\bibfield  {journal} {\bibinfo  {journal} {\apj}\ }\textbf {\bibinfo {volume} {598}},\ \bibinfo {pages} {1076} (\bibinfo {year} {2003})},\ \Eprint {https://arxiv.org/abs/astro-ph/0308356} {arXiv:astro-ph/0308356 [astro-ph]} \BibitemShut {NoStop}%
\bibitem [{\citenamefont {{Weidner}}\ \emph {et~al.}(2013)\citenamefont {{Weidner}}, \citenamefont {{Kroupa}}, \citenamefont {{Pflamm-Altenburg}},\ and\ \citenamefont {{Vazdekis}}}]{2013MNRAS.436.3309W}%
  \BibitemOpen
  \bibfield  {author} {\bibinfo {author} {\bibfnamefont {C.}~\bibnamefont {{Weidner}}}, \bibinfo {author} {\bibfnamefont {P.}~\bibnamefont {{Kroupa}}}, \bibinfo {author} {\bibfnamefont {J.}~\bibnamefont {{Pflamm-Altenburg}}},\ and\ \bibinfo {author} {\bibfnamefont {A.}~\bibnamefont {{Vazdekis}}},\ }\bibfield  {title} {\bibinfo {title} {{The galaxy-wide initial mass function of dwarf late-type to massive early-type galaxies}},\ }\href {https://doi.org/10.1093/mnras/stt1806} {\bibfield  {journal} {\bibinfo  {journal} {\mnras}\ }\textbf {\bibinfo {volume} {436}},\ \bibinfo {pages} {3309} (\bibinfo {year} {2013})},\ \Eprint {https://arxiv.org/abs/1309.6634} {arXiv:1309.6634 [astro-ph.CO]} \BibitemShut {NoStop}%
\bibitem [{\citenamefont {{Fontanot}}\ \emph {et~al.}(2017)\citenamefont {{Fontanot}}, \citenamefont {{De Lucia}}, \citenamefont {{Hirschmann}}, \citenamefont {{Bruzual}}, \citenamefont {{Charlot}},\ and\ \citenamefont {{Zibetti}}}]{2017MNRAS.464.3812F}%
  \BibitemOpen
  \bibfield  {author} {\bibinfo {author} {\bibfnamefont {F.}~\bibnamefont {{Fontanot}}}, \bibinfo {author} {\bibfnamefont {G.}~\bibnamefont {{De Lucia}}}, \bibinfo {author} {\bibfnamefont {M.}~\bibnamefont {{Hirschmann}}}, \bibinfo {author} {\bibfnamefont {G.}~\bibnamefont {{Bruzual}}}, \bibinfo {author} {\bibfnamefont {S.}~\bibnamefont {{Charlot}}},\ and\ \bibinfo {author} {\bibfnamefont {S.}~\bibnamefont {{Zibetti}}},\ }\bibfield  {title} {\bibinfo {title} {{Variations of the stellar initial mass function in semi-analytical models: implications for the mass assembly and the chemical enrichment of galaxies in the GAEA model}},\ }\href {https://doi.org/10.1093/mnras/stw2612} {\bibfield  {journal} {\bibinfo  {journal} {\mnras}\ }\textbf {\bibinfo {volume} {464}},\ \bibinfo {pages} {3812} (\bibinfo {year} {2017})},\ \Eprint {https://arxiv.org/abs/1606.01908} {arXiv:1606.01908 [astro-ph.GA]} \BibitemShut {NoStop}%
\bibitem [{Note4()}]{Note4}%
  \BibitemOpen
  \bibinfo {note} {We calculated the effects of non-universal IMFs on the WD mass function, but not yet on SFHs. We are not aware of published SFH models assuming non-universal IMFs.}\BibitemShut {Stop}%
\bibitem [{\citenamefont {{Umeda}}\ \emph {et~al.}(1999)\citenamefont {{Umeda}}, \citenamefont {{Nomoto}}, \citenamefont {{Yamaoka}},\ and\ \citenamefont {{Wanajo}}}]{1999ApJ...513..861U}%
  \BibitemOpen
  \bibfield  {author} {\bibinfo {author} {\bibfnamefont {H.}~\bibnamefont {{Umeda}}}, \bibinfo {author} {\bibfnamefont {K.}~\bibnamefont {{Nomoto}}}, \bibinfo {author} {\bibfnamefont {H.}~\bibnamefont {{Yamaoka}}},\ and\ \bibinfo {author} {\bibfnamefont {S.}~\bibnamefont {{Wanajo}}},\ }\bibfield  {title} {\bibinfo {title} {{Evolution of 3-9 M$_{solar}$ Stars for Z=0.001-0.03 and Metallicity Effects on Type Ia Supernovae}},\ }\href {https://doi.org/10.1086/306887} {\bibfield  {journal} {\bibinfo  {journal} {\apj}\ }\textbf {\bibinfo {volume} {513}},\ \bibinfo {pages} {861} (\bibinfo {year} {1999})},\ \Eprint {https://arxiv.org/abs/astro-ph/9806336} {arXiv:astro-ph/9806336 [astro-ph]} \BibitemShut {NoStop}%
\bibitem [{\citenamefont {{Zahid}}\ \emph {et~al.}(2014)\citenamefont {{Zahid}}, \citenamefont {{Dima}}, \citenamefont {{Kudritzki}}, \citenamefont {{Kewley}}, \citenamefont {{Geller}}, \citenamefont {{Hwang}}, \citenamefont {{Silverman}},\ and\ \citenamefont {{Kashino}}}]{2014ApJ...791..130Z}%
  \BibitemOpen
  \bibfield  {author} {\bibinfo {author} {\bibfnamefont {H.~J.}\ \bibnamefont {{Zahid}}}, \bibinfo {author} {\bibfnamefont {G.~I.}\ \bibnamefont {{Dima}}}, \bibinfo {author} {\bibfnamefont {R.-P.}\ \bibnamefont {{Kudritzki}}}, \bibinfo {author} {\bibfnamefont {L.~J.}\ \bibnamefont {{Kewley}}}, \bibinfo {author} {\bibfnamefont {M.~J.}\ \bibnamefont {{Geller}}}, \bibinfo {author} {\bibfnamefont {H.~S.}\ \bibnamefont {{Hwang}}}, \bibinfo {author} {\bibfnamefont {J.~D.}\ \bibnamefont {{Silverman}}},\ and\ \bibinfo {author} {\bibfnamefont {D.}~\bibnamefont {{Kashino}}},\ }\bibfield  {title} {\bibinfo {title} {{The Universal Relation of Galactic Chemical Evolution: The Origin of the Mass-Metallicity Relation}},\ }\href {https://doi.org/10.1088/0004-637X/791/2/130} {\bibfield  {journal} {\bibinfo  {journal} {\apj}\ }\textbf {\bibinfo {volume} {791}},\ \bibinfo {eid} {130} (\bibinfo {year} {2014})},\ \Eprint {https://arxiv.org/abs/1404.7526} {arXiv:1404.7526 [astro-ph.GA]} \BibitemShut {NoStop}%
\bibitem [{\citenamefont {{Srivastav}}\ \emph {et~al.}(2022)\citenamefont {{Srivastav}}, \citenamefont {{Smartt}}, \citenamefont {{Huber}}, \citenamefont {{Chambers}}, \citenamefont {{Angus}}, \citenamefont {{Chen}}, \citenamefont {{Callan}}, \citenamefont {{Gillanders}}, \citenamefont {{McBrien}}, \citenamefont {{Sim}} \emph {et~al.}}]{2022MNRAS.511.2708Smax10}%
  \BibitemOpen
  \bibfield  {author} {\bibinfo {author} {\bibfnamefont {S.}~\bibnamefont {{Srivastav}}}, \bibinfo {author} {\bibfnamefont {S.~J.}\ \bibnamefont {{Smartt}}}, \bibinfo {author} {\bibfnamefont {M.~E.}\ \bibnamefont {{Huber}}}, \bibinfo {author} {\bibfnamefont {K.~C.}\ \bibnamefont {{Chambers}}}, \bibinfo {author} {\bibfnamefont {C.~R.}\ \bibnamefont {{Angus}}}, \bibinfo {author} {\bibfnamefont {T.~W.}\ \bibnamefont {{Chen}}}, \bibinfo {author} {\bibfnamefont {F.~P.}\ \bibnamefont {{Callan}}}, \bibinfo {author} {\bibfnamefont {J.~H.}\ \bibnamefont {{Gillanders}}}, \bibinfo {author} {\bibfnamefont {O.~R.}\ \bibnamefont {{McBrien}}}, \bibinfo {author} {\bibfnamefont {S.~A.}\ \bibnamefont {{Sim}}}, \emph {et~al.},\ }\bibfield  {title} {\bibinfo {title} {{SN 2020kyg and the rates of faint Iax supernovae from ATLAS}},\ }\href {https://doi.org/10.1093/mnras/stac177} {\bibfield  {journal} {\bibinfo  {journal} {\mnras}\ }\textbf {\bibinfo {volume} {511}},\ \bibinfo {pages} {2708} (\bibinfo {year} {2022})},\ \Eprint
  {https://arxiv.org/abs/2111.09491} {arXiv:2111.09491 [astro-ph.HE]} \BibitemShut {NoStop}%
\bibitem [{\citenamefont {{Takaro}}\ \emph {et~al.}(2020)\citenamefont {{Takaro}}, \citenamefont {{Foley}}, \citenamefont {{McCully}}, \citenamefont {{Fong}}, \citenamefont {{Jha}}, \citenamefont {{Narayan}}, \citenamefont {{Rest}}, \citenamefont {{Stritzinger}},\ and\ \citenamefont {{McKinnon}}}]{2020MNRAS.493..986T}%
  \BibitemOpen
  \bibfield  {author} {\bibinfo {author} {\bibfnamefont {T.}~\bibnamefont {{Takaro}}}, \bibinfo {author} {\bibfnamefont {R.~J.}\ \bibnamefont {{Foley}}}, \bibinfo {author} {\bibfnamefont {C.}~\bibnamefont {{McCully}}}, \bibinfo {author} {\bibfnamefont {W.-f.}\ \bibnamefont {{Fong}}}, \bibinfo {author} {\bibfnamefont {S.~W.}\ \bibnamefont {{Jha}}}, \bibinfo {author} {\bibfnamefont {G.}~\bibnamefont {{Narayan}}}, \bibinfo {author} {\bibfnamefont {A.}~\bibnamefont {{Rest}}}, \bibinfo {author} {\bibfnamefont {M.}~\bibnamefont {{Stritzinger}}},\ and\ \bibinfo {author} {\bibfnamefont {K.}~\bibnamefont {{McKinnon}}},\ }\bibfield  {title} {\bibinfo {title} {{Constraining Type Iax supernova progenitor systems with stellar population age dating}},\ }\href {https://doi.org/10.1093/mnras/staa294} {\bibfield  {journal} {\bibinfo  {journal} {\mnras}\ }\textbf {\bibinfo {volume} {493}},\ \bibinfo {pages} {986} (\bibinfo {year} {2020})},\ \Eprint {https://arxiv.org/abs/1901.05461} {arXiv:1901.05461 [astro-ph.HE]}
  \BibitemShut {NoStop}%
\bibitem [{\citenamefont {{Moll}}\ \emph {et~al.}(2014)\citenamefont {{Moll}}, \citenamefont {{Raskin}}, \citenamefont {{Kasen}},\ and\ \citenamefont {{Woosley}}}]{2014ApJ...785..105M}%
  \BibitemOpen
  \bibfield  {author} {\bibinfo {author} {\bibfnamefont {R.}~\bibnamefont {{Moll}}}, \bibinfo {author} {\bibfnamefont {C.}~\bibnamefont {{Raskin}}}, \bibinfo {author} {\bibfnamefont {D.}~\bibnamefont {{Kasen}}},\ and\ \bibinfo {author} {\bibfnamefont {S.~E.}\ \bibnamefont {{Woosley}}},\ }\bibfield  {title} {\bibinfo {title} {{Type Ia Supernovae from Merging White Dwarfs. I. Prompt Detonations}},\ }\href {https://doi.org/10.1088/0004-637X/785/2/105} {\bibfield  {journal} {\bibinfo  {journal} {\apj}\ }\textbf {\bibinfo {volume} {785}},\ \bibinfo {eid} {105} (\bibinfo {year} {2014})},\ \Eprint {https://arxiv.org/abs/1311.5008} {arXiv:1311.5008 [astro-ph.HE]} \BibitemShut {NoStop}%
\bibitem [{\citenamefont {{Bhalerao}}\ \emph {et~al.}(2022)\citenamefont {{Bhalerao}}, \citenamefont {{Sawant}}, \citenamefont {{Pai}}, \citenamefont {{Tendulkar}}, \citenamefont {{Vadawale}}, \citenamefont {{Bhattacharya}}, \citenamefont {{Rana}}, \citenamefont {{Adalja}}, \citenamefont {{Anupama}}, \citenamefont {{Bala}} \emph {et~al.}}]{2022arXiv221112052Bmax10}%
  \BibitemOpen
  \bibfield  {author} {\bibinfo {author} {\bibfnamefont {V.}~\bibnamefont {{Bhalerao}}}, \bibinfo {author} {\bibfnamefont {D.}~\bibnamefont {{Sawant}}}, \bibinfo {author} {\bibfnamefont {A.}~\bibnamefont {{Pai}}}, \bibinfo {author} {\bibfnamefont {S.}~\bibnamefont {{Tendulkar}}}, \bibinfo {author} {\bibfnamefont {S.}~\bibnamefont {{Vadawale}}}, \bibinfo {author} {\bibfnamefont {D.}~\bibnamefont {{Bhattacharya}}}, \bibinfo {author} {\bibfnamefont {V.}~\bibnamefont {{Rana}}}, \bibinfo {author} {\bibfnamefont {H.~K.~L.}\ \bibnamefont {{Adalja}}}, \bibinfo {author} {\bibfnamefont {G.~C.}\ \bibnamefont {{Anupama}}}, \bibinfo {author} {\bibfnamefont {S.}~\bibnamefont {{Bala}}}, \emph {et~al.},\ }\bibfield  {title} {\bibinfo {title} {{Science with the Daksha High Energy Transients Mission}},\ }\href {https://doi.org/10.48550/arXiv.2211.12052} {\bibfield  {journal} {\bibinfo  {journal} {arXiv e-prints}\ ,\ \bibinfo {eid} {arXiv:2211.12052}} (\bibinfo {year} {2022})},\ \Eprint {https://arxiv.org/abs/2211.12052}
  {arXiv:2211.12052 [astro-ph.HE]} \BibitemShut {NoStop}%
\bibitem [{\citenamefont {{Hui}}\ and\ \citenamefont {{MoonBEAM Team}}(2021)}]{2021AAS...23731502H}%
  \BibitemOpen
  \bibfield  {author} {\bibinfo {author} {\bibfnamefont {C.}~\bibnamefont {{Hui}}}\ and\ \bibinfo {author} {\bibnamefont {{MoonBEAM Team}}},\ }\bibfield  {title} {\bibinfo {title} {{MoonBEAM: A Beyond Earth-orbit Gamma-ray Burst Detector for Multi-Messenger Astronomy}},\ }in\ \href@noop {} {\emph {\bibinfo {booktitle} {American Astronomical Society Meeting Abstracts}}},\ \bibinfo {series} {American Astronomical Society Meeting Abstracts}, Vol.~\bibinfo {volume} {53}\ (\bibinfo {year} {2021})\ p.\ \bibinfo {pages} {315.02}\BibitemShut {NoStop}%
\bibitem [{\citenamefont {{Hui}}(2022)}]{2022HEAD...1930505H}%
  \BibitemOpen
  \bibfield  {author} {\bibinfo {author} {\bibfnamefont {M.}~\bibnamefont {{Hui}}},\ }\bibfield  {title} {\bibinfo {title} {{Moon Burst Energetics All-sky Monitor: A Beyond Earth-orbit Gamma-ray Burst Detector for Multi-Messenger Astronomy}},\ }in\ \href@noop {} {\emph {\bibinfo {booktitle} {AAS/High Energy Astrophysics Division}}},\ \bibinfo {series} {AAS/High Energy Astrophysics Division}, Vol.~\bibinfo {volume} {54}\ (\bibinfo {year} {2022})\ p.\ \bibinfo {pages} {305.05}\BibitemShut {NoStop}%
\bibitem [{\citenamefont {{Jung}}\ and\ \citenamefont {{Kim}}(2020)}]{2020PhRvR...2a3113J}%
  \BibitemOpen
  \bibfield  {author} {\bibinfo {author} {\bibfnamefont {S.}~\bibnamefont {{Jung}}}\ and\ \bibinfo {author} {\bibfnamefont {T.}~\bibnamefont {{Kim}}},\ }\bibfield  {title} {\bibinfo {title} {{Gamma-ray burst lensing parallax: Closing the primordial black hole dark matter mass window}},\ }\href {https://doi.org/10.1103/PhysRevResearch.2.013113} {\bibfield  {journal} {\bibinfo  {journal} {Physical Review Research}\ }\textbf {\bibinfo {volume} {2}},\ \bibinfo {eid} {013113} (\bibinfo {year} {2020})},\ \Eprint {https://arxiv.org/abs/1908.00078} {arXiv:1908.00078 [astro-ph.CO]} \BibitemShut {NoStop}%
\bibitem [{\citenamefont {{Gawade}}\ \emph {et~al.}(2024)\citenamefont {{Gawade}}, \citenamefont {{More}},\ and\ \citenamefont {{Bhalerao}}}]{2024MNRAS.527.3306G}%
  \BibitemOpen
  \bibfield  {author} {\bibinfo {author} {\bibfnamefont {P.}~\bibnamefont {{Gawade}}}, \bibinfo {author} {\bibfnamefont {S.}~\bibnamefont {{More}}},\ and\ \bibinfo {author} {\bibfnamefont {V.}~\bibnamefont {{Bhalerao}}},\ }\bibfield  {title} {\bibinfo {title} {{On the feasibility of primordial black hole abundance constraints using lensing parallax of GRBs}},\ }\href {https://doi.org/10.1093/mnras/stad3336} {\bibfield  {journal} {\bibinfo  {journal} {\mnras}\ }\textbf {\bibinfo {volume} {527}},\ \bibinfo {pages} {3306} (\bibinfo {year} {2024})},\ \Eprint {https://arxiv.org/abs/2308.01775} {arXiv:2308.01775 [astro-ph.CO]} \BibitemShut {NoStop}%
\end{thebibliography}%
